\pdfoutput=1

\def\displayArxivTexts{}

\def\enableLlncsCode{}

\ifdefined\displayArxivTexts\else
  \def\cameraReadyVersion{}
\fi

\def\enableLlncsCodeSlightlyImproved{}

\ifdefined\displayArxivTexts
  \def\nicerThanLlncs{}
\fi

\documentclass[\ifdefined\cameraReadyVersion runningheads,a4paper,\else\ifdefined\nicerThanLlncs ,11pt,a4paper,\fi\fi\ifdefined\enableLlncsCodeSlightlyImproved envcountsect, envcountsame\fi]{llncs}

\IfFileExists{personal_changes.tex}{\input{personal_changes}}{}
\input{include.tex}

\title{Oblivious Transfer from Zero-Knowledge Proofs}
\subtitle{Or How to Achieve Round-Optimal Quantum Oblivious Transfer and Zero-Knowledge Proofs on Quantum States\footnote{\textcopyright{} IACR 2023. This article is the full version of the work published by Springer-Verlag (ASIACRYPT 2023).}}
\usepackage{orcidlink}
\author{
  \anonymizedVsNot{}{%
    Léo Colisson\,\scalebox{.7}{\faIcon{envelope}}\,\orcidlink{0000-0001-8963-4656}\hspace*{0.05em}\inst{1,3}, %
    Garazi Muguruza\inst{2,3}, %
    Florian Speelman\,\orcidlink{0000-0003-3792-9908}\hspace*{0.05em}\inst{2,3}\\%
    \href{mailto:leo.colisson@cwi.nl}{leo.colisson@cwi.nl}, \href{mailto:g.muguruzalasa@uva.nl}{g.muguruzalasa@uva.nl}, \href{mailto:f.speelman@uva.nl}{f.speelman@uva.nl}
  }
}
\institute{
  \anonymizedVsNot{}{%
    Centrum Wiskunde \& Informatica, Netherlands
    \and
    Informatics Institute, University of Amsterdam, Netherlands
    \and
    QuSoft, Netherlands
  }%
}

\date{}

\begin{document}
{\def\addcontentsline#1#2#3{}\maketitle} 

\begin{abstract}
  We provide a generic construction to turn any classical Zero-Knowledge (ZK) protocol into a composable (quantum) oblivious transfer (OT) protocol, mostly lifting the round-complexity properties and security guarantees (plain-model/statistical security/unstructured functions…) of the ZK protocol to the resulting OT protocol. Such a construction is unlikely to exist classically as Cryptomania is believed to be different from Minicrypt.\\

  In particular, by instantiating our construction using Non-Interactive ZK (NIZK), we provide the first round-optimal (2-message) quantum OT protocol secure in the random oracle model, and round-optimal extensions to string and $k$-out-of-$n$ OT.\\
  
  At the heart of our construction lies a new method that allows us to prove properties on a received quantum state without revealing additional information on it, even in a non-interactive way, without public-key primitives, and/or with statistical guarantees when using an appropriate classical ZK protocol. We can notably prove that a state has been partially measured (with arbitrary constraints on the set of measured qubits), without revealing any additional information on this set. This notion can be seen as an analog of ZK to quantum states, and we expect it to be of independent interest as it extends complexity theory to \emph{quantum} languages, as illustrated by the two new complexity classes we introduce, $\ZKstatesQIP{}{}$ and $\ZKstatesQMA{}$.
\end{abstract}
\keywords{Quantum Cryptography, Oblivious Transfer, Zero-Knowledge on Quantum States, Multi-Party Computing, Zero-Knowledge}
\arxivOnly{\newpage \tableofcontents \newpage}

\cameraVsOther{
  \pratendSetGlobal{text link={The proof in the full version~\cite{CMS23_ObliviousTransferZeroKnowledge}.}}
}{}
\section{Introduction}
\pratendSetLocal{category=preliminaries}

Oblivious Transfer (OT) is an extremely powerful primitive, as it was shown~\cite{Kil88_FoundingCrytpographyOblivious} to be sufficient to perform multi-party computing (MPC), allowing multiple parties to jointly compute any function while keeping the input of each party secret.
Since the introduction of $2$-party computing in the seminal article of Yao~\cite{Yao82_ProtocolsSecureComputations}, followed by the famous generalisation to arbitrary many parties of Goldreich, Micali and Wigderson~\cite{GMW87_HowPlayANY}, OT and MPC received a tremendous amount of attention~\cameraVsOther{%
  \cite{Wie83_ConjugateCoding,PVW08_FrameworkEfficientComposable,Rab05_HowExchangeSecrets,EGL85_RandomizedProtocolSigning,CGS02_SecureMultipartyQuantum,DGJ+20_SecureMultipartyQuantum,KP17_MultipartyDelegatedQuantum,LT22_ReviewStateArt}%
}{%
  \cite{Wie83_ConjugateCoding,PVW08_FrameworkEfficientComposable,Rab05_HowExchangeSecrets,EGL85_RandomizedProtocolSigning,CGS02_SecureMultipartyQuantum,DGJ+20_SecureMultipartyQuantum,KP17_MultipartyDelegatedQuantum,LT22_ReviewStateArt,YAVV22_SurveyObliviousTransfer}%
}.

However, all classical OT protocols need to use some structured computational assumptions providing trapdoors. Said differently, OT (classically) lives in \emph{Cryptomania}~\cite{Imp95_PersonalViewAveragecase}, a world where public-key cryptography exists. On the other hand, it was recently shown~\cite{GLSV21_ObliviousTransferMiniQCrypt,BCKM21_OneWayFunctionsImply} that quantumly, OT lives in \emph{MiniQCrypt}, meaning that it is possible to obtain OT protocols using a much weaker assumption, based only on (unstructured) one-way functions.

There are many reasons to avoid using trapdoor functions. For instance, this additional structure can often be exploited by quantum computers, leading to attacks. As a result, many OT protocols (based on RSA, quadratic residue, elliptic curves…) are vulnerable against quantum adversaries. While some proposals~\cite{PVW08_FrameworkEfficientComposable,BD18_TwoMessageStatisticallySenderPrivate,Qua20_UCSecureOTLWE} based on post-quantum assumptions like the Learning-With-Errors problem (\LWE{}) still seem to resist against quantum adversaries, minimizing assumptions is an important safety-guard against potential future attacks on the computational assumptions. Understanding the minimal required assumptions is also an active field of research, with the recent introduction of the notion of pseudo-random quantum states~\cite{JLS18_PseudorandomQuantumStates}, which is an even weaker assumption than one-way functions.

However, while we know (even classical) $2$-message OT protocols\===optimal in term of round complexity\===achievable using trapdoors~\cite{PVW08_FrameworkEfficientComposable,BD18_TwoMessageStatisticallySenderPrivate}, there is no known round-optimal protocol requiring no structure (such protocol would necessary be quantum unless Cryptomania collapses to MiniCrypt). The original proposal~\cite{CK88_AchievingObliviousTransfer} for quantum OT (studied and improved in a long line of research~\cite{BBCS92_PracticalQuantumOblivious,MS94_QuantumObliviousTransfer,Yao95_SecurityQuantumProtocols,DFL+09_ImprovingSecurityQuantum,Unr10_UniversallyComposableQuantum,BF10_SamplingQuantumPopulation,GLSV21_ObliviousTransferMiniQCrypt,BCKM21_OneWayFunctionsImply}, see also this review~\cite{SMP22_QuantumObliviousTransfer} for quantum OT protocols based on physical assumptions, that we will not cover here) requires $7$ messages, and~\cite{ABKK22_NewFrameworkQuantum} managed to obtain a $3$-message protocol (computationally secure, in the random oracle model). However, they left the following question open:

\begin{center}
  \emph{Does there exist two-message quantum chosen-input bit OT, that allows both parties to choose inputs?}
\end{center}

They also raise the question of the existence of a $2$-message string OT, even when the bit chosen by the receiver is random. The main bottle-neck to further reduce the communication complexity of these protocols is the use of a ``cut-and-choose'' approach, where the receiver sends a quantum state and some commitments on the description of this state, gets a challenge from the sender to ensure that the quantum states were honestly prepared, and opens some commitments. Classically, we can avoid cut-and-choose by using Non-Interactive Zero-Knowledge proofs (NIZK) in order to prove an \NP{} statement on a classical string without revealing anything on that string except the fact that the statement is true. However, defining NIZK proofs on quantum states is challenging as any measurement on a quantum state will irremediably alter it. While NIZK proofs on Quantum States (NIZKoQS) have been recently introduced~\cite{CGK21_NonDestructiveZeroKnowledgeProofs} and can be used to prove really advanced properties, they rely on trapdoor functions (\LWE{}), and therefore live in Cryptomania, and are moreover fundamentally only computationally secure. \cite{CGK21_NonDestructiveZeroKnowledgeProofs} actually raised two open questions:

\begin{center}
  \emph{Is it possible to do NIZKoQS without relying on LWE? Or with statistical security?}
\end{center}

  \begin{figure}
    \centering
    \begin{autoFit}
      { \def\cno{red8}
        \def\cyes{green9}
        \def\cmid{yellow9}
        \catcode`\&=4 
        \begin{tblr}{colspec={ccccccc}}
          Article                                                      & Classical              & Setup                       & Messages                                                           & MiniQCrypt                    & Composable                                                                                       & Statistical                 \\
          \hline
          \cite{PVW08_FrameworkEfficientComposable}                    & \SetCell{bg=\cyes} Yes & \SetCell{bg=\cmid} CRS              & \SetCell{bg=\cyes}$2$                                              & \SetCell{bg=\cno} No (\LWE{}) & \SetCell{bg=\cyes} Yes                                                                           & \SetCell{bg=\cyes} Either \\ 
          \cite{BD18_TwoMessageStatisticallySenderPrivate}             & \SetCell{bg=\cyes} Yes & \SetCell{bg=\cyes} Plain M.         & \SetCell{bg=\cyes}$2$                                              & \SetCell{bg=\cno} No (\LWE{}) & \SetCell{bg=\cmid} Sender                                                                        & \SetCell{bg=\cyes} Receiver \\ 
          \cite{CK88_AchievingObliviousTransfer} + later works         & \SetCell{bg=\cno} No   & \SetCell{bg=\cmid} Depends          & \SetCell{bg=red7}$7$                                               & \SetCell{bg=\cyes} Yes        & \SetCell{bg=\cyes} Yes~\cite{DFL+09_ImprovingSecurityQuantum,Unr10_UniversallyComposableQuantum} & \SetCell{bg=\cyes} Either  \\ 
          \cite{GLSV21_ObliviousTransferMiniQCrypt}                    & \SetCell{bg=\cno} No   & {\SetCell{bg=\cyes} Plain M./\\CRS} & {\SetCell{bg=red7}$\textsf{poly}$/\\\textsf{cte} $\geq$ 7} & \SetCell{bg=\cyes} Yes        & \SetCell{bg=\cyes} Yes                                                                           & \SetCell{bg=\cno} No        \\ 
          \cite{BCKM21_OneWayFunctionsImply}                           & \SetCell{bg=\cno} No   & {\SetCell{bg=\cyes} Plain M./\\CRS} & {\SetCell{bg=red7}$\textsf{poly}$/\\\textsf{cte} $\geq$ 7} & \SetCell{bg=\cyes} Yes        & \SetCell{bg=\cyes} Yes                                                                           & \SetCell{bg=\cyes} Sender   \\ 
          \cite{ABKK22_NewFrameworkQuantum}                            & \SetCell{bg=\cno} No   & \SetCell{bg=\cmid} RO               & \SetCell{bg=\cno}$3$                                               & \SetCell{bg=\cyes} Yes        & \SetCell{bg=\cyes} Yes                                                                           & \SetCell{bg=\cno} No        \\ 
          \cite{BKS23_SecureComputationShared}                         & \SetCell{bg=\cno} No   & {\SetCell{bg=red7} RO + Shared EPR} & \SetCell{bg=\cyes}$2$                                               & \SetCell{bg=\cyes} Yes        & \SetCell{bg=\cyes} Yes                                                                           & \SetCell{bg=\cyes} Yes       \\
          This work + \cite{Unr15_NonInteractiveZeroKnowledgeProofs} & \SetCell{bg=\cno} No     & \SetCell{bg=\cmid} RO               & \SetCell{bg=\cyes}$2$                                              & \SetCell{bg=\cyes} Yes        & \SetCell{bg=\cyes} Yes                                                                           & \SetCell{bg=\cno} No        \\ 
          This work + \cite{HSS11_ClassicalCryptographicProtocols}     & \SetCell{bg=\cno} No   & \SetCell{bg=\cyes} Plain M.         & \SetCell{bg=\cno}$> 2 $                                            & \SetCell{bg=\cno} No (\LWE{}) & \SetCell{bg=\cyes} Yes                                                                           & \SetCell{bg=\cno} No        \\ 
          This work + S-NIZK                                           & \SetCell{bg=\cno} No   & \SetCell{bg=\cmid} Like ZK          & \SetCell{bg=\cyes} $2$                                             & \SetCell{bg=\cmid} Like ZK    & \SetCell{bg=\cyes} Yes                                                                           & \SetCell{bg=\cyes} Sender   \\ 
          This work + NIZK proof                                       & \SetCell{bg=\cno} No   & \SetCell{bg=\cmid} Like ZK          & \SetCell{bg=\cyes} $2$                                             & \SetCell{bg=\cmid} Like ZK    & \SetCell{bg=\cyes} Yes                                                                           & \SetCell{bg=\cyes} Receiver \\ 
          This work + ZK                                               & \SetCell{bg=\cno} No   & \SetCell{bg=\cmid} Like ZK          & \SetCell{bg=\cmid} ZK $+ 1$ or $2$\footnotemark{}                  & \SetCell{bg=\cmid} Like ZK    & \SetCell{bg=\cyes} Yes                                                                           & \SetCell{bg=\cyes} Like ZK  \\ 
        \end{tblr}
      }
    \end{autoFit}
    \caption{Comparison with related works. ``RO'' stands for Random Oracle, ``Plain M.'' stands for ``plain model'', ``Like ZK'' means that the properties (mostly) inherit from the property of the underlying ZK protocol, the party in the ``statistical'' column represents the malicious party allowed to be unbounded to get statistical security. Note that using \cite{WW06_ObliviousTransferSymmetric} we can get statistical security against the other party (of course we lose the statistical security against the first party~\cite{Lo97_InsecurityQuantumSecure}), at the cost of an additional message. This list only considers standard bit or string OT (notably \cite{BKS23_SecureComputationShared} also provides a $1$-message protocol in the (strong) shared-EPR model, but for a randomized-version of OT).}
    \label{tab:comparisonRelatedWorks}
  \end{figure}
  \footnotetext{$+1$ in the Common Random String model, $+2$ in the plain model.}


\subsection{Contributions}

In this work, we answer positively all these open questions. We first state our results on OT protocols (see also \cref{tab:comparisonRelatedWorks} for a table comparing existing works):
\begin{theorem}[informal]\label{thm:informalOT}
  There exists a (non-black-box\footnote{Our protocol requires the use of a hash function $h$: since we need to prove statements on preimages of $h$ in a ZK protocol, this makes our protocol non-black-box with respect to $h$ since the circuit of $h$ must be known to the verifier. Therefore, even if the assumptions on $h$ (collision-resistant and hiding) are trivially true if $h$ is modelled as a random oracle, we cannot directly run the ZK protocol on an oracle since the source code of $h$ cannot efficiently be sent to the verifier. For this reason, we do not model $h$ itself as an oracle (this assumption is required by the ZK protocol), and only assume that $h$ is collision-resistant and hiding.}) $2$-message string OT (even $k$-out-of-$n$ string OT) quantum protocol composably secure in the random oracle model, assuming the existence of a collision-resistant hiding\footnote{Informally, a hiding function $h$ is a function such that it is not possible to get any information on $x$ given $h(x\|r)$ for sufficiently large random $r$ (this is used for instance in commitments). Actually, we use in practice a weaker assumption called ``second-bit hardcore'' (the function must only hide the second bit of $x$), since we believe that we could use the hardcore-bit construction of Goldreich-Levin to weaken the assumptions further by only assuming that the function is one-way.} function.
\end{theorem}

Actually, we provide a much more generic construction that allows us to obtain a variety of quantum OT protocols, depending on whether we want to optimize the round-complexity, the security (against unbounded sender, or unbounded verifier), the setup model (plain-model, Common Reference String (CRS), Random Oracle), or the computational assumptions (one-way functions, \LWE{}, etc.).

\begin{theorem}[informal]\label{lem:informalOT}
  Assuming the existence of a collision-resistant hiding one-way function, given any $n$-message ZK proof (or argument) of knowledge, we can obtain a $n+1$-message OT\footnote{This holds for all variations of OT: bit OT, string OT, and $k$-out-of-$n$ OT.} protocol (or $n+2$ in the plain model\footnote{The model of security is the same as the ZK protocol if we want a $n+2$-message protocol, and if we add the Common (uniform) Reference String assumption (weaker than the Random Oracle model) to provide the hash function, we can obtain a protocol with $n+1$ messages.}).

  Moreover, if the ZK protocol is secure against any unbounded verifier (resp.\ prover) and if the function is statistically hiding (resp.\ injective), the resulting OT protocol is secure against any unbounded sender (resp.\ receiver).
\end{theorem}

Note that classical ZK is a widely studied primitive as it turns out to be extremely useful in many applications, including in MPC, authentication, blockchain protocols~\cite{ELE_ZcashPrivacyprotectingDigital}, and more. Trapdoors are not necessary to build ZK as they can be built using only hash functions, and therefore live in Minicrypt. Many candidates have been proposed to achieve various ZK flavors: statistical security against malicious prover or malicious verifier, non-interactive or constant rounds protocols, security in the plain model, CRS, or random oracle~\cite{GMR85_KnowledgeComplexityInteractive,Lin13_NoteConstantRoundZeroKnowledge,Unr15_NonInteractiveZeroKnowledgeProofs,PVW08_FrameworkEfficientComposable,BD18_TwoMessageStatisticallySenderPrivate,HSS11_ClassicalCryptographicProtocols,PS19_NoninteractiveZeroKnowledge}… In this paper, we notably consider the non-interactive ZK protocol of Unruh~\cite{Unr15_NonInteractiveZeroKnowledgeProofs}, proven secure in the random oracle model, together with the ZK protocol of Hallgren, Smith and Song~\cite{HSS11_ClassicalCryptographicProtocols}, proven secure in the plain-model assuming the hardness of \LWE{}, but much work has been done to study ZK under many other assumptions~\cite{Wat09_ZeroKnowledgeQuantumAttacks,AL20_SecureQuantumExtraction,Unr12_QuantumProofsKnowledge,BS20_PostquantumZeroKnowledge,LMS21_PostQuantumZeroKnowledge}.

At the heart of our approach lies the first creation of a (potentially statistically secure when instantiated correctly) ZK protocol on \emph{quantum} states, that can be seen as an extension of ZK and complexity theory to \emph{quantum} languages:
\begin{theorem}[informal]
  Under the same assumptions as \cref{lem:informalOT}, a receiver can obtain a quantum state while being sure that a subset $T$ of the qubits has been measured, without getting any information on $T$ beside the fact that it fulfils some arbitrary fixed constraints.

  The resulting protocol is $n$-message ($n+1$ in the plain model), and can in particular be non-interactive when using a NIZK protocol. Statistical security can also be obtained under the conditions described in \cref{lem:informalOT} (the receiver playing the role of the prover, and the sender the verifier).
\end{theorem}

We also extend the concept of ZK on Quantum State (ZKoQS), together with the notion of \emph{quantum} languages and we define the first two ``quantum-language'' based complexity classes $\ZKstatesQIP{}{}$ and $\ZKstatesQMA{}$. Finally, we prove relations between ZKoQS and various ideal functionalities, we prove that we can realize them, and we show examples of quantum languages belonging to $\ZKstatesQIP{}{}$ and $\ZKstatesQMA{}$.

\subsection{Overview of the main contributions.}\label{subsec:overview}

In this section, we provide a quick, informal, overview of our approach. The OT functionality can be described as follows: a sender, Bob, owns two bits\footnote{Our approach also works for strings or $k$-out-of-$n$ OT.} $m_0$ and $m_1$, and Alice wants to learn $m_b$ where the bit $b$ is provided as an input. Importantly, a malicious Bob should be unable to learn the value $b$ of Alice, and a malicious Alice should be unable to get information on both $m_0$ and $m_1$.

\paragraph{First attempt: a naive OT protocol.}

A first remark we can make is that if we are given a state in the computational basis $\ket{l}$ for some bit $l$, rotating it by applying a $Z^{m}$ gate for some bit $m$ will leave the state unchanged (up to a global phases). On the other hand, if we are given a state in the Hadamard basis $H\ket{r}$ for some bit $r$, applying a $Z^{m}$ gate will flip the encoded bit if $m = 1$, giving the state $H\ket{r \xor m}$. Therefore, we can imagine a naive protocol for OT: Alice could prepare two states $\ket{\psi^{(b)}} \eqdef H\ket{r^{(b)}}$ and $\ket{\psi^{(1-b)}} \eqdef \ket{l}$ for some random bits $r^{(b)}$ and $l$, send $\ket{\psi^{(0)}}$ and $\ket{\psi^{(1)}}$ to Bob, Bob could rotate the $i$-th qubit according to $Z^{m_i}$, and measure them in the Hadamard basis, getting outcomes $z^{(i)}$ that will be sent back to Alice. In the light of the above comment, it is easy to see that $z^{(b)} = m_b \xor r^{(b)}$ while $z^{(1-b)}$ is a random bit, uncorrelated with $m_{1-b}$. Therefore, Alice can easily recover $m_b = z^{(b)} \xor r^{(b)}$ while she is unable to recover $m_{1-b}$. Moreover, because the density matrix of $\frac{1}{2} (\ketbra{0}{0}+ \ketbra{1}{1}) = \frac{1}{2} (\ketbra{+}{+}+ \ketbra{-}{-})$ is the completely mixed state, Bob cannot recover any information on $b$…

Unfortunately, this protocol is not secure: Alice can easily cheat by sending two $\ket{+}$ states to learn both $m_0$ and $m_1$.

\paragraph{The need for ZK on quantum state.}

To avoid this trivial cheating strategy, we would like, informally, to prove to Bob that at least one of the received states is in the computational basis… without revealing the position of this qubit, and without destroying that state. So in a sense, we would like a quantum equivalent of ZK, except that the statement is on a quantum state instead of on a classical bit string.

As a first sight, this might seems to contradict laws of physics: it is impossible to learn the basis of a random state, and anyway any measurement would certainly disturb the state. However, we can change a bit the procedure to send $\ket{\psi^{(0)}}$ and $\ket{\psi^{(1)}}$, by sending instead bigger, more structured states encoding the original qubit: Bob would then do some (non-destructive) tests on this large state in order to check that the encoding is valid, and that at least one state is not in superposition, before collapsing it to a $2$-qubit system.

At a high level, it is handy to define the encoded state as a superposition of pre-images of multiple (publicly known) images of a given hash function $h$: To control the number of elements allowed in the superposition, the key idea is to prove (using this time classical ZK), that the sender knows pre-images to all the publicly known images, where some of them are tagged as \emph{dummy}, i.e.\ forbidden (e.g.\ by making sure they start with a $0$). This way, if we prove that one of the two states admits only a single non-dummy preimage (without revealing which state), this state cannot be in superposition of multiple elements, or it would be possible to extract a collision of the hash function. Of course, this assume that the receiver performs some checks to ensure that the quantum state is a valid encoding and only contains non-dummy preimages of $h$: this can be done for instance by checking in superposition that all elements are non-dummy (e.g.\ by measuring the first bit and checking that it's one), and by computing $h$ and checking (in superposition) that it belongs to the set of allowed images. This way, ZK is used on a classical string to verify, indirectly, properties on the quantum state.

More formally, instead of sending $\ket{l}$, we sample a random bit string $w^{(1-b)}_l$ starting with a $0$ (this will be important later, but informally this indicates that this is a valid, non-dummy element) and send $\ket{\psi^{(1-b)}} \eqdef \ket{l}\ket{w^{(1-b)}_l}$, together with the hash $h^{(1-b)}_l \eqdef h(l \| w^{(1-b)}_l)$.
Similarly, we can apply this idea on states in superposition: instead of sending $\ket{0}+(-1)^{r^{(b)}}\ket{1}$, we sample similarly $w^{(b)}_0$ and $w^{(b)}_1$, and send $\ket{\psi^{(b)}} \eqdef \ket{0}\ket{w^{(b)}_0}+(-1)^{r^{(b)}}\ket{1}\ket{w^{(b)}_1}$, together with the hashes $h^{(b)}_0 \eqdef h(0 \| w^{(b)}_0)$ and $h^{(b)}_1 \eqdef h(1 \| w^{(b)}_1)$.
Of course, now, it is relatively easy to distinguish both qubits, as the qubit in the computational basis comes with a single classical hash, while the other comes with two hashes.
To avoid this issue, we add a ``dummy'' hash by sampling a random $w^{(1-b)}_{1-l}$ starting with a $1$ (indicating that the hash is dummy), and defining $h^{(1-b)}_{1-l} \eqdef h(l \| w^{(1-b)}_{1-l})$.
Importantly, given a hash, it is impossible to see if it is a dummy hash, as the hash function is hiding its input.
However, Alice can prove to Bob, using classical ZK, that at least one of the provided hashes is a dummy hash, without revealing its position.
Therefore, to sum-up, Alice sends the hashes, proves that she knows a preimage for all of them and that one of them is a dummy hash (i.e.\ its preimage has a $1$ in its second position), before sending the states $\ket{\psi^{(0)}}$ and $\ket{\psi^{(1)}}$ to Bob (if the ZK proof is non-interactive, she can send everything in a single message).

Then, after verifying the ZK proof, Bob will verify that $\ket{\psi^{(0)}}$ and $\ket{\psi^{(1)}}$ are in a superposition of valid, non-dummy, preimages. More precisely, for $i \in \{0,1\}$, he applies the unitary $\ket{x}\ket{w}\ket{0} \rightarrow \ket{x}\ket{w}\ket{w[1] = 0 \land h(x \| w) \in \{h^{(i)}_0,h^{(i)}_1\}}$ on the $i$-th qubit (after adding an auxiliary qubit), and measures the last register to check if it is equal to $1$. Note that for honestly prepared state, this measurement will not alter the state, as the last registers always contains a $\ket{1}$ and can therefore be factored out as the state is separable. Once the check is performed, we can shrink both states to obtain a $2$-qubit state by measuring the second register containing the $w$'s in the Hadamard basis, getting two outcomes $s^{(i)}$'s. One can easily check that since $\ket{\psi^{(1-b)}}$ is already in the computational basis, it will not alter the first qubit, resulting in the $\ket{l}$ state, i.e.\ a qubit in the computational basis. On the other hand, it is not hard to see that the qubit $\ket{\psi^{(b)}}$ will be turned into $\ket{0}\ket{w^{(b)}_0}+(-1)^{r^{(b) \xor \langle s, w_0^{(b)} \xor w_1^{(b)}\rangle}}\ket{1}$, i.e.\ the final state will be in the Hadamard basis (the encoded bit might be flipped, but Alice can easily recover that bit flip knowing the outcomes of the measurements).

This way, we are back to the original requirement of the naive oblivious transfer described above: Bob can rotate each qubit $i$ using $Z^{m_i}$, measure them in the Hadamard basis, and send the outcomes $z^{(i)}$ to Alice, together with the measurements $s^{(0)}$ and $s^{(1)}$. Alice will then be able to recover the final bit $m_b$ by computing $r^{(b)} \xor \langle s, w_0^{(b)} \xor w_1^{(b)}\rangle \xor z^{(b)}$.

This protocol is summarized in \cref{protoc:2messOT}, and can easily be generalized to string OT or $k$-out-of-$n$ OT by sending one ``hashed qubit'' per bit to transmit, and proving via ZK the wanted properties on the number and position of the dummy hashes (e.g.\ either the first half of hashes are dummy, or the second half). This will be described in more details below.

\paragraph{Sketch of security proof.}

Interestingly, this method is significantly simpler to analyse than the interactive cut-and-choose approach used in previous works, as illustrated by the long line of research trying to prove the security of the original proposal~\cite{BBCS92_PracticalQuantumOblivious,MS94_QuantumObliviousTransfer,Yao95_SecurityQuantumProtocols,DFL+09_ImprovingSecurityQuantum,Unr10_UniversallyComposableQuantum,BF10_SamplingQuantumPopulation}. Of course, part of this analysis is offloaded to the ZK protocol, but we like to see it as a feature: this allows us to have a more modular protocol (any improvement on ZK directly implies an improvement on OT), and the analysis only needs to be done once for the \emph{classical} ZK protocol.

At a very high level, since the ZK protocol leaks no information on the witness, and because the hash is hiding\footnote{In practice, we ask for $h$ to be ``second-bit hardcore'', meaning that it is not possible to learn the second bit of $x$ given $h(x)$, but we could also certainly extend the construction to work for any one-way function using the Goldreich-Levin construction and rejection sampling.}, Bob learns no information on $b$. Note that the quantum state does not help as one can see that for any bit string $x_0$, $x_1$ the density matrix of $\ket{x}$ where $x \sample \{x_0,x_1\}$ is equal to the density matrix of $\ket{x_0} \pm \ket{x_1}$, where the sign is randomly chosen. To translate this informal argument into a composable security proof, we design our simulator by first replacing the ZK proof with a simulated proof (that does not need access to the witness), then we turn the dummy hash into a non-dummy hash (indistinguishable since $h$ is hiding), and we sample $\ket{\psi^{(1-b)}}$ like $\ket{\psi^{(b)}}$ (indistinguishable by the above argument on density matrices). This way, the simulator can extract both $m_0$ and $m_1$, and provide them to the ideal functionality for OT, that will be in charge of discarding $m_{1-b}$ and outputting $m_b$. See \cref{thm:realizesOT} for more details.

On the other hand, to learn information about both $m_0$ and $m_1$, Alice needs to produce two non-collapsed states. But the tests performed by Bob force Alice to send a superposition of non-dummy preimages (in case she does not, the test might pass with some probability, but the state will be anyway projected on a superposition of non-dummy valid preimages in that case). However, by the ZK property, at least one of the classical hashes must be a dummy hash, and therefore if the corresponding qubit contains a superposition of multiple valid preimages, one of them must either collide with the dummy hash, or with the non-dummy one. This collision can even be obtained with non-negligible probability by measuring the state in the computational basis and comparing the outcome with the preimages extracted by the simulator during the ZK protocol. More details can be found in the proof of \cref{thm:realizesOT}.

Note that if all the properties hold against an unbounded Alice (resp.\ Bob), notably by instantiating the protocol with a ZK \emph{proof} of knowledge and an injective function $h$ (resp.\ a statistical ZK and a statistically hiding function) our OT protocol is secure against an unbounded receiver (resp.\ sender). Note also that since our adversaries are non-uniform, we need to find a way to distribute the function $h$ in such a way that the non-uniform advice cannot depend on $h$ (or it might hardcode a collision). By relying on the CRS assumption (actually a uniformly random string is enough), the hash function can be distributed non-interactively by the CRS (or heuristically replaced with a fixed hash function). If we want to stay in the plain model we can instead ask Bob to sample the function and send it to Alice at the beginning of the protocol, adding an additional message (providing a $(n+2)$-message OT protocol instead of $n+1$, where $n$ is the number of messages of the ZK protocol).

\paragraph{ZKoQS and quantum language.}

The above protocol internally proves a statement on a quantum state, suggesting a quantum analogue to classical Zero-Knowledge and languages. While this notion was introduced in \cite{CGK21_NonDestructiveZeroKnowledgeProofs} (\cite{CGK21_NonDestructiveZeroKnowledgeProofs} actually relies on the Learning-With-Error (LWE) problem while we do not require such structure, and they are fundamentally only computationally secure), we extend their definition of ZK, notably introducing the notion of subclass needed when the protocol is composed into other protocols, and we provide a second, MPC-based point of view.

At a high level, a quantum language is, similarly to classical language $\lang \subseteq \{0,1\}^*$, described by a set of quantum states $\lang_\cQ$. Analogously to classical proof systems, where a proof should be accepted only if $x \in L$, quantumly we expect the proof to be accepted only if $\rho \in \lang_\cQ$, where $\rho$ is the obtained quantum state. Classically, we also divide $\lang$ into subsets $\lang_w$ where $w$'s are called witnesses: during an honest run of the protocol we expect $x \in \lang_w$. Similarly, quantumly we divide $\lang_\cQ$ into subsets $\lang_{\omega,\omega_s}$, where $(\omega,\omega_s)$ are classical elements\footnote{For instance, you can think of $\omega$ as the basis of $\rho$, and $\omega_s$ as the bits encoded in these basis.} (say bit strings, we will explain later why we need two elements): like classically\footnote{Note that in the formal definitions, we actually formalize them using the more general notion of simulators for various reasons, to be compatible with simulation-based proofs, but also since quantumly it is not possible to physically check if a state belongs to a set, since some distributions of quantum states are different but still indistinguishable.}, we expect to have $\rho \in \lang_{\omega,\omega_s} \subseteq \lang_{\omega}$ during an honest run of the protocol. $\omega$ and $\omega_s$ can therefore be seen as a partial classical description of $\rho$. Finally, classically, the ZK property states that a malicious receiver should not learn $w$: quantumly we expect a malicious receiver to be unable to learn $\omega$.

\begin{remark}\label{rk:classicalVsQuantum}
  Despite the similarities of ZKoQS with the corresponding classical notions, there are still a few differences with the classical setting:
  \begin{itemize}
  \item First, \cameraVsOther{}{as pictured in \cref{fig:classicalZKvsquantumZK}, }classical ZK is typically defined in a ``mono-directional'' way, where the prover gets as input $x$ and $w$, and where the verifier learns $x$ and whether $x$ belongs to $\lang$. Quantumly, the prover does get $\omega$ as input (analog of $w$), but instead of receiving the classical description of $\rho$ (the analog of $x$), it \emph{outputs} $\omega_s$, so that $(\omega,\omega_s)$ (partially) describes $\rho$. One might wonder why $\omega_s$ is not sent as an \emph{input}: While this would certainly be possible, because of the fundamental non-deterministic nature of quantum mechanics, the qubit obtained by the receiver will typically \emph{not} belong to $\lang_{\omega,\omega_s}$ after a single round of interaction (typically, while the basis is always the same, the encoded bit is random), so we would need another round of communication to correct the quantum state. In practice, the exact $\omega_s$ (encoded bit) does not really matter (but we still want to know its value of course), but we do want to optimize the number of rounds of communications.
  \item The second question that one might ask is why we only describe \emph{partially} $\rho$ with $(\omega,\omega_s)$ instead of describing the full classical description of $\rho$ (in practice we do not reveal the bit encoded in the qubit in the computational basis). This can be explained since if we send the full description of $\rho$, this gives too much information to the adversary (distinguisher), to the point that we are unable to prove the security of the protocol. However, in practice this is not an issue, since the discarded information on $\rho$ is typically a useless random value, not needed in the rest of the protocol.
  \end{itemize}  
\end{remark}

\paragraph{Extensions, and formalisation of ZKoQS and quantum language.}

In the rest of the article, we formalize the notion of quantum language (\cref{def:quantumLanguage}) and Zero-Knowledge on Quantum states (ZKoQS, \cref{def:NIZKoQS}). We define the corresponding complexity classes \ZKstatesQIP{S}{k} and \ZKstatesQMA{S} (\cref{def:ZKstatesQIP}). While ZKoQS is quite generic, it does not translate naturally to an ideal functionality, useful to prove the security of protocols in the simulation-based and composable quantum standalone framework~\cite{HSS11_ClassicalCryptographicProtocols}. As a result, we define a relatively generic ideal functionality that is in charge of applying some measurement operators (\cref{def:Fpm}), and we prove that under some assumptions on the measurement operators (called postponable measurements, \cref{def:postponableOperator}), this functionality implies ZKoQS (\cref{thm:FpmImpliesZKoQS}). While for now we do not know a realization of this functionality for any measurement operator, we consider a particular case (\cref{def:fsemicol}) where the functionality is in charge of measuring a subset $T$ of qubits (such that $\Pred(T) = \top$ for an arbitrary predicate $\Pred$) and rotating randomly the other qubits. We show in \cref{thm:fsemicol} how to realize this functionality, and we prove in \cref{cor:ZKoQSForSemiCol} that it is a ZKoQS functionality for the language $\langSemCol{\Pred}$ of semi-collapsed states (\cref{def:langSemCol}). We provide in \cref{cor:existingZKstatesQMA} the implications in term of complexity theory (e.g.\ $\langSemCol{\Pred}$ is in $\ZKstatesQMA*{}[\mathsf{RO}]$). We also show in \cref{thm:semicolImpliesOtPred} that this functionality can be used to realize a very generic notion of OT protocol that we call $\Pred$-OT, and in particular string-OT and $k$-out-of-$n$ OT (\cref{cor:stringOT}). Finally, since our result requires the use of (NI)ZK protocols, we prove in \cref{sec:unr15_is_composable} that the non-interactive protocol of \cite{Unr15_NonInteractiveZeroKnowledgeProofs} (proven secure in the RO model) can be expressed in the quantum standalone framework, and can therefore be used in our protocol (\cite{HSS11_ClassicalCryptographicProtocols} already provides another interactive protocol in the plain-model).

\subsection{Concurrent work}

A few months after releasing our article online, a related and independent article was posted on the arXiv~\cite{BKS23_SecureComputationShared}, but as noted in \cite{BKS23_SecureComputationShared}, our contributions are orthogonal, with completely different methods. They indeed assume that adversaries share EPR pairs before starting the protocol (which is a strong assumption), but they show that in this sufficient to obtain $1$-message OT assuming the hardness of (sub-exponential) \LWE{} (requiring public-key cryptography), and a $2$-message OT in the random oracle setting. See \cref{tab:comparisonRelatedWorks} for a detailed comparison.

\subsection{Open problems and ongoing works}

We expect our method used to build non-interactive OT to be of independent interest, which also raises a number of open questions\publishedVsArxiv{. In particular, we do not know if $2$-message OT without structure is possible without multi-qubit entanglement, if we can build round-efficient OT from even weaker assumptions, or what are the quantum languages that belong to $\ZKstatesQMA{}$. More details can be found in \cameraVsOther{the full version~\cite{CMS23_ObliviousTransferZeroKnowledge}}{\cref{sec:openproblems}}.}{:}

\publishedVsArxiv{\inAppendixIfPublished{\subsection{Open problems and ongoing works}\label{sec:openproblems}We detail here some remaining open questions:}}{}
\inAppendixIfPublished{
  \begin{itemize}
  \item \textbf{Reducing entanglement}: our protocols require the preparation of states representing a superposition of bit strings, and the application of a hash function $h$ in superposition. For practical considerations, it would be great to see if we could get $2$-message quantum OT protocols and/or ZKoQS with single-qubit operations (or prove impossibility results).
  \item \textbf{Universal composability}: the model of security we are using allows sequential composability but not parallel composability. A priori, we expect our proof method to extend to a general composability framework like Composable Cryptography or Universal Composability, but we also need to find ZK protocols secure in this stronger model of security (note that \cite{Unr15_NonInteractiveZeroKnowledgeProofs} already provides online extractability and is therefore certainly a good starting point).
  \item \textbf{Characterization of $\ZKstatesQIP{}{}$ and $\ZKstatesQMA{}$}: For now we have only proven the belonging of a small class of quantum languages in $\ZKstatesQMA{}[\mathsf{RO}]$ and $\ZKstatesQIP{S}[\mathsf{pm}]{}$, but it would be thrilling to study the set of quantum languages that belong (or does not belong) to the various classes $\ZKstatesQIP{}{}$ and $\ZKstatesQMA{}$. For instance it would be interesting to see if it is possible to prove that states belong to the Hadamard basis or to the computational basis (methods inspired by quantum money might be useful).
  \item \textbf{ZK for statistical security}: While our approach states that we can get quantum OT with statistical security assuming the existence of statistical ZK argument of knowledge (for unbounded verifier/sender) or ZK proof of knowledge (for unbounded prover/receiver), it is important to check that such protocols exist (for now the protocols we analyse only bring computational security, which results in a computationally secure OT, like~\cite{ABKK22_NewFrameworkQuantum}). There are countless classical candidates and ways to analyse them quantumly~(\cite{Wat09_ZeroKnowledgeQuantumAttacks,AL20_SecureQuantumExtraction,BS20_PostquantumZeroKnowledge,LMS21_PostQuantumZeroKnowledge}, especially with the recent breakthrough of~\cite{LMS21_PostQuantumZeroKnowledge}, but each construction often uses their own slightly different definitions of ZK. Therefore, a proper analysis is needed to see which protocol fits in the quantum standalone framework. Similarly, finding a ZK in the plain model not based on trapdoors could provide a simpler proof for the results of \cite{GLSV21_ObliviousTransferMiniQCrypt,BCKM21_OneWayFunctionsImply}, and even if candidates exists, we have not yet analysed them properly to see if they fit in the quantum standalone framework. Finally, the ZK construction~\cite{Unr15_NonInteractiveZeroKnowledgeProofs} that we use to get $2$-message OT is in the random oracle model, and it would be great to obtain a similar ZK construction in the CRS model.
  \item \textbf{Even weaker assumptions}: the hash function needs to be hiding (our actual assumption is actually slightly weaker), but we don't know if we can reduce this assumption to use only one-way functions (we sketch a construction based on the Goldreich-Levin theorem, but this still need to be analysed formally). Moreover, pseudo-random states~\cite{JLS18_PseudorandomQuantumStates} were introduced to provide an ever lower assumption compared to one-way functions. They are known to imply OT~\cite{BCKM21_OneWayFunctionsImply,AQY22_CryptographyPseudorandomQuantum}, but it is unclear if our approach could lead to more efficient protocols.
  \item \textbf{Reducing complexity}: for now, when doing string OT of size $n$ we sample $2n$ random $w$'s, and therefore we need to do $2n$ ZK proofs on them. However, it might seem reasonable to use the same randomness for the first $n$ bits, and a second randomness for the last $n$ bits, leading to a much shorter ZK proof. It could also be great to see if it is possible somehow to re-use the same quantum register containing the randomness to also lower the quantum complexity for string OT.
  \item \textbf{Reducing communication in the plain model}: while our approach can get us to the optimal round-complexity ($2$ messages), such optimal complexity cannot be obtained in the plain-model, at least in a composable framework. It would be interesting to study the minimum number of rounds in the plain-model (but staying in MiniQCrypt), possibly giving up on composable proofs.
  \item \textbf{Applications}: While OT is definitely an important application for the ZKoQS protocol, we expect ZKoQS to find applications in other fields. Exploring the potential applications would therefore be an interesting line of research.
  \item \textbf{Weaker ZK protocols}: For now we assume the existence of a ZK protocol for \NP{}, but we informally mainly want to prove that some classical languages contain few elements, which might be more efficient to realize than with a fully fledged ZK protocol for \NP{}. Studying the links with witness indistinguishability or witness elimination~\cite{KZ09_ZeroKnowledgeProofsWitness} might also be nice to see how we can weaken the assumptions. Moreover, interestingly we don't need the PoK property to extract the $m_b$'s, only to get the value of $b$. It might be interesting to see if we can get rid of the PoK assumption of the ZK protocol.
  \item \textbf{Comparison of quantum communication}: while our approach potentially needs non-trivial quantum operations on the server side (notably applying $h$ in superposition, note that all ZK operations are fully classical), the quantum communication seems relatively low compared to other works like~\cite{ABKK22_NewFrameworkQuantum}. The reason is that we only send the randomness $w$, so if we take a randomness of size $160$ (that should be enough to avoid brute-force attacks and quadratic improvement in grover-like attacks), we can transmit $2 \times 161 = 322$ qubits to get $80$ bits of security.  Instead, our understanding of~\cite{ABKK22_NewFrameworkQuantum} is that we need to send $3200\lambda = 256\,000$ qubits for a similar security guarantee. However, a proper analysis should be made.
  \end{itemize}
}
\section{Preliminaries}
\pratendSetLocal{category=preliminaries}

\subsection{Notations}

We assume basic familiarities with quantum computing~\cite{NC10_QuantumComputationQuantum}. For any Hermitian matrix $A$, we denote its trace norm as $\|A\|_1 \eqdef \Tr(\sqrt{A^\dagger A}) = \sum_i |\lambda_i|$ where $\lambda_i$'s are the eigen-values of $A$ (considered with there multiplicity). We denote the trace distance between two density matrices $\rho$ and $\sigma$ as $\TD(\rho, \sigma) \eqdef \frac{1}{2} \|\rho-\sigma\|_1$. A bipartite state between two registers or parties $\Alice$ and $\Bob$ will be denoted $\rho^{\Alice,\Bob}$. For any bit string $x$ and $x'$, $x[i]$ is the $i$-th element of $x$, starting from $1$, and $\langle x, x'\rangle \eqdef \xor_i x[i]x'[x]$. For a gate $Z$ and a quantum state $\ket{\psi}$, $Z^{\Bob,i}\ket{\psi}_{\Bob,\cE}$ represents the state obtained after applying $Z$ on the $i$-th qubit of the register $\Bob$ of $\psi$ (we might omit the register when it is clear from the context). We might abuse notations and consider that outputting true is the same as outputting $1$, but for more complex formulas $P$ it can be handy to define $\delta_P \in \{0,1\}$ such that $\delta_P = 1$ iff $P$ is true.

\subsection{Model of security}

We follow the quantum stand-alone security model defined in \cite{HSS11_ClassicalCryptographicProtocols} that we quickly summarize here.

\begin{figure}
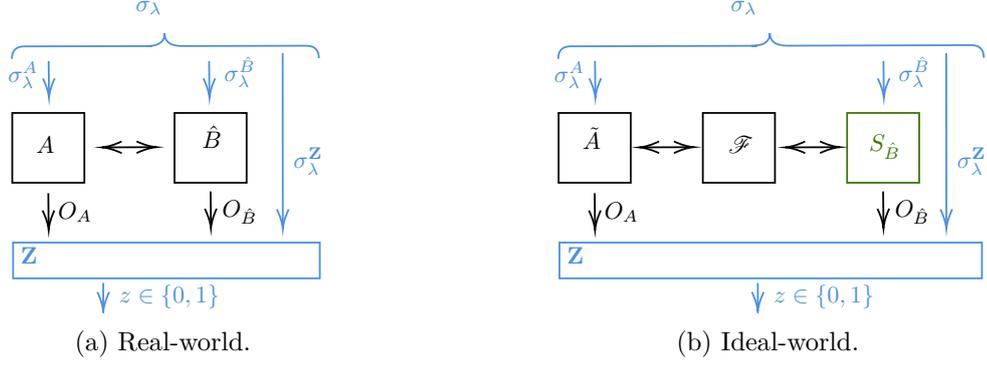

  \centering
  \begin{subfigure}[b]{0.5\textwidth}\centering
  	\includegraphics[scale=0.9]{figures/real.tex}
  	\caption{Real-world.}
  \end{subfigure}%
  \begin{subfigure}[b]{0.5\textwidth}\centering
	\includegraphics[scale=0.9]{figures/ideal.tex}
	\caption{Ideal-world.}
  \end{subfigure}
  \caption{Real-world and ideal-world executions when Bob is malicious.}
  \label{fig:realIdeal}
\end{figure}

\publishedVsArxiv{
  This model of security follows the usual real-world/ideal-world paradigm, where a protocol $\Pi$ is said to be quantum-standalone (\QSA) secure\footnote{If the security holds against a set of unbounded parties $S$, we denote it as $\CSQSA{S}$.} if no environment $\Zenv$ can distinguish this real world, where corrupted adversaries are replaced with an arbitrary adversary $\cA$, from a so-called ideal-world where the distinguisher interacts (through a simulator $\Sim_\cA$) with an ideal functionality $\cF$ playing the role of a trusted third-party. More formally, we expect to have $\Real^\sigma_{\Pi, \cA, \Zenv} \approx \Ideal^{\sigma,\cF}$, with $\Real^\sigma_{\Pi, \cA, \Zenv} \eqdef \Zenv ((\Pi \interacts \cA) \otimes I)\sigma$ and $\Ideal^{\sigma,\cF}_{\tilde{\Pi}, \Sim_\cA, \Zenv} \eqdef \Zenv ((\tilde{\Pi} \interactsF{\cF} \Sim_\cA) \otimes I)\sigma$, $\interacts$ (resp.\ $\interactsF{\cF}$) being the interaction between multiple parties (resp.\ through the functionality $\cF$), and $\sigma$ being a non-uniform advice. This is pictured in \cref{fig:realIdeal}, more details on the framework are available in \cameraVsOther{the full version~\cite{CMS23_ObliviousTransferZeroKnowledge}}{\cref{subsec:modelSecurity}}.

  \inAppendixIfPublished{\subsection{Model of security}\label{subsec:modelSecurity}}
}{}

\inAppendixIfPublished{
  \paragraph{Quantum Interactive Machines (QIM).}

  In this model, a quantum interactive machine (QIM) $\Alice = \{A_\lambda\}_{\lambda \in \N}$ is a sequence of quantum circuits $A_\lambda$ indexed by the security parameter $\lambda$ working on an input, output and network register. Two machines can interact by sharing their network register while they are activated alternately. A (two-party) protocol $\Pi = (\Alice,\Bob)$ is a couple of QIM. We denote by $\Alice \interacts \Bob$ the sequence of quantum maps (indexed by $\lambda \in \N$) representing the interaction between $A_\lambda$ and $B_\lambda$: Namely this map takes as input a quantum state on two registers $S_A$ and $S_B$, provides to $A_\lambda$ (resp. $B_\lambda$) the input $S_A$ (resp. $S_B$), let $A_\lambda$ and $B_\lambda$ interact and outputs at the end of the interaction the two registers containing the outputs of $A_\lambda$ and $B_\lambda$. We might also write $z \gets \OUT_\Bob{}(\Alice_\lambda(x) \interacts \Bob_\lambda(y))$ instead of $(\_, z) \gets (\Alice_\lambda(x) \interacts \Bob_\lambda(y))$ to denote the output of the party $\Bob$. A protocol is said to be \emph{poly-time} if all the parties run in polynomial time. The security of a protocol is expressed with respect to a \emph{functionality} $\cF$ (having no input) playing the role of a trusted third party. A functionality is a QIM interacting with all parties: for two QIM $\Alice$ and $\Bob$, we similarly denote as $\Alice \interactsF{\cF} \Bob$ the quantum map that forwards the two input registers to $\Alice$ and $\Bob$ and that returns their outputs after letting both of them interact (only) with $\cF$, as pictured in \cref{fig:realIdeal}. Note that we might provide access to oracles $H$ (QIM that answer queries to functions, e.g.\ a random oracle), in which case we will either denote it as $\Alice^H \interacts \Bob^H$ or $\Alice \interactsF{H} \Bob$ (in this case $H$ is the functionality that answers queries and forwards other messages). Moreover, for two sequences of quantum maps $\Alice = \{A_\lambda\}_{\lambda \in \N}$ and $\Bob = \{B_\lambda\}_{\lambda \in \N}$, we also define naturally their sequential composition as $\Alice \Bob \eqdef \{A_\lambda B_\lambda\}_{\lambda \in \N}$.

  \paragraph{Adversaries.}

  An adversary $\cA$ is a QIM able to corrupt parties (i.e.\ $\cA$ will replace the corrupted parties). We consider only \emph{static} adversaries, meaning that $\cA \in \{\hat{\Alice},\hat{\Bob}\}$ chooses before the beginning of the protocol the set of corrupted party. In particular, we denote by $\hat{\Alice}$ the adversary that corrupts (and replaces) $\Alice$ (similarly $\hat{\Bob}$ would corrupt $\Bob$). We define $\Pi \interacts \cA$ as the quantum map obtained when the protocol $\Pi$ is run in the presence of the adversary $\cA$: Notably, $\Pi \interacts \hat{\Alice} = \hat{\Alice} \interacts \Bob$ and $\Pi \interacts \hat{\Bob} = \Alice \interacts \hat{\Bob}$.

  \paragraph{Real and ideal worlds.}

  The security relies on the usual \emph{simulation} paradigm involving a real-world and an ideal-world, where the real-world represents a run of the protocol where some parties can potentially be corrupted while the ideal-world paradigm represents an idealized version of the protocol where the parties are only allowed to interact through the trusted ideal functionality. A QIM $\Zenv$ called \emph{environment} will be in charge of distinguishing these two worlds. Informally, if both worlds are indistinguishable, the protocol is said secure as any attack doable in the real-world would apply in the ideal-world (otherwise it would provide a way to distinguish both worlds) and therefore on the ideal functionality $\cF$, which is secure by definition. In order to ``fake'' a transcript from the real world during an execution of the ideal world, we replace any honest party $\Alice$ by a idealized party\footnote{This is the analogue of filters in constructive cryptography.} $\tilde{\Alice}$ that honestly interact with $\cF$ (it is typically trivially interacting with $\cF$ by forwarding the inputs and outputs to/from $\cF$ and is therefore often omitted), and we write $\tilde{\Pi} \eqdef (\tilde{A},\tilde{B})$ to denote this dummy protocol. Moreover, to deal with the corrupted parties, we introduce a special kind of adversary $\Sim_\cA$ called a \emph{simulator}, that must corrupt the same party as the adversary $\cA$ and whose goal is to fake the transcript outputted by $\cA$ (i.e.\ simulate $\cA$, hence its name).

  We formalize now this concept:
  \begin{definition}
    Let $\Pi = (\Alice,\Bob)$ be a two-party protocol, $\cA$ be a static adversary as defined above, $\Sim_\cA$ be a simulator, $\sigma = \{ \sigma_\lambda \in \Sim_A(\lambda) \otimes \Sim_B(\lambda) \otimes \cW(\lambda) \}_{\lambda \in \N}$ be a sequence of quantum states and $\Zenv$ be a QIM called \emph{environment} outputting a single classical bit. We denote by $\Real^\sigma_{\Pi, \cA, \Zenv} \eqdef \Zenv ((\Pi \interacts \cA) \otimes I)\sigma$ the (sequence of) binary random variables outputted by the environment $\Zenv$ at the end of an interaction where the adversary $\cA$ corrupts some parties in $\Pi$. We define similarly $\Ideal^{\sigma,\cF}_{\tilde{\Pi}, \Sim_\cA, \Zenv} \eqdef \Zenv ((\tilde{\Pi} \interactsF{\cF} \Sim_\cA) \otimes I)\sigma$ as the (sequence of) binary random variables outputted by the environment $\Zenv$ at the end of an interaction where the simulator can corrupt some dummy parties interacting with the ideal functionality $\cF$.
  \end{definition}

  \begin{definition}[Indistinguishable random variables]\label{def:indistinguishable}
    Two sequences of random variables $\mathbf{X} = \{X_\lambda\}_{\lambda \in \N}$ and $\mathbf{Y} = \{Y_\lambda\}_{\lambda \in \N}$ are said to be $\eps$-\emph{indistinguishable}, denoted $\mathbf{X} \approxRV_\eps \mathbf{Y}$, if $|\pr{X_n = 1} - \pr{Y_n = 1}| \leq \eps(\lambda)$. In particular, if $\eps = \negl[\lambda]$, $\mathbf{X}$ and $\mathbf{Y}$ are said to be indistinguishable, denoted $\mathbf{X} \approxRV \mathbf{Y}$.
  \end{definition}

  \begin{definition}[Indistinguishable quantum maps]\label{def:indistinguishableQuantumMaps}
    Two sequences of quantum maps $\mathbf{X} = \{X_\lambda\}_{\lambda \in \N}$ and $\mathbf{Y} = \{Y_\lambda\}_{\lambda \in \N}$ are said to be \emph{computationally} (resp.\ \emph{statistically}) \emph{indistinguishable}, denoted $\mathbf{X} \approxRVC \mathbf{Y}$ (resp.\ $\mathbf{X} \approxRVS \mathbf{Y}$), if for any poly-time (resp.\ unbounded) $\Zenv = \{\Zenv_\lambda\}_{\lambda \in \N}$ and any sequence of bipartite advices $\sigma = \{\sigma_\lambda\}_\lambda$, $\Zenv (\mathbf{X} \otimes I) \sigma \approxRV \Zenv (\mathbf{Y} \otimes I) \sigma$.
  \end{definition}

  \begin{definition}[Quantum stand-alone ({\StarQSA*{}}) realization of a functionality~\cite{HSS11_ClassicalCryptographicProtocols}]\label{def:QSA}
    Let $\cF$ be a poly-time two-party functionality and $\Pi$ be a poly-time two-party protocol. We say that $\Pi$ \emph{computationally quantum-stand-alone} (\CQSA*) (resp.\ \emph{statistically} quantum-stand-alone (\SQSA*)) realizes $\cF$ if for any poly-time (resp. unbounded) adversary $\cA$ there is a poly-time (in the time taken by $\cA$) simulator $\Sim_{\cA}$ such that for any poly-time (resp. unbounded) environment $\Zenv$ and family of states $\sigma = \{\sigma_\lambda\}_{\lambda \in \N}$, $\Real^{\sigma}_{\Pi,\cA,\Zenv} \approxRV \Ideal^{\sigma,\cF}_{\tilde{\Pi},\Sim_\cA,\Zenv}$.

    Moreover, we extend this definition by saying that $\Pi$ $\CSQSA*{S}$ (where $S$ is a set of subset of parties realizes $\cF$ when \emph{statistical} security holds only if the adversary corrupts\footnote{Remember that the adversary is static, and therefore determines the set of parties to corrupt before the beginning of the protocol. Note that we will omit in the proof the case where $\cA$ corrupts all parties as this case is trivial (the simulator can just run the adversary and ignore the functionality).} a set of parties in $S$ (i.e.\ if $\cA$ corrupts a set of party in $S$ then $\cA$ and $\Zenv$ are allowed to be unbounded, otherwise they are poly-time). In particular, if the protocol has two parties $A$ and $B$, $\CSQSA*{\emptyset} = \CQSA*$ and $\CSQSA*{\{\emptyset, \{A\}, \{B\}, \{A,B\}\}} = \SQSA*$. Note that because it is always possible to turn malicious parties into honest parties, $\CSQSA*{S}$ implies $\CSQSA*{S \cup \{X\}}$ for any $X \subseteq s$ such that $s \in S$ (for instance $\CSQSA*{\{A,B\}}$) implies $\CSQSA*{\{\emptyset, \{A\}, \{B\}, \{A,B\}\}})$. For this reason, we will consider from now only maximal sets $S$ with respect to this augmentation procedure and we will often only write the larger set: We will notably be particularly interested in statistical security against a malicious Alice ($\CSQSA*{\{\emptyset, \{A\}\}}$, or $\CSQSA*{A}$ for short) or a malicious Bob ($\CSQSA*{\{\emptyset, \{B\}\}}$, or $\CSQSA*{B}$ for short).
  \end{definition}
}

\paragraph{Some functionalities.} We present here some ideal functionalities used later, starting with the main OT functionality:

\begin{definition}[Functionality for bit oblivious transfer $\Fot*$~\cite{HSS11_ClassicalCryptographicProtocols}]\label{def:Fot}
  We define the ideal functionality $\Fot*$ for oblivious transfer as follows:
  \begin{itemize}
  \item it receives two messages $m_0$ and $m_1$ from Bob's interface, or an abort message
  \item it receives one bit $b \in \{0,1\}$ from Alice's interface, or an abort message
  \item if no party decided to abort, it sends $m_b$ to Alice.
  \end{itemize}
  We define trivially the dummy parties $\tilde{\Pi} = (\tilde{\Alice},\tilde{\Bob})$ that forward the inputs/outputs to/from $\Fot$.
\end{definition}

We will then prove that our protocol can trivially be extended to more advanced OT functionalities. First, we define a generic functionality where the statements can be proven on any predicate on the bits of the message, we will then consider particular cases like string OT (to receive strings instead of bits) or $k$-out-of-$m$ string OT (to receive $k$ strings among $n$):
\begin{definition}[Functionality for predicate oblivious transfer $\FotPred*$]\label{def:Fotpred}
  Let $n \in \N$ and $\Pred\colon \cP([n]) \rightarrow \{0,1\}$ be a predicate\footnote{This predicate might depend on a secret witness $w$ known only to the prover, in which case we always replace $\Pred(\cdots)$ with $\Pred(w,\cdots)$, $w$ being sent to the ideal functionalities and used in the ZK proofs. For simplicity, we will omit the witness from now.} on any subset of bits. We define the ideal functionality $\FotPred*$ for predicate oblivious transfer as follows:
  \begin{itemize}
  \item It receives $n$ bits $(m_i)_{i \in [n]}$ from Bob's interface, or an abort message.
  \item It receive a subset $B \subseteq [n]$ from Alice's interface (we might also encode $B$ as a bit string, where $B[x] = 1$ iff $x \in B$), or an abort message.
  \item If $B = \bot$ or $\Pred(B) = \bot$, it sends an abort message to Bob.
  \item If no party decided to abort and $\Pred(B) = \top$, it sends $(m_i)_{i \in B}$ to Alice. Otherwise it sends $\bot$ to all parties.
  \end{itemize}
  We define trivially the dummy parties $\tilde{\Pi} = (\tilde{\Alice},\tilde{\Bob})$ that forward the inputs/outputs to/from $\Fot$.

  We define particular cases of interest:
  \begin{itemize}
  \item \textbf{String OT}: If $n= 2m$ and $\Pred(B)$ is true iff $B \in \{ 1^m0^m, 0^m1^m\}$ then we call this functionality \emph{string OT}, denoted $\FotStr$ (to send the two messages $m_a$ and $m_b$, we define $m = m_a \| m_b$).
  \item \textbf{$k$-out-of-$m$ string OT}: If $n= lm$ and $\Pred(B)$ is true iff $B = B_1 \| \dots B_m$ with $\forall i, B_i \in \{0^l, 1^l\}$, such that the number of $B_i$'s equal to $1^l$ is equal to $k$, then we call this functionality \emph{$k$-out-of-$m$ string OT}, denoted $\FotKoutN$ (to sent the $m$ messages $m_a$ and $m_b$, we define $m = m_a \| m_b$).
  \end{itemize}
\end{definition}

Classical Zero-Knowledge (ZK) proofs allow a party (the \emph{prover}) to prove a statement to another party (the \emph{verifier}) without revealing anything beyond the fact that this statement is true. Our protocols use a ZK protocol as a blackbox. We define now the functionality corresponding to ZK.

\begin{definition}[{Functionality for zero-knowledge $\Fzk*{\cR}$~\cite{HSS11_ClassicalCryptographicProtocols}}]\label{def:Fzk}
  We define the ideal functionality $\Fzk*{\cR}$ for zero-knowledge, where $\cR$ is a relation describing a given language $\cL$ ($x \in \cL \Leftrightarrow \exists w, x \cR w$):
  \begin{itemize}
  \item it receives $(x,w)$ from the prover's (a.k.a. Alice) interface or an abort message $\bot$,
  \item if $x \cR w$ then the verifier (a.k.a. Bob) receives $x$ otherwise it receives $\bot$.
  \end{itemize}
\end{definition}

This functionality also implies that the ZK protocol is a \emph{proof of knowledge} protocol (PoK, quantumly it is also know as \emph{state-preserving} as extracting the witness should not disturb the state of the adversary) as the functionality can extract the witness. But our protocol could be proven secure in different ways:
\begin{itemize}
\item One of them is to assume that the protocol is a state-preserving PoK (PoK is not needed to extract $m_0$ and $m_1$ from a malicious Bob, but is handy to extract $b$ from a malicious Alice). That's the approach taken in this paper since it has the advantage of applying also in the plain model.
\item It should also be possible to obtain similar guarantees without state-preserving PoK, notably by assuming that the simulator can extract the queries made to the oracle (either by relying on Common Reference String (CRS) or on the random oracle model (ROM)). However, this approach is less modular and seems to rely heavily on CRS/RO and is therefore harder to generalize to the plain model. Moreover, we already know state-preserving NIZK PoK in the RO model~\cite{Unr15_NonInteractiveZeroKnowledgeProofs}, so this second approach seems less interesting and will not be explored in this article.
\end{itemize}

Moreover, we often make the distinction between ZK arguments (computational soundness against malicious prover), ZK proofs (statistical soundness against malicious prover) and statistical ZK (ZK also holds against a malicious unbounded verifier). In the quantum stand-alone formalism, ZK proofs are protocols that \CSQSA{P} realize $\Fzk{\cR}$ and statistical ZK are protocols that \CSQSA{V} realize $\Fzk{\cR}$.

Note that nearly all the properties of our protocol reduce to the properties of the ZK scheme. If we use a Non-Interactive ZK (NIZK) protocol secure in the Quantum Random Oracle (OT) model or in the Common Reference String (CRS) model, then our final protocols will be optimal in term of round complexity ($2$-message OT, or $1$-message NIZKoQS) but will rely on the RO or CRS assumption. On the other hand, we may prefer to use a $n$-message NIZK protocol in the plain model: in that case our protocols will be secure in the plain model, and the communication complexity will be $n$ for the NIZKoQS protocol, resulting in an $n+1$-message OT protocol.

There are multiple protocols realising the $\Fzk{\cR}$ functionality, either in the plain model~\cite{HSS11_ClassicalCryptographicProtocols} or  non-interactively in the random-oracle model~\cite{Unr15_NonInteractiveZeroKnowledgeProofs} (this last work is not expressed in the quantum stand-alone model, but we prove in \cref{sec:unr15_is_composable} that it can be reformulated in this framework).

Because we are dealing with non-uniform adversaries, we need to sample hash functions independently of the non-uniform advice, and this is usually done via a Common-Reference-String (CRS) assumption. CRS assumes that a string, honestly sampled according to a fixed procedure, can be shared among all parties (this is typically not counted in the communication as in practice we can often heuristically take a publicly known string instead, for instance by feeding the generation procedure with a known uniformly sampled string… unless the sampling needs trapdoor which is not our case here). While this adds an assumptions, it can be practical sometimes to obtain more efficient protocols (in term of communication complexity), and often can be heuristically replaced by a publicly known string (e.g.\ if the string contains the description of a collision resistant function like in our case, we might pick the well known SHA-256 hash function instead). Note that our protocol can also be realized without a CRS assumption at the cost of an additional message as discussed in \cref{sec:securityProof} and in \cref{lem:cFcrPlainModel}. We model CRS as an ideal functionality:
\begin{definition}\label{def:cFcrs}
  Let $\Gen$ be a PPT sampling procedure. Then the ideal functionality $\cFcrs*{\Gen}$ samples $x \gets \Gen(1^\lambda)$ and outputs $x$ to all parties.
\end{definition}

\inAppendixIfPublished{
  \paragraph{Hybrid models.}

  For the sake of modularity, it is often handy to express a protocol realizing a functionality $\cF$ assuming that there exists another (unspecified) protocol realizing a more primitive functionality $\cG$ ($\cG$ might also be considered as a setup assumption, like in the CRS model). To that end, we denote as $\Pi^\cG$ a protocol where each party can interact with a trusted party running $\cG$ (of course each party having only access to their respective interface), and we say that we are in the \emph{$\cG$-hybrid} model. The $\cFcrs{}$-hybrid model is also called the \emph{CRS model}, and if not such assumption is made, we say that we are in the \emph{plain model}. Importantly, if a protocol realizes $\cG$ and if a $\cG$-hybrid protocol realizes $\cF$, then combining both protocols in the natural way gives a (non-hybrid) protocol realizing $\cG$.
}

\subsection{Cryptographic requirements.}

Before stating our security guarantees, we need to define some security definitions. A function is said to have a hardcore second-bit if it is hard to find the second bit of $x$ given $h(x)$ (note that this notion is weaker than the more standard notion of hiding as we only need to hide a single bit). More formally:

\begin{definition}[Hardcore second-bit]\label{def:firstBitHardcore}
  We say that a function $h$ has a computational (resp. statistical) hardcore second-bit property if there exists two polynomials $n$ and $m$, such that for any $l \in \{0,1\}$, any QPT (resp. unbounded) adversary $\cA$ and for any advice $\sigma = \{\sigma_\lambda\}_{\lambda \in \N}$:
  \begin{align}
    \begin{split}
      &\left|\pr[x \sample \{l\} \times \{0\} \times \{0,1\}^{n(\lambda)}]{\cA(\lambda, \sigma_\lambda, h(x)) = 1} \right. \\& - \left.\pr[x \sample \{l\} \times \{1\} \times \{0,1\}^{n(\lambda)}]{\cA(\lambda, \sigma_\lambda, h(x)) = 1} \right| \leq \negl[\lambda]
    \end{split}
  \end{align}
  We extend this definition to a family of functions $\{h_k \colon \{0,1\}^{n(\lambda)} \rightarrow \{0,1\}^{m(\lambda)}\}_{k \in \cK}$ if for any $k \in \cK$, $h_k$ has a computational hardcore second-bit property, and if one can efficiently check for any $k$ whether $k \in \cK$ or not.
\end{definition}

We note that many functions have (or are expected to have) a hardcore second-bit property, in particular since it can be seen as a special case of hiding. It is the case for random functions (e.g. in the RO model), where it is even possible to get statistical security if the function is lossy 
(i.e. many inputs map to the same output), and we expect it to be true for hash functions used nowadays since they are believed to be hiding. We note that people often consider a weaker assumption called hardcore bit predicate (even achievable from any one-way function thanks to the Goldreich-Levin construction~\cite{GL89_HardcorePredicateAll}), where the unknown bit is a fixed predicate $b(x)$ instead of the second bit of $x$. While we believe that our construction could be adapted to that setting (by doing a rejection sampling to find $x$ such that $b(x)$ has the right value), this complicates the constructions, so we leave this extension for further work. We will therefore keep this construction for future works.

\begin{definition}[Collision resistance]\label{def:collisionResistance}
  A family of functions $\{h_k \colon \{0,1\}^{l(\lambda)} \rightarrow \{0,1\}^{m(\lambda)}\}_{k \in \cK}$ is said to be (computationally) collision-resistant if there exists a polynomial generation algorithm $k \gets \Gen_h(1^\lambda)$ such that for any $k \in \cK$, $h_k$ can be classically evaluated in polynomial time, and for any (potentially non-uniform) \QPT{} adversary $\cA$ and advice $\{\sigma_\lambda\}_{\lambda \in \N}$:
  \begin{align}
    \pr[k \gets \Gen_h(1^\lambda), (x,x') \gets \cA(k,\sigma_\lambda)]{x \neq x' \land h_k(x) = h_k(x')} \leq \negl
  \end{align}
\end{definition}

\begin{remark}
  Note that we do not directly require the functions to be collapsable~\cite{Unr16_ComputationallyBindingQuantum}\===which is often required when considering quantum adversaries\===as we can show that any attack leads to the finding of a collision. However, we do require the existence of a ZK proof of knowledge scheme, that may, in turn, require the existence of such a function. Moreover, when considering unbounded provers, the function is expected to be statistically collision-resistant, i.e.\ injective, and is therefore collapsing.
\end{remark}

Note that even if we heuristically expect the protocol to stay secure when we replace $h_k$ with a fixed hash function like SHA-256, to prove the security we need to sample the function $h_k$ after the beginning of the protocol. The reason is that the adversaries are non-uniform (i.e.\ get an arbitrary advice), and the advice could contain a collision if it was chosen after $h_k$. As a result, one needs to decide who is going to sample $h_k$, leading to various tradeoffs:
\begin{itemize}
\item If we let a user\footnote{Only Bob can sample the function as collision resistance must hold against Alice and a malicious Alice could cheat when generating the function.} sample the function, then we need to send an additional message from Bob to Alice, but on the other side we are in the plain-model.
\item Otherwise, we can assume that the circuit of $h_k$ is provided by a CRS, which requires no additional round of communication, but we are not anymore in the plain-model. 
\end{itemize}
In order to keep the proof independent of this choice, we abstract the distribution of the value of $h_k$ in an ideal functionality:

\begin{definition}\label{def:Fh}
  Let $\{h_k \colon \{0,1\}^{l(\lambda)} \rightarrow \{0,1\}^{m(\lambda)}\}_{k \in \cK}$ be a family of collision resistant functions generated by $\Gen$, with a hardcore second-bit property. Then, we define the ideal functionality $\Fh*{\Gen}$ as follows. $\Fh*{\Gen}$ receives an input $c$ from Bob's interface, if $c=\top$, the functionality samples $k \gets \Gen(1^\lambda)$ and sends $k$ to both parties, otherwise if $c \in \cK$, it forwards $c$ to Alice's interface. The ideal party $\Alice_I$ just forwards the received $k$, while the ideal party $\Bob_I$ sends $c = \top$ to the functionality and outputs the received $k$.
\end{definition}

We prove now that this functionality can be realized in the plain-model with one message or non-interactively in the CRS model.
\begin{lemmaE}[{$\Fh{}$ in the CRS model}][end]\label{lem:CRFBinCRS}
  In the CRS model (a.k.a. $\cFcrs{\Gen}$-hybrid model), the trivial $0$-message protocol where both Alice and Bob output the value given by $\cFcrs{\Gen}$ realizes the functionality $\Fh{\Gen}$.
\end{lemmaE}
\inAppendixIfPublished{\subsection{Proofs}}
\begin{proofE}
  The proof is trivial: we can split the proof in multiple cases. If no party is corrupted (correctness), then the outputs of both parties are always distributed according to $\cFcrs{\Gen}$, so no environment can (even statistically) distinguish between the ideal and real world. If Alice gets corrupted, we define the simulator exactly as the malicious Alice $\hat{\Alice}$, and both worlds are identical (up to the name we give to the different parts of the world) and therefore indistinguishable. If Bob ($\hat{Bob}$) is malicious, then we define the simulator as outputting $c = \top$ to the functionality, and then feeding the received advice and the received $k$ to $\hat{Bob}$. Again, the ideal and real worlds are equal, which concludes the proof.
\end{proofE}

\begin{lemmaE}[{$\Fh{}$ in the plain model}][end]\label{lem:cFcrPlainModel}
  The $1$-message protocol where Bob samples $x \gets \Gen(1^\lambda)$ and sends $x$ to Alice, and Alice outputs $x$ only if $x \in \cK$ realizes the functionality $\Fh{\Gen}$ in the plain model.
\end{lemmaE}
\begin{proofE}
  The proof is trivial and very similar to \cref{lem:CRFBinCRS}. The only case where it differs is when Bob is malicious. Then the simulator runs $\hat{\Bob}$ and forwards the output to the ideal functionality: the test performed by the functionality is exactly the test performed by Alice in the real world, so both worlds are equal.
\end{proofE}

\section{Protocol for bit OT}
\pratendSetLocal{category=bitOT}

\subsection{The protocol}

While we will define formally ZKoQS later, together with more advanced OT protocols (string-OT, $k$-out-of-$n$ OT…), in this section we provide a self-contained description and security proof of our bit-OT protocol. For an intuitive explanation of our protocol, we refer to the overview in \cref{subsec:overview}. The bit OT protocol is described in \publishedVsArxiv{\cref{protoc:2messOT}\cameraVsOther{}{ (see also the picture version in \cref{protoc:2messOTPicture})}{\cref{protoc:2messOT}}}.

\publishedVsArxiv{
  \begin{protocol}[!htbp]
    \caption{Protocol for (possibly $2$-message) bit Oblivious Transfer}\label{protoc:2messOT}
    \textbf{Inputs}: Alice gets $b \in \{0,1\}$ as input, Bob gets $(m_0,m_1) \in \{0,1\}^2$\\
    \textbf{Assumption}: $(\Azk,\Bzk)$ is a $n$-message ZK protocol (\cref{def:Fzk}), $h$ is a collision-resistant (\cref{def:collisionResistance}) and second-bit hardcore (\cref{def:firstBitHardcore}) function distributed using $\Fh{}$ (\cref{def:Fh}), either non-interactively via a CRS, heuristically using a fixed hash function, or sent by Bob, adding an additional message (\cref{lem:cFcrPlainModel}).\\
    \textbf{Protocol}:
    \begin{enumerate}
    \item \textbf{Alice} samples $l \in \{0,1\}$, $(w^{(b)}_0,w^{(b)}_1,w^{(1-b)}_{l}) \sample (\{0\} \times \{0,1\}^n)^3$ and $w^{(1-b)}_{1-l} \sample \{1\} \times \{0,1\}^n$, and computes for all $(c,d) \in \{0,1\}^2$, $h^{(c)}_d \eqdef h(d \| w^{(c)}_d)$. Then, she sends $(h^{(c)}_d)_{c \in \{0,1\},d \in \{0,1\}}$ to Bob (if the ZK protocol is non-interactive she can send it later in a single message with the NIZK proof and the quantum states) and runs the ZK protocol $\Azk$ with Bob (running $\Bzk$) to prove that:
      \begin{align}
        \exists (w_d^{(c)})_{c \in \{0,1\},d \in \{0,1\}}, \forall c, d, h_d^{(c)} = h(d\| w_d^{(c)}))\text{ and }\exists c, d \text{ s.t. } w_d^{(c)}[1] = 1
      \end{align}
      Then, she samples $r^{(b)} \sample \{0,1\}$ together with:
      \begin{align}
        \ket{\psi^{(b)}} \assign \ket{0}\ket{w_0^{(b)}}+(-1)^{r^{(b)}}\ket{1}\ket{w_1^{(b)}} \qquad
        \ket{\psi^{(1-b)}} \assign \ket{l}\ket{w_l^{(1-b)}}
      \end{align}
      Finally, she sends $(\ket{\psi^{(0)}}, \ket{\psi^{(1)}})$ to Bob.
    \item \textbf{Bob} verifies the ZK proof. Then, he verifies that the quantum state is honestly prepared by adding an auxiliary qubit and running the unitary:
      \begin{align}
        \ket{x}\ket{w}\ket{0} \mapsto \ket{x}\ket{w}\ket{w[1] \neq 1 \land h(x\|w) = \{h_0^{(c)}, h_1^{(c)}\}}
      \end{align}
      and measuring the last auxiliary register, checking if it is equal to $1$. If not, he aborts (and sends an abort message to Alice), otherwise he measures the second registers of $\ket{\psi^{(0)}}$ and $\ket{\psi^{(1)}}$ in the Hadamard basis (getting outcomes $(s^{(0)},s^{(1)})$). Note that at that step, the first register of $\ket{\psi^{(b)}}$ contains a $\ket{\pm}$ state while $\ket{\psi^{(1-b)}}$ contains $\ket{l}$: this fact will be called ZKoQS later. Then, for any $c \in \{0,1\}$, Bob applies $Z^{m_c}$ on $\ket{\psi^{(c)}}$ and measures it in the Hadamard basis ($\{\ket{+},\ket{-}\}$), getting outcome $z^{(c)}$. Bob sends back $(s^{(c)}, z^{(c)})_{c \in \{0,1\}}$ to Alice.
    \item \textbf{Alice} computes $\alpha \assign r^{(b)} \xor \bigoplus_i s^{(b)}[i] (w^{(b)}_0 \xor w^{(b)}_1)[i]$ and outputs $\alpha \xor  z^{(b)}$ (that should be equal to $m_b$).
    \end{enumerate}
  \end{protocol}
}{}

\inAppendixIfPublished{
\begin{protocol}[!htbp]
  \caption{Protocol for $2$-message chosen bit Oblivious Transfer}\publishedVsArxiv{\label{protoc:2messOTPicture}}{\label{protoc:2messOT}}
  \begin{autoFit}
    \pseudocodeblock{
      \textbf{Alice($b \in \{0,1\}$)} \< \< \textbf{Bob($(m_0,m_1) \in \{0,1\}^2$)} \\[][\hline]
      \\[-.5\baselineskip]
      \pclinecomment{Witness for $\cL^{(b)} = \{0,1\}$:}%
      \\
      \forall d \in \{0,1\}, w^{(b)}_d \sample \{0\} \times \{0,1\}^n\\
      \pclinecomment{Witness for $\cL^{(1-b)} = \{l\}$:}\\
      l \sample \{0,1\}\\
      w^{(1-b)}_l \sample \{0\} \times \{0,1\}^n\\
      w^{(1-b)}_{1-l} \sample \{1\} \times \{0,1\}^n\\
      \pclinecomment{Compute the characterization}\\
      \pclinecomment{of the languages:}\\
      \forall (c,d) \in \{0,1\}^2, h_d^{(c)} \assign h(d \| w_d^{(c)})\begin{tikzpicture}[overlay,remember picture]
        \node[name=start,anchor=south west,align=left,text width=8cm,blue,rounded corners,draw,yshift=1cm,xshift=2cm]{$h$ is a collision-resistant (\cref{def:collisionResistance}) and second-bit hardcore (\cref{def:firstBitHardcore}) function distributed using $\Fh{}$ (\cref{def:Fh}), either non-interactively via a CRS, heuristically using a fixed hash function, or sent by Bob, adding an additional message (\cref{lem:cFcrPlainModel})};
        \draw[blue,->] (start) -- (2mm,0);
      \end{tikzpicture}\\
      \pclinecomment{Proof that at least one language}\\
      \pclinecomment{contains a single element}\\  
      \pi \assign \text{(NI)ZK proof that:}\begin{tikzpicture}[overlay,remember picture]
        \node[name=startPi,align=left,text width=5cm,anchor=west,yshift=1mm,xshift=3cm,blue,rounded corners,draw]{If the ZK proof is interactive, then we actually run the ZK protocol (before sending the quantum state) instead of sending the proof (of course this adds additional rounds of communication).};
        \draw[blue,->](startPi) -- (2mm,1mm);
      \end{tikzpicture}\\
      \t \exists (w_d^{(c)})_{c,d}, \forall c, d, h_d^{(c)} = h(d\| w_d^{(c)})) \\
      \t \t \text{and } \exists c, d \text{ s.t. } w_d^{(c)}[1] = 1.\\
      \pclinecomment{Define the quantum states:}\\
      r^{(b)} \sample \{0,1\}\\
      \ket{\psi^{(b)}} \assign \ket{0}\ket{w_0^{(b)}}+(-1)^{r^{(b)}}\ket{1}\ket{w_1^{(b)}}\\
      \ket{\psi^{(1-b)}} \assign \ket{l}\ket{w_l^{(1-b)}}
      \< \sendmessage{->}{topstyle={overlay},length=2.8cm,top={$\forall (c,d): h^{(c)}_d, \pi, \ket{\psi^{(0)}},\ket{\psi^{(1)}}$},style={transform canvas={yshift=-5mm}}}\\
      \<\< \pclinecomment{Check that one language has size $\leq 1$:} \\
      \<\< \text{Check (or run if interactive proof) $\pi$.}\\
      \<\<\pclinecomment{Check that the state contains a superposition}\\
      \<\<\pclinecomment{of (valid) elements of $\cL^{(0)}$ and $\cL^{(1)}$:}\\
      \<\<\text{$\forall c$, apply on $\ket{\psi^{(c)}}\ket{0}$ the unitary:}\\
      \<\<\t x,w \mapsto w[1] \neq 1 \land \exists d, h(x\|w) = h_d^{(c)},\\
      \<\<\t \text{measure the last (output) register}\\
      \<\<\t \text{and check that the outcome is $1$.}\\
      \<\<\forall c, \text{measure the second register of $\ket{\psi^{(c)}}$}\\
      \<\<\t \begin{tikzpicture}[overlay,remember picture]
        \node[name=start,align=left,text width=5cm,anchor=east,yshift=8mm,xshift=-1.5cm,blue,rounded corners,draw]{At that step, $\ket{\psi^{(b)}} = \ket{0} \pm \ket{1}$ and $\ket{\psi^{(1-b)}} = \ket{l}$, but Bob does not know $b$ (NIZKoQS).};
        \draw[blue,->](start)--(-2mm,-1mm);
      \end{tikzpicture}\text{in the Hadamard basis (outcome $s^{(c)}$).}\pclb
      \pcintertext[dotted]{End of NIZKoQS}
      \<\< \forall c, \text{apply $Z^{m_c}$ on $\ket{\psi^{(c)}}$ and measure it}\\
      \<\<\t \text{in the Hadamard basis (outcome $z^{(c)}$).}\\
      \<\sendmessage{<-}{topstyle={overlay},length=2.8cm,top={$\forall c, s^{(c)}, z^{(c)}$}}\\
      \text{Compute } \alpha \assign r^{(b)} \xor \langle s^{(b)}, w^{(b)}_0 \xor w^{(b)}_1\rangle\\
      \pcreturn \alpha \xor  z^{(b)}\pccomment{Should be $m_b$}
    }
  \end{autoFit}
\end{protocol}
}
\subsection{Security proof}\label{sec:securityProof}

We prove now our main theorem, i.e.\ that \cref{protoc:2messOT} securely realizes the OT functionality.

\begin{theoremE}[Security and correctness][end,text link=]\label{thm:realizesOT}
  Let $\{h_k\}_{k \in \cK}$ be a family of collision resistant functions sampled by $\Gen$, having the hardcore second-bit property (\cref{def:firstBitHardcore}). Let $\Pi_h = (\Alice_h, \Bob_h)$ be a protocol\footnote{As a reminder, this protocol is sampling and distributing a function $h$ according to $\Gen$, and can either be done without communication in the CRS model (or heuristically if we replace $h$ with a well known collision-resistant hash function), or with one message in the plain model.} $\CSQSA{S_h}$ realizing $\cFcrs{\Gen}$ and $\Pizk = (\Azk,\Bzk)$ be a protocol that \CSQSA{S} realizes the ZK functionality $\Fzk{\cR}$, where $(h_0^0,h_1^0,h_0^1,h_1^1) \cR (w_0^0,w_1^0,w_0^1,w_1^1) \Leftrightarrow \forall c,d, h(d \| w_d^c) = h_d^c$ and  $\exists c,d$ such that $w_d^c[1] = 1$.

  Then the \cref{protoc:2messOT}, in which $h$ is obtained by first running $\Pi_h$, \CQSA{} realizes the functionality $\Fot$. More precisely, it \CSQSA{S'} realizes $\Fot$ for any set $S'$ of unbounded parties such that:
  \begin{itemize}
  \item $S' \subseteq S \cap S_h$,
  \item $\{ \Bob \} \in S'$ only if $h$ has the statistical hardcore second-bit property,
  \item $\{ \Alice \} \in S'$ only if for any $k \in \cK$, $h_k$ is injective (i.e.\ statistically collision resistant).
  \end{itemize}
\end{theoremE}
\begin{proof}[Sketch of proof]
  For a first intuitive proof of the correctness and security, we refer to the corresponding paragraph in \cref{subsec:overview}. We provide here only a sketch of the proof, and we refer the reader to the \cameraVsOther{full security proof in the full version~\cite{CMS23_ObliviousTransferZeroKnowledge}}{\hyperref[proof:prAtEnd\pratendcountercurrent]{full security proof} in \pratendSectionlikeCref{}}.

  \paragraph{Malicious sender (Bob).}
  We consider the case where the adversary $\cA = \hat{\Bob}$ corrupts the sender Bob. Informally the goal of the simulator $\Sim_{\hat{\Bob}}$ is to extract the two values $m_0$ and $m_1$ from $\hat{\Bob}$ to provide these two values to the ideal functionality. To that end, at a high level, the simulator will interact with $\hat{\Bob}$ by providing a transcript that an honest Alice could provide, except that $\ket{\psi^{(1-b)}}$ is sampled like $\ket{\psi^{(b)}}$: since the state is now in the Hadamard basis, it can also recover $m_{1-b}$ following the procedure used by Alice to recover $m_{b}$. However, because it is now impossible to run the ZK proof (because the statement is not even true!) the simulator will run instead the simulator of the ZK proof to convince the distinguisher that the statement is true while it is not. To prove that this simulator is valid, we write a series of hybrid games: we start from the protocol where Alice is honest, then we replace the ZK proof with the simulated proof (indistinguishable by the ZK property). In the next step we sample $w_{1-b}$ as a non-dummy witness (i.e. starting with a $0$, indistinguishable because the function $h$ is hiding). Then we set $\ket{\psi^{(1-b)}} = \ket{0}\ket{w_0^{(1-b)}} + (-1)^{r^{(1-b)}} \ket{1}\ket{w_1^{(1-b)}}$ where $r^{(1-b)} \gets \{0,1\}$ is sampled uniformly at random (indistinguishable because the density matrices are equal: for any (potentially known) string $x$ and $y$, $ \frac{1}{2} (\ketbra{x} + \ketbra{y}) = \frac{1}{4} \sum_{r \in \{0,1\}}(\ket{x}+(-1)^r\ket{y})(\bra{x}+(-1)^r\bra{y})$). Note that one might be worried that the output of Alice leaks additional information on this quantum state: however, the output of Alice is linked with the \emph{other}, non-dummy, quantum state and any additional information regarding this dummy state are anyway discarded. Finally, we can now apply the decoding performed by Alice on both outputs and output only the one corresponding to $m_b$: this is exactly the role of the ideal functionality. Since nothing depends on any secret (except this very last step where the functionality discards $m_{1-b}$ and outputs $m_b$), the simulator can fully run this procedure. \cameraVsOther{See the full version for details.}{See the \hyperref[proof:prAtEnd\pratendcountercurrent]{full security proof} in \pratendSectionlikeCref{} for more details.}

  \paragraph{Malicious receiver (Alice).}
  We consider now the case where the adversary $\cA = \hat{\Alice}$ corrupts the receiver Alice.

  Informally the goal of the simulator $\Sim_{\hat{\Alice}}$ is to extract the value $b$ from Alice in order to provide this value to the ideal functionality, and to appropriately use the $m_b$ provided by the functionality to fake measurement outcomes expected by Alice. At a high level, since the ZK protocol is a (state-preserving) proof (or argument) of knowledge (PoK), we can use this property to extract the witnesses $(w_d^{(c)})_{c,d}$. From this witness we can find a $w_d^{(b)}$ that starts with a $1$ in order to learn $b$. Then, to fake the measurement outcomes, the simulator can apply exactly the same quantum operations as the one done by the honest Bob, using the $m_b$ given by the functionality, except that the simulator will choose $m_{1-b} = 0$. Note that if the malicious Alice really sent a state $\ket{\psi^{(1-b)}}$ in the computational basis, then the $Z^{m_{1-b}}$ rotation does nothing, irrespective of the value of $m_{1-b}$. Now, if Alice sent a state that is in superposition of two pre-images with non-negligible amplitude, since it must pass the test checking that it contains non-dummy preimage of $h$, then it means that Alice ``knows'' a collision for $h$\dots{} or rather, we can measure the state to get a first preimage and compare it with the preimages extracted during the ZK protocol to get another preimage: with non-negligible probability (on the measurement outcome) they will be different, breaking the collision resistant property of $h$ which contradicts our assumption. Note that some care must be taken as the probability of finding a collision differs across runs, but we can formalize this argument as shown is the full proof. In practice, we will define a few hybrid games, by first replacing the distribution of $h$ and the ZK protocol by their simulated versions (since the ZK is a PoK, the simulator can learn $b$ and the preimages of $h$), then we remove the $Z^{m_{1-b}}$ rotation (indistinguishable or the state is far from a state in the computational basis, in which case we can recover a collision). Finally, since this does not depend on the secret $m_{1-b}$, we can reorganize the elements to recover the ideal word. See the \cameraVsOther{full security proof in the full version~\cite{CMS23_ObliviousTransferZeroKnowledge}}{\hyperref[proof:prAtEnd\pratendcountercurrent]{full security proof} in \pratendSectionlikeCref{}} for more details.
\end{proof}
\begin{proofE}
  Let $\cA$ be a static adversary. To prove that the above protocol realizes the above functionality, we will split the proof depending on the parties that the static adversary $\cA$ can corrupt.

  \paragraph{Case 1: correctness (no corrupted party).}

  If no party is corrupted, we are proving that the protocol is correct. First, we can cut $\Alice$ and $\Bob$ in four parts: the part that runs $\Alice_h$, the part that generates $h_c^{(c)}$, the part that runs the ZK proof, and the rest. Because of the correctness of the protocol distributing $h$, and because of the completeness of the ZK protocol, we can indistinguishably replace the first and third parts of $\Alice$ and $\Bob$ with the ideal dummy parties and the corresponding functionalities. Now, because $w_{1-l}^{(1-b)}[1] = 1$, the statement that we prove in the NIZK proof is true, so by the completeness of the NIZK protocol the check succeeds. Then, during the second step we apply the unitary that maps $\ket{x}\ket{w}\ket{0}$ to $\ket{x}\ket{w}\ket{1}$ only if $w[1] \neq 1$ (the witness is valid) and if $h(x\|w)$ appears in the list of hashes, which is true for all terms appearing in $\ket{\psi^{(b)}}$ and $\ket{\psi^{(1-b)}}$ by construction, so after this step the two states become:
  \begin{align}
    \ket{\psi^{(b)}} &\rightsquigarrow \ket{0}\ket{w_0^{(b)}}\ket{1} + (-1)^{r^{(b)}}\ket{1}\ket{w_1^{(b)}}\ket{1} = \ket{\psi^{(b)}}\ket{1}\\
    \ket{\psi^{(1-b)}} &\rightsquigarrow \ket{l}\ket{w_l^{(b)}}\ket{1} = \ket{\psi^{(1-b)}}\ket{1}
  \end{align}
  Because the last register (let us call them $\cT^{(b)}$ and $\cT^{(1-b)}$) of each state is not entangled with their respective first two registers, measuring $\cT^{(b)}$ and $\cT^{(1-b)}$ will output $1$ in both cases and will not disturb the states on the first two registers. Then, we measure the second register of each state in the Hadamard basis, i.e.\ we first apply Hadamard gates on all qubits and then we measure in the computational basis. After the Hadamard gates, the state $\ket{\psi^{(b)}}$ is turned into (omitting the constants):
  \begin{align}
    &(I \otimes H^{n+1})\ket{\psi^{(b)}}\\
    &= \ket{0} H^{n+1}\ket{w_0^{(b)}} + (-1)^{r^{(b)}} \ket{1}H^{n+1}\ket{w_1^{(b)}}\\
    &=\ket{0} \sum_{s^{(b)} \in \{0,1\}^{n+1}} (-1)^{\langle s^{(b)},w_0^{(b)}\rangle} \ket{s^{(b)}} + (-1)^{r^{(b)}} \ket{1} \sum_{s^{(b)} \in \{0,1\}^{n+1}} (-1)^{\langle s^{(b)},w_1^{(b)}\rangle} \ket{s^{(b)}}\\
    &= \ket{0} + \ket{1}\left(\sum_{s^{(b)} \in \{0,1\}^{n+1}} (-1)^{r^{(b)}\xor \langle s^{(b)},w_0^{(b)} \xor w_1^{(b)}\rangle} \ket{s^{(b)}}\right)\label{eq:correctnessComputationH}
  \end{align}
  where $\langle a,b \rangle \eqdef \xor_i a[i]b[i]$ is the standard inner product of bit strings. If we measure then an outcome $s^{(b)}$ and define $\alpha \eqdef r^{(b)} \xor \langle s^{(b)},w_0^{(b)} \xor w_1^{(b)}\rangle$ the above state collapses to:
  \begin{align}
    \ket{\psi^{\prime(b)}} \eqdef \ket{0} + (-1)^\alpha \ket{1}\label{eq:correctnessFirstState}
  \end{align}
  (note that the state is in the Hadamard basis.)

  For the second state $\ket{\psi^{(1-b)}} = \ket{l}\ket{w_l^{(1-b)}}$, the first register is not entangled with the second, so measuring the second register does not disturb the first one. So we end up with the state
  \begin{align}
    \ket{\psi^{\prime(1-b)}} \eqdef \ket{l}\label{eq:correctnessSecondState}
  \end{align}
  (i.e.\ the state is in the computational basis).
  
  In the final step, Bob rotates the two states:
  \begin{align}
    Z^{m_b}\ket{\psi^{\prime (b)}} = \ket{0} + (-1)^{\alpha \xor m_b}\ket{1} \qquad Z^{m_{1-b}}\ket{\psi^{\prime (1-b)}} = Z^{m_{1-b}} \ket{l} = \ket{l}  
  \end{align}
  (note that the rotation does nothing in the second case, hence all information about $m_{1-b}$ is lost). Then, Bob measures both states in the Hadamard basis: therefore we have $z^{(b)} =\alpha \xor m_b$ and $z^{(1-b)}$ is just a random bit since we measure a qubit in the computational basis. So at the end, Alice outputs $\alpha \xor z^{(b)} = m_b$ which ends the proof of correctness.

  \paragraph{Case 2: malicious sender (Bob).}

  We consider now the case where the adversary $\cA = \hat{\Bob}$ corrupts the sender Bob. For an intuitive overview of the proof, see \cref{subsec:overview}, together with the sketch of proof after \cref{thm:realizesOT}.

  First, we notice that since the functionality $\Fzk{}$ sends the word $x$ (in our case the hashes $h_d^{(c)}$) to the verifier, we do not need to send $x$ another time before (if needed we can assume that the ZK protocol starts by sending $x$). Let $\hat{\Bob}$ be an adversary. Without loss of generality\footnote{If the ZK is non-interactive, $\hat{\Bob}_0$ would just store the proof $\pi$, the hashes $(h_{i,j})$ and advice $\sigma$ and forward it to $\hat{\Bob}_1$.}, we can decompose $\hat{\Bob}$ into $\hat{\Bob}_0$ and $\hat{\Bob}_1$, where $\hat{\Bob}_0$ is the QIM running during the ZK protocol (so it receives an arbitrary advice $\sigma$ and interacts with $\Alice$ during the ZK protocol), forwarding its final internal state to $\hat{\Bob}_1$ that runs the rest of the protocol (in particular $\hat{\Bob}_1$ receives the quantum state and is supposed to output some measurement outcomes). For any $\Zenv$ and family of states $\sigma = \{\sigma_\lambda\}_{\lambda \in \N}$, we define below the following hybrids:
  \begin{itemize}
  \item $\World_0 \eqdef \Real^{\sigma}_{\Pi,\hat{\Bob},\Zenv}$ is the real world (where Alice runs internally the ZK protocol $\Azk$ with $\hat{\Bob}$) as pictured in \cref{fig:world0}.
  \item $\World_1$ is like $\World_0$ except that the $\Pi_h$ protocol (in charge of sharing $h$) is replaced by the simulated version as pictured in \cref{fig:world1}.
  \item $\World_2$ is like $\World_0$ except that the ZK protocol is replaced by the simulated version as pictured in \cref{fig:world2}.
  \item $\World_3$ is like $\World_2$ except that we remove $\Fzk{}$ and always forward $(h^c_d)_{c,d}$ to $\Sim_{\hat{\Bob}_0}$ as pictured in \cref{fig:world3}.
  \item $\World_4$ is like $\World_3$ except that we sample $w^{(1-b)}_{1-l} \sample \{0\} \times \{0,1\}^{n}$ as pictured in \cref{fig:world4}.
  \item $\World_5$ is like $\World_4$ except that we define instead $\ket{\psi^{(1-b)}} \eqdef \ket{0}\ket{w_0^{(1-b)}} + (-1)^{r^{(1-b)}} \ket{1}\ket{w_1^{(1-b)}}$ where $r^{(1-b)} \sample \{0,1\}$ is a random bit as pictured in \cref{fig:world5}.
  \item $\World_6$, pictured in \cref{fig:world6}, is like $\World_5$ except that we reorder some operations and we cut Alice in three parts: a simulator $\Sim_{\hat{\Bob}}$ (the simulator will also absorb $\Sim_{\hat{\Bob}_0}$ and $\hat{\Bob}_1$ by simply forwarding the input $\sigma$ to $\Sim_{\hat{\Bob}_0}$ and the output of $\hat{\Bob}_1$ to $\Zenv$), the ideal functionality $\Fot$, and the dummy party $\tilde{\Alice}$ that forwards $b$ to $\Fot$ and outputs the answer $m_b$ of $\Fot$ (in \cref{fig:world6} $\Fot$ and $\tilde{\Alice}$ are drawn together for to save space). More precisely, we see that all the messages sent to $\hat{\Bob}$ are sampled exactly in the same way, irrespective of the value of $b$, so we can push that outside of Alice into the simulator. The only part that still depends on $b$ is the output message. To avoid this dependency, the simulator will compute the two outputs (when $b=0$ and when $b=1$) and send them to $\Fot$ that will be in charge of outputting the appropriate value. This way, we see that $\World_6 = \Ideal^{\sigma,\Fot}_{\tilde{\Pi},\Sim_{\hat{\cB}},\Zenv}$.
  \end{itemize}

  \begin{figure}
    \centering
    \begin{autoFit}
      \pseudocodeblock{
        \textbf{$\Alice(b \in \{0,1\})$} \< \< \textbf{$\hat{\Bob}(\sigma_{\hat{\Bob}})$} \< \< \textbf{$\Zenv(\sigma_{\Zenv})$} \\[][\hline]
        \\[-.5\baselineskip]
        \text{Run $\Alice_h$ to obtain $h$.} \< \sendmessage{<->}{} \< \hat{\Bob}_0\\
        \text{$\forall c,d$, sample $w^{(c)}_d$ and $h^{(c)}_d$} \\
        \text{like in the main protocol}\\
        \text{Run $\Azk(\forall c,d, w^{(c)}_d, h^{(c)}_d)$} \< \sendmessage{<->}{} \< \hat{\Bob}_1\\
        \text{Sample $\ket{\psi^{(0)}}$,$\ket{\psi^{(1)}}$ like}\\
        \text{in the main protocol} \< \sendmessage{->}{topstyle={overlay},top={$\ket{\psi^{(0)}},\ket{\psi^{(1)}}$}} \< \\
        \text{Compute $\alpha$ like} \< \sendmessage{<-}{topstyle={overlay},top={$\forall c, s^{(c)}, z^{(c)}$}}  \< \hat{\Bob}_2 \< \sendmessage{->}{length=1cm,topstyle={overlay},top={state}} \<\\
        \text{in the main protocol}  \< \sendmessagerightx{5}{\alpha \xor  z^{(b)}} \< \pcreturn \text{anything}
      }
    \end{autoFit}
    \caption{$\World_0$}
    \label{fig:world0}
  \end{figure}

  \begin{figure}
    \centering
    \begin{autoFit}
      \pseudocodeblock{
        \textbf{$\Alice(b \in \{0,1\})$} \< \< \textbf{$\hat{\Bob}(\sigma_{\hat{\Bob}})$} \< \< \textbf{$\Zenv(\sigma_{\Zenv})$} \\[][\hline]
        \\[-.5\baselineskip]
        \text{\newStuff{Run $\Fh{\Gen}$ to obtain $h$.}} \< \sendmessage{<-}{topstyle={overlay},top={\text{\newStuff{$h$}}}} \< \text{\newStuff{$\Sim_{h,\hat{\Bob}_0}$}}\\
        \text{\newStuff{i.e.\ check if $h \in \cK$.}}\\
        \text{$\forall c,d$, sample $w^{(c)}_d$ and $h^{(c)}_d$} \\
        \text{like in the main protocol}\\
        \text{Run $\Azk(\forall c,d, w^{(c)}_d, h^{(c)}_d)$} \< \sendmessage{<->}{} \< \hat{\Bob}_1\\
        \text{Sample $\ket{\psi^{(0)}}$,$\ket{\psi^{(1)}}$ like}\\
        \text{in the main protocol} \< \sendmessage{->}{topstyle={overlay},top={$\ket{\psi^{(0)}},\ket{\psi^{(1)}}$}} \< \\
        \text{Compute $\alpha$ like} \< \sendmessage{<-}{topstyle={overlay},top={$\forall c, s^{(c)}, z^{(c)}$}}  \< \hat{\Bob}_2 \< \sendmessage{->}{length=1cm,topstyle={overlay},top={state}} \<\\
        \text{in the main protocol} \\
        \< \sendmessagerightx{5}{\alpha \xor  z^{(b)}} \< \pcreturn \text{anything}
      }
    \end{autoFit}
    \caption{$\World_1$}
    \label{fig:world1}
  \end{figure}

  \begin{figure}
    \centering
    \begin{autoFit}
      \pseudocodeblock{
        \textbf{$\Alice(b \in \{0,1\})$} \< \< \textbf{$\hat{\Bob}(\sigma_{\hat{\Bob}})$} \< \< \textbf{$\Zenv(\sigma_{\Zenv})$} \\[][\hline]
        \\[-.5\baselineskip]
        \text{Run $\Fh{\Gen}$ to obtain $h$,} \< \sendmessage{<-}{topstyle={overlay},top={h}} \< \Sim_{h,\hat{\Bob}_0}\\
        \text{i.e.\ check if $h \in \cK$.}\\
        \text{$\forall c,d$, sample $w^{(c)}_d$ and $h^{(c)}_d$} \\
        \text{like in the main protocol}\\
        \text{\newStuff{Run $\Fzk{\cR}$, i.e.\ check if}}\\
        \text{\newStuff{$\exists c',d'$, $h(d' \| w^{(c')}_{d'}) = h^{(c')}_{d'}$}}\\
        \text{\newStuff{and outputs all $h^{(c)}_d$}} \< \sendmessage{->}{topstyle={overlay},top={\text{\newStuff{$\forall c,d, h^{(c)}_d$}}}} \< \text{\newStuff{$\Sim_{zk,\hat{\Bob}_1}$}}\\
        \text{Sample $\ket{\psi^{(0)}}$,$\ket{\psi^{(1)}}$ like}\\
        \text{in the main protocol} \< \sendmessage{->}{topstyle={overlay},top={$\ket{\psi^{(0)}},\ket{\psi^{(1)}}$}} \< \\
        \text{Compute $\alpha$ like} \< \sendmessage{<-}{topstyle={overlay},top={$\forall c, s^{(c)}, z^{(c)}$}}  \< \hat{\Bob}_2 \< \sendmessage{->}{length=1cm,topstyle={overlay},top={state}} \<\\
        \text{in the main protocol} \\
        \< \sendmessagerightx{5}{\alpha \xor  z^{(b)}} \< \pcreturn \text{anything}
      }
    \end{autoFit}
    \caption{$\World_2$}
    \label{fig:world2}
  \end{figure}

  \begin{figure}
    \centering
    \begin{autoFit}
      \pseudocodeblock{
        \textbf{$\Alice(b \in \{0,1\})$} \< \< \textbf{$\hat{\Bob}(\sigma_{\hat{\Bob}})$} \< \< \textbf{$\Zenv(\sigma_{\Zenv})$} \\[][\hline]
        \\[-.5\baselineskip]
        \text{Run $\Fh{\Gen}$ to obtain $h$,} \< \sendmessage{<-}{topstyle={overlay},top={h}} \< \Sim_{h,\hat{\Bob}_0}\\
        \text{i.e.\ check if $h \in \cK$.}\\
        \text{$\forall c,d$, sample $w^{(c)}_d$ and $h^{(c)}_d$} \\
        \text{like in the main protocol}\\
        \text{\newErasedStuff{Check …}} \< \sendmessage{->}{topstyle={overlay},top={\text{$\forall c,d, h^{(c)}_d$}}} \< \Sim_{zk,\hat{\Bob}_1}\\
        \text{Sample $\ket{\psi^{(0)}}$,$\ket{\psi^{(1)}}$ like}\\
        \text{in the main protocol} \< \sendmessage{->}{topstyle={overlay},top={$\ket{\psi^{(0)}},\ket{\psi^{(1)}}$}} \< \\
        \text{Compute $\alpha$ like} \< \sendmessage{<-}{topstyle={overlay},top={$\forall c, s^{(c)}, z^{(c)}$}}  \< \hat{\Bob}_2 \< \sendmessage{->}{length=1cm,topstyle={overlay},top={state}} \<\\
        \text{in the main protocol} \\
        \< \sendmessagerightx{5}{\alpha \xor  z^{(b)}} \< \pcreturn \text{anything}
      }
    \end{autoFit}
    \caption{$\World_3$}
    \label{fig:world3}
  \end{figure}

  \begin{figure}
    \centering
    \begin{autoFit}
      \pseudocodeblock{
        \textbf{$\Alice(b \in \{0,1\})$} \< \< \textbf{$\hat{\Bob}(\sigma_{\hat{\Bob}})$} \< \< \textbf{$\Zenv(\sigma_{\Zenv})$} \\[][\hline]
        \\[-.5\baselineskip]
        \text{Run $\Fh{\Gen}$ to obtain $h$,} \< \sendmessage{<-}{topstyle={overlay},top={h}} \< \Sim_{h,\hat{\Bob}_0}\\
        \text{i.e.\ check if $h \in \cK$.}\\
        \text{$\forall c,d$, sample $w^{(c)}_d$ and $h^{(c)}_d$ like} \\
        \text{in the main protocol, \newStuff{except for}}\\
        \text{\newStuff{$w^{(1-b)}_{1-l} \sample \{0\} \times \{0,1\}^{n}$.}} \< \sendmessage{->}{topstyle={overlay},top={$\forall c,d, h^{(c)}_d$}} \< \Sim_{zk,\hat{\Bob}_1}\\
        \text{Sample $\ket{\psi^{(0)}}$,$\ket{\psi^{(1)}}$ like}\\
        \text{in the main protocol} \< \sendmessage{->}{topstyle={overlay},top={$\ket{\psi^{(0)}},\ket{\psi^{(1)}}$}} \< \\
        \text{Compute $\alpha$ like} \< \sendmessage{<-}{topstyle={overlay},top={$\forall c, s^{(c)}, z^{(c)}$}}  \< \hat{\Bob}_2 \< \sendmessage{->}{length=1cm,topstyle={overlay},top={state}} \<\\
        \text{in the main protocol} \\
        \< \sendmessagerightx{5}{\alpha \xor  z^{(b)}} \< \pcreturn \text{anything}
      }
    \end{autoFit}
    \caption{$\World_4$}
    \label{fig:world4}
  \end{figure}

  \begin{figure}
    \centering
    \begin{autoFit}
      \pseudocodeblock{
        \textbf{$\Alice(b \in \{0,1\})$} \< \< \textbf{$\hat{\Bob}(\sigma_{\hat{\Bob}})$} \< \< \textbf{$\Zenv(\sigma_{\Zenv})$} \\[][\hline]
        \\[-.5\baselineskip]
        \text{Run $\Fh{\Gen}$ to obtain $h$,} \< \sendmessage{<-}{topstyle={overlay},top={h}} \< \Sim_{h,\hat{\Bob}_0}\\
        \text{i.e.\ check if $h \in \cK$.}\\
        \text{$\forall c,d$, sample $w^{(c)}_d$ and $h^{(c)}_d$ like} \\
        \text{in the main protocol, except for}\\
        \text{$w^{(1-b)}_{1-l} \sample \{0\} \times \{0,1\}^{n}$.} \< \sendmessage{->}{topstyle={overlay},top={$\forall c,d, h^{(c)}_d$}} \< \Sim_{zk,\hat{\Bob}_1}\\
        \text{Sample $\ket{\psi^{(\text{\newStuff{$b$}})}}$ like in the main}\\
        \text{protocol, \newStuff{$r^{(1-b)} \sample \{0,1\}$ and}}\\
        \text{\newStuff{$\ket{\psi^{(1-b)}} \eqdef \ket{0}\ket{w_0^{(1-b)}}$}}\\
        \text{\newStuff{$+ (-1)^{r^{(1-b)}} \ket{1}\ket{w_1^{(1-b)}}$}} \< \sendmessage{->}{topstyle={overlay},top={$\ket{\psi^{(0)}},\ket{\psi^{(1)}}$}} \< \\
        \text{Compute $\alpha$ like} \< \sendmessage{<-}{topstyle={overlay},top={$\forall c, s^{(c)}, z^{(c)}$}}  \< \hat{\Bob}_2 \< \sendmessage{->}{length=1cm,topstyle={overlay},top={state}} \<\\
        \text{in the main protocol} \\
        \< \sendmessagerightx{5}{\alpha \xor  z^{(b)}} \< \pcreturn \text{anything}
      }
    \end{autoFit}
    \caption{$\World_5$}
    \label{fig:world5}
  \end{figure}

  \begin{figure}
    \centering
    \begin{autoFit}
      \pseudocodeblock{
        \textbf{$\Alice_I(b)$ and $\Fot$} \< \< \textbf{$\Sim_{\hat{\Bob}}(\sigma_{\hat{\Bob}})$} \< \< \textbf{$\Zenv(\sigma_{\Zenv})$} \\[][\hline]
        \\[-.5\baselineskip]
        \< \< (h, \mathsf{state}_0) \gets \Sim_{h,\hat{\Bob}_0}(\sigma_{\hat{\Bob}})\\
        \< \< \text{Check if $h \in \cK$.}\\
        \< \<  \text{$\forall c,d$, sample $w^{(c)}_d$ and $h^{(c)}_d$ like} \\
        \< \<  \text{in the main protocol, except for}\\
        \< \<  \text{$w^{(1-b)}_{1-l} \sample \{0\} \times \{0,1\}^{n}$.}\\
        \< \< (\mathsf{state}_1) \gets \Sim_{zk,\hat{\Bob}_1}(\forall c,d, h^{(c)}_d,\mathsf{state}_0)\\
        \< \< \text{Sample $\ket{\psi^{(\text{\newStuff{$b$}})}}$ like in the main}\\
        \< \< \text{protocol, \newStuff{$r^{(1-b)} \sample \{0,1\}$ and}}\\
        \< \< \text{\newStuff{$\ket{\psi^{(1-b)}} \eqdef \ket{0}\ket{w_0^{(1-b)}}$}}\\
        \< \< \text{\newStuff{$+ (-1)^{r^{(1-b)}} \ket{1}\ket{w_1^{(1-b)}}$}}\\
        \< \< (\forall c, s^{(c)},z^{(c)}, \mathsf{state}_2) \gets \hat{\Bob}_2(\ket{\psi^{(0)}},\ket{\psi^{(1)}}, \mathsf{state}_1) \\
        \< \sendmessage{<-}{topstyle={overlay},length=1cm,top={$m^{(0)}, m^{(1)}$}} \< \forall c, m^{(c)} = (r^{(c)} \xor \bigoplus_i s^{(c)}[i] (w^{(c)}_0 \xor w^{(c)}_1)[i]) \xor z^{(c)}\\    
        \text{$\Fot$ outputs $m^{(b)}$} \<  \< \< \sendmessage{->}{length=1cm,topstyle={overlay},top={$\mathsf{state}_2$}} \<\\
        \< \sendmessagerightx{5}{m^{(b)}} \< \pcreturn \text{anything}
      }
    \end{autoFit}
    \caption{$\World_6$}
    \label{fig:world6}
  \end{figure}

  First, we see that $\World_0 \approxRV \World_2$ because we assumed that the underlying protocol \CSQSA{S} realizes $\Fzk{}$: If it were not the case, then we could easily break the \CSQSA{S'} property of $\Fzk{}$ (and therefore the \CSQSA{S} property of $\Fzk{}$ since $S' \subseteq S$) by merging the classical sampling procedure inside $\sigma$ to get a new $\sigma'$ (the value of the witness being kept as a side-information for $\Zenv$ in $\sigma'$) and the rest of the procedure (preparation of the quantum state and running $\hat{B}_1$) inside $\Zenv$ to produce a new $\Zenv'$ able to attack $\Fzk{}$ with exactly the same probability.

  Then, $\World_2 = \World_3$: by construction, there always exists a witness starting with a $0$, so $\Fzk{}$ will always forward $(h^c_d)_{c,d}$.

  We also have $\World_3 \approxRV \World_4$ because of the hardcore second-bit property (see \cref{def:firstBitHardcore}). Otherwise, we can easily break the hardcore second-bit property by defining $\cA$ as $\World_3$ except that the $w^{(1-b)}_{1-l}$ is sampled externally by the ``challenger'' that only provides $h^{(1-b)}_{1-l}$ to $\cA$ (note that $w^{(1-b)}_{1-l}$ is not needed here except to compute $h^{(1-b)}_{1-l}$). Note that this attacker is only valid if the computational power of $\hat{\Bob}$ is lower than the computational power defined in the hardcore second-bit property, but this is fine if $\{ \Bob \} \in S'$ only if $h$ has the statistically hardcore second-bit property as assumed.

  $\World_4 = \World_5$, because for any $x$ and $x'$, it is statistically impossible to distinguish $\ket{x} + (-1)^{r^{(1-b)}} \ket{x'}$ from $\ket{x''}$ where $x''$ equals $x$ with probability $\frac{1}{2}$ and $x'$ otherwise, even given $x$ and $x'$ (the random sign $r^{(1-b)}$ must however stay hidden\footnote{It is also possible to remove the sign, but then the proof is harder and only applies statistically if $h$ is lossy.}). Indeed, we can simply study the density matrices, first of $\ket{x} \pm \ket{x'}$ (the first sum is on the random choice $r^{(1-b)}$ of the sign):
  \begin{align}
    &\frac{1}{2} (\frac{\ket{x}+\ket{x'}}{\sqrt{2}}\frac{\bra{x} + \bra{x'}}{\sqrt{2}} + \frac{\ket{x}-\ket{x'}}{\sqrt{2}}\frac{\bra{x} - \bra{x'}}{\sqrt{2}})\\
    &= \frac{1}{4} (\ketbra{x}{x} + \ketbra{x}{x'} + \ketbra{x'}{x} + \ketbra{x'}{x'} + \ketbra{x}{x} - \ketbra{x}{x'} - \ketbra{x'}{x} + \ketbra{x'}{x'})\\
    &= \frac{1}{2} (\ketbra{x}{x} + \ketbra{x'}{x'})
  \end{align}
  But this last expression is exactly the density matrix of $\ket{x''}$ where $x''$ equals $x$ with probability $1/2$ and $x'$ otherwise.

  In our case, we have $x = 0 \| w^{(1-b)}_0$, $x' = 1 \| w^{(1-b)}_1$ and $x'' = l \| w^{(1-b)}_{l}$ since $l$ plays the role of the random coin determining the element to send in $\World_4$ (you can see that $l$ plays no role in $\World_5$ since we can re-order the steps in the sampling procedure). Since $l$ and $r^{(1-b)}$ are only used to determine $\ket{\psi^{(1-b)}}$ and since both worlds are equal beside the choice of $\ket{\psi^{(1-b)}}$, we can conclude that $\World_4 = \World_5$ as otherwise we could ``factor out'' $\ket{\psi^{(1-b)}}$ from the worlds and use the remaining part to distinguish between $\ket{x} + (-1)^{r^{(1-b)}} \ket{x'}$ from $\ket{x''}$ which is physically impossible as we just saw.

  Finally, $\World_5 = \World_6$ since we just reordered the sampling procedure, delocalized their computation to $\Sim_{\hat{\Bob}}$ and used $\Fot$ to discard the message corresponding to $m_{1-b}$ in order to keep only $m_b$ as in $\World_5$.

  Therefore, by transitivity, $\World_0 \approxRV \World_6$, i.e.\ $\Real^{\sigma}_{\Pi,\hat{\Bob},\Zenv} \approxRV \Ideal^{\sigma,\Fot}_{\tilde{\Pi},\Sim_{\hat{\cB}},\Zenv}$ which concludes this part of the proof.
 
  \paragraph{Case 3: malicious receiver (Alice).}

  We consider now the case where the adversary $\cA = \hat{\Alice}$ corrupts the receiver Alice. For an intuitive overview of the proof, see \cref{subsec:overview}, together with the sketch of proof after \cref{thm:realizesOT}.

  Like in case 2, without loss of generality we can decompose $\hat{\Alice}$ into three parts: $\hat{\Alice}_0$ will be the part running the $\Pi_h$ protocol, $\hat{\Alice}_1$ will be the part running the ZK protocol, forwarding its internal state to $\hat{\Alice}_2$ that will be in charge of the rest of the protocol as pictured in \cref{fig:worldCase3:0}. In the following, we define, for any bit $b$, $\cX^{(b)}$ as the register containing the state $\ket{\psi^{(b)}}$, $\cW^{(1-b)}$ as the sub-register of $\cX^{(b)}$ containing the witness $w$ (so all but the first qubit) and $\cW^{(1-b)}[1]$ as the first qubit in the register $\cW^{(1-b)}$.
  
  We formalize this reasoning by defining the following hybrid worlds:
  \begin{itemize}
  \item $\World_0 \eqdef \Real^{\sigma}_{\Pi,\hat{\Alice},\Zenv}$ is the real world (where Bob runs internally the ZK protocol $\Bzk$ with $\hat{\Alice}$) as pictured in \cref{fig:worldCase3:0}.
  \item $\World_1$ is like $\World_0$ except that we replace $\Pi_h$ with the ideal world ($\hat{\Alice}_0$ is replaced with $\Sim_{\hat{\Alice}_0}$, and $\Bob_h$ is now replaced with the ideal resource $\Fh{\Gen}$ and the idealized party $\Bob_{h,I}$ sampling honestly $h \gets \Gen(1^\lambda)$), as pictured in \cref{fig:worldCase3:1}.
  \item $\World_2$ is like $\World_1$ except that we replace $\hat{\Alice}_1$ with the simulator of the ZK protocol interacting with $\Fzk{}$, and integrate the ideal resource checking if there exists $w^{1-b}$ starting with a $1$ in $\Bob$ as pictured in \cref{fig:worldCase3:2}.
  \item $\World_3$ is like $\World_2$ except that it does not perform the $Z^{m_{1-b}}$ rotation, as pictured in \cref{fig:worldCase3:3}.
  \item For $\World_4$, we can realize that in $\World_3$, the code does not depend on $m_{1-b}$ anymore. So we can reorganise the elements of $\World_3$ to let the functionality provide $m_b$ and discard $m_{1-b}$: we split $\Bob$ into three parts: the functionality $\Fot$, a dummy Bob that just forwards $m_0$ and $m_1$ to $\Fot$, and a simulator $\Sim_{\hat{\Alice}}$ that runs the code of $\Bob$ in the previous World (extracting $b$ in the same way), except that it sends $b$ to $\Fot$ to get $m_b$. We also let $\Sim_{\hat{\Alice}}$ absorb $\hat{\Alice}$, $\Bob_{h,I}$,$\Sim_{\hat{\Alice}_0}$ and $\Sim_{\hat{\Alice}_1}$ (appropriately forwarding their input/outputs): this way $\World_4 = \Ideal^{\sigma,\Fot}_{\tilde{\Pi},\Sim_{\hat{\Alice}},\Zenv}$. This is pictured \cref{fig:worldCase3:4}.
  \end{itemize}
  
  \begin{figure}
    \centering
    \begin{autoFit}
      \pseudocodeblock{
        \textbf{$\hat{\Alice}(\sigma_{\hat{\Alice}})$} \< \< \textbf{$\Bob(m_0,m_1)$} \< \< \textbf{$\Zenv(\sigma_{\Zenv})$} \\[][\hline]
        \\[-.5\baselineskip]
        \hat{\Alice}_0 \< \sendmessage{<->}{} \< \text{Run $\Bob_{h}$ to obtain $h$}\\
        \hat{\Alice}_1 \< \sendmessage{<->}{} \< \text{Run $\Bob_{zk}$ to obtain $\forall c,d, h^{(c)}_d$}\\
        \hat{\Alice}_2 \< \sendmessage{->}{topstyle={overlay},top={$\ket{\psi}^{(0)}, \ket{\psi}^{(1)}$}} \< \text{Do the measurements and}\\
        \<\<\text{rotations of the real protocol.}\\
        \hat{\Alice}_3 \< \sendmessage{<-}{topstyle={overlay},top={$\forall c, s^{(c)}, z^{(c)}$}} \< \\    
        \< \sendmessagerightx[7.7cm]{5}{\mathsf{state}} \< \pcreturn \text{anything}
      }
    \end{autoFit}
    \caption{Case 3 (malicious Alice): $\World_0$}
    \label{fig:worldCase3:0}
  \end{figure}

  \begin{figure}
    \centering
    \begin{autoFit}
      \pseudocodeblock{
        \textbf{$\hat{\Alice}(\sigma_{\hat{\Alice}})$} \< \< \textbf{$\Bob(m_0,m_1)$} \< \< \textbf{$\Zenv(\sigma_{\Zenv})$} \\[][\hline]
        \\[-.5\baselineskip]
        \text{\newStuff{$\Sim_{h,\hat{\Alice}_0}$}} \< \sendmessage{<-}{topstyle={overlay},top={\newStuff{$h$}}} \< \text{\newStuff{$h \gets \Gen(1^\lambda)$}}\\
        \hat{\Alice}_1 \< \sendmessage{<->}{} \< \text{Run $\Bob_{zk}$ to obtain $\forall c,d, h^{(c)}_d$}\\
        \hat{\Alice}_2 \< \sendmessage{->}{topstyle={overlay},top={$\ket{\psi}^{(0)}, \ket{\psi}^{(1)}$}} \< \text{Do the measurements and}\\
        \<\<\text{rotations of the real protocol}\\
        \hat{\Alice}_3 \< \sendmessage{<-}{topstyle={overlay},top={$\forall c, s^{(c)}, z^{(c)}$}} \< \\    
        \< \sendmessagerightx[7.7cm]{5}{\mathsf{state}} \< \pcreturn \text{anything}
      }
    \end{autoFit}
    \caption{Case 3 (malicious Alice): $\World_1$}
    \label{fig:worldCase3:1}
  \end{figure}

  \begin{figure}
    \centering
    \begin{autoFit}
      \pseudocodeblock{
        \textbf{$\hat{\Alice}(\sigma_{\hat{\Alice}})$} \< \< \textbf{$\Bob(m_0,m_1)$} \< \< \textbf{$\Zenv(\sigma_{\Zenv})$} \\[][\hline]
        \\[-.5\baselineskip]
        \Sim_{h,\hat{\Alice}_0} \< \sendmessage{<-}{topstyle={overlay},top={$h$}} \< h \gets \Gen(1^\lambda)\\
        \text{\newStuff{$\Sim_{zk,\hat{\Alice}_1}$}} \< \sendmessage{->}{top={$\forall c, d, w^{(c)}_d, h^{(c)}_d$}} \< \text{\newStuff{Run $\Fzk{}$, i.e.\ check if $w$'s}}\\
        \< \< \text{\newStuff{are valid witnesses for $\cR$}}\\
        \< \< \text{(notably $\exists b, h(l \| w^{(b)}_l) = h^{(b)}_l$).}\\
        \hat{\Alice}_2 \< \sendmessage{->}{topstyle={overlay},top={$\ket{\psi}^{(0)}, \ket{\psi}^{(1)}$}} \< \text{Do the measurements and}\\
        \<\<\text{rotations of the real protocol.}\\
        \hat{\Alice}_3 \< \sendmessage{<-}{topstyle={overlay},top={$\forall c, s^{(c)}, z^{(c)}$}} \< \\    
        \< \sendmessagerightx[7.7cm]{5}{\mathsf{state}} \< \pcreturn \text{anything}
      }
    \end{autoFit}
    \caption{Case 3 (malicious Alice): $\World_2$}
    \label{fig:worldCase3:2}
  \end{figure}
  
  \begin{figure}
    \centering
    \begin{autoFit}
      \pseudocodeblock{
        \textbf{$\hat{\Alice}(\sigma_{\hat{\Alice}})$} \< \< \textbf{$\Bob(m_0,m_1)$} \< \< \textbf{$\Zenv(\sigma_{\Zenv})$} \\[][\hline]
        \\[-.5\baselineskip]
        \Sim_{h,\hat{\Alice}_0} \< \sendmessage{<-}{topstyle={overlay},top={$h$}} \< h \gets \Gen(1^\lambda)\\
        \text{$\Sim_{zk,\hat{\Alice}_1}$} \< \sendmessage{->}{top={$\forall c, d, w^{(c)}_d, h^{(c)}_d$}} \< \text{Run $\Fzk{}$, i.e.\ check if $w$'s}\\
        \< \< \text{are valid witnesses for $\cR$}\\
        \< \< \text{(notably $\exists b, h(l \| w^{(b)}_l) = h^{(b)}_l$).}\\
        \hat{\Alice}_2 \< \sendmessage{->}{topstyle={overlay},top={$\ket{\psi}^{(0)}, \ket{\psi}^{(1)}$}} \< \text{Do the measurements and}\\
        \<\<\text{rotations of the real protocol,}\\
        \hat{\Alice}_3 \< \sendmessage{<-}{topstyle={overlay},top={$\forall c, s^{(c)}, z^{(c)}$}} \< \text{\newStuff{except $Z^{m_{1-b}}$.}} \\    
        \< \sendmessagerightx[7.7cm]{5}{\mathsf{state}} \< \pcreturn \text{anything}
      }
    \end{autoFit}
    \caption{Case 3 (malicious Alice): $\World_3$}
    \label{fig:worldCase3:3}
  \end{figure}

  \begin{figure}
    \centering
    \begin{autoFit}
      \pseudocodeblock{
        \textbf{$\Sim_{\hat{\Alice}(\sigma_{\hat{\Alice}})}$} \< \< \textbf{$\tilde{\Bob}(m_0,m_1)$ \& $\Fot$} \< \< \textbf{$\Zenv(\sigma_{\Zenv})$} \\[][\hline]
        \\[-.5\baselineskip]
        h \gets \Gen(1^\lambda)\\
        \mathsf{state}_0 \gets \Sim_{h,\hat{\Alice}_0}(\sigma_{\hat{\Alice}},h)\\
        \text{$(\forall c, d, w^{(c)}_d, h^{(c)}_d, \mathsf{state}_1) \gets \Sim_{zk,\hat{\Alice}_1}(\mathsf{state}_0)$}\\
        \text{Run $\Fzk{}$, i.e.\ check if $w$'s}\\
        \text{are valid witnesses for $\cR$}\\
        \text{(notably $\exists b, h(l \| w^{(b)}_l) = h^{(b)}_l$).} \< \sendmessage{->}{length=1cm,topstyle={overlay},top={$b$}} \< \\
        (\ket{\psi}^{(0)}, \ket{\psi}^{(1)}, \mathsf{state}_2) \gets \hat{\Alice}_2(\mathsf{\state}_1) \< \sendmessage{<-}{length=1cm,topstyle={overlay},top={$m_b$}}\\
        \text{Do the measurements and}\\
        \text{rotations of the real protocol,}\\
        \text{except $Z^{m_{1-b}}$.}\\
        \mathsf{state}_3 \gets \hat{\Alice}_3(\forall c, s^{(c)}, z^{(c)},\mathsf{state}_2)
        \< \sendmessagerightx[4.3cm]{5}{\mathsf{state}_3} \< \pcreturn \text{anything}
      }
    \end{autoFit}
    \caption{Case 3 (malicious Alice): $\World_4$}
    \label{fig:worldCase3:4}
  \end{figure}

  First, we see that $\World_0 \approxRV \World_1$ because we assumed that the $\Pi_h$ protocol \CSQSA{S_h} realizes $\Fzk{}$: If it were not the case, then we could easily break the \CSQSA{S'} property of $\Pi_h$ (and therefore the \CSQSA{S_h} property since $S' \subseteq S_h$) by merging basically all the procedure after the $\Pi_h$ protocol inside $\Zenv$ to produce a new $\Zenv'$ able to attack $\Fzk{}$ with exactly the same probability.

  We have also that $\World_1 \approxRV \World_2$ because we assumed that the underlying protocol \CSQSA{S} realizes $\Fzk{}$: If it were not the case, then we could easily break the \CSQSA{S'} property of this underlying protocol (and therefore its \CSQSA{S} property since $S' \subseteq S$) by merging basically all the procedure after the ZK protocol inside $\Zenv$ to produce a new $\Zenv'$ able to attack $\Fzk{}$ with exactly the same probability.

  Then, we prove that $\World_2 \approxRV \World_3$.
  \begin{subproof}
    This part is slightly more technical, but intuitively the goal is to show that the state sent by Alice is close in trace distance to a state in the computational basis (otherwise we can break the collision resistance property of $h$), and therefore a $Z$ rotation does not significantly disturb the state.

    To formalize this intuition, we will first associate a quantity $\beta$ to each run, show that this $\beta$ is linked to the probability of finding a collision and to the trace distance to the measured state, and finally we show that the average\footnote{We could also, equivalently, prove that the probability of having a non-negligible $\beta$ is negligible, but this introduce two polynomials, leading to a less elegant proof.} value of $\beta$ must be negligible, like that average trace distance between the two worlds. To that aim, it is handy to consider, for a fixed run, a (normalized) purification $\ket{\psi}_{\cT,\Bob,E} = \ket{t} \sum_{x,y} \beta^{(1-b)}_{x,y}\ket{x}\ket{y}$ of the states $\ket{\psi^{(0)}}$ and $\ket{\psi^{(1)}}$ sent by $\hat{\Alice}$ and partially measured\footnote{We are referring to the operations where Bob extracts the witness, where it checks that the quantum register containing the witnesses in superposition starts with a $0$ and that they are valid pre-images. The result of these tests ($t = 1$ iff all tests pass) is put in a new register $\cT$.} by $\Bob$ with the result of the tests it the register $\cT$, including any potential entanglement with the adversary or environment by adding a third register $\cE$ (we also put in this register the internal memory of $\hat{\Alice}$ right after she sent the state to $\Bob$). Moreover, we can assume without loss of generality that $\hat{\Alice}$ does nothing before receiving the measurement outcomes sent by Bob, by simply postponing in time its actions.

    Then, we can first study the simplest case: if $h$ is injective (a.k.a.\ statistically collision resistant), then $\World_2 = \World_3$:
    \begin{subproof}
      First, if $t = 0$ (invalid test), then the remaining actions in $\World_2$ and $\World_3$ are identical. Now, if $t = 1$, because $h$ is injective there exists at most one $x^{(1-b)}$ such $x^{(1-b)}$ has a $0$ at the second position and $h(x^{(1-b)}) \in \{h^{(1-b)}_0, h^{(1-b)}_1\}$ (we already extracted above one pre-image of $h^{(1-b)}_0$ or $h^{(1-b)}_1$ with a $1$ at the second position in the ZK protocol, and by injectivity of $h$ there are at most two pre-images to this set, so a single image can have a $0$ at the second position). Because Bob measured exactly that the first register is a pre-image of $\{h^{(1-b)}_0, h^{(1-b)}_1\}$ (since $t = 1$), the state $\ket{\psi}_{\Bob,E}$ will collapse into $\ket{x^{(1-b)}}(\sum_{y} \beta^{(1-b)}_{x,y} \ket{y})$ (up to a renormalisation factor). Therefore, since $Z^{m}\ket{x} = \ket{x}$, applying $Z^{m_{1-b}}_2$ or not does not change the state at all (this is true for any run), leading to $\World_2 = \World_3$.
    \end{subproof}
    Now, we consider the case where $h$ is only computationally collision-resistant. During a valid extraction of the witness, we found an element $x^{(1-b)}_{1-l}$ whose witness part starts with a $1$ ($x^{(1-b)}_{1-l}[2] = 1$) such that $h(x^{(1-b)}_{1-l}) \in \{h^{(1-b)}_0, h^{(1-b)}_1\}$: let $x^* \eqdef x^{(1-b)}_{l}$ be the other part of the witness such that $h(x^*) \in \{h^{(1-b)}_0, h^{(1-b)}_1\}$ (if both of them start with a $1$, we can choose arbitrarily). Then, there exists $\beta \in [0,1]$ (possibly equal to $1$ if $x^*$ starts with a $1$), a normalized pure state $\ket{\phi^*}$, and a normalized pure state $\ket{\phi}$ such that $\Tr((\ketbra{x^*}\otimes I)\ketbra{\phi}) = 0$ such that:
    \begin{align}
      \ket{\psi}_{\cT,\Bob,E} = \ket{t}(\sqrt{1-\beta}\ket{x^*}\ket{\phi^*}+\sqrt{\beta}\ket{\phi})
    \end{align}
    \begin{subproof}
      This can easily be seen by rewriting $\ket{\psi}_{\Bob,E}$ appropriately:
      \begin{align}
        \ket{\psi}_{\Bob,E}
        &= \sum_{x,y} \beta^{(1-b)}_{x,y}\ket{x}\ket{y}\\
        &= \ket{x^*}(\sum_{y}\beta^{(1-b)}_{x^*,y}\ket{y}) + \sum_{x \neq x^*,y} \beta^{(1-b)}_{x,y}\ket{x}\ket{y}\\
        &= (\sum_y |\beta^{(1-b)}_{x^*,y}|^2) \ket{x^*}(\frac{1}{\sum_y |\beta^{(1-b)}_{x^*,y}|^2} \sum_{y}\beta^{(1-b)}_{x^*,y}\ket{y}) + \sum_{x \neq x^*,y} \beta^{(1-b)}_{x,y}\ket{x}\ket{y}
      \end{align}
      Then, because $\ket{\psi}_{\Bob,E}$ is normalized, we have $0 \leq \sum_y |\beta^{(1-b)}_{x^*,y}|^2 \leq 1$, so by defining
      \begin{align*}
        \beta &\eqdef 1-(\sum_y |\beta^{(1-b)}_{x^*,y}|^2)^2 \quad
        \ket{\phi^*} \eqdef \frac{1}{\sum_y |\beta^{(1-b)}_{x^*,y}|^2} \sum_{y}\beta^{(1-b)}_{x^*,y}\ket{y}\\
        \ket{\phi}& \eqdef \frac{1}{\sqrt{\beta}} (\ket{\psi}_{\Bob,E} - \sqrt{1-\beta} \ket{\phi^*})
      \end{align*}
      (if $\beta = 0$, $\ket{\phi} = 0$) we have that $\beta \in [0,1]$, $\ket{\phi^*}$ is normalized, $\ket{\psi}_{\Bob,E} = \sqrt{1-\beta}\ket{x^*}\ket{\phi^*}+\sqrt{\beta}\ket{\phi}$, $\Tr((\ketbra{x^*}\otimes I)\ketbra{\phi}) = 0$ (since $\ket{\phi} \propto \ket{\psi}_{\Bob,E} - \sqrt{1-\beta} \ket{\phi^*} = \sum_{x \neq x^*,y} \beta^{(1-b)}_{x,y}\ket{x}\ket{y}$, i.e.\ a sum on terms $x \neq x^*$), and $\ket{\phi}$ is also normalized since $\ket{\psi}_{\Bob,E}$ is also normalized.
    \end{subproof}
    We observe now that the probability of finding a collision is greater than $t\beta$:
    \begin{subproof}
      First, if $t=0$ (the test fails), then $t\beta = 0$ so this is obviously true. Now, if $t = 1$, this can be seen by first remarking that if we measure the register $\Bob$ of $\ket{\psi}_{\Bob,E}$, we get an outcome $x$: since $\Tr((\ketbra{x^*}\otimes I)\ketbra{\phi}) = 0$, $x$ equals $x^*$ with probability $\sqrt{1-\beta}^2 = 1-\beta$, i.e.\ $x \neq x^*$ with probability $\beta$. Moreover, because $t = 1$, we know that $x[2] = 0$ and $h(x) \in \{h^{(1-b)}_0, h^{(1-b)}_1\}$. In the first case, if $h(x) = h^{(1-b)}_{1-l}$, then $x \neq x^{(1-b)}_{1-l}$ since $x[2] = 1$ and $x^{(1-b)}_{1-l}[2] = 0$ so $(x, x^{(1-b)}_{1-l})$ is a collision (reminder: $x^{(1-b)}_{1-l}$ was extracted by the simulator during the ZK protocol). In the second case, if $h(x) = h^{(1-b)}_{l}$, because with probability $\beta$ we have $x \neq x^*$, $(x,x^*)$ forms a collision with probability $\beta$. Therefore we can find a collision with probability greater than $\beta = t\beta$.
    \end{subproof}
    Moreover, we observe that the trace distance between $\ket{\psi}_{\cT,\Bob,E}$ and $Z^{t m_{1-b}}_{\Bob,2}\ket{\psi}_{\Bob,E}$ is smaller than $2t\sqrt{\beta}$:
    \begin{subproof}
      First, if $t = 0$, both states are strictly equal, so their trace distance is $0$. If $t = 1$, this can be seen using the triangle inequality (first inequality), the fact that on states in the computational basis, $Z\ket{x} = \ket{x}$, \cref{lem:smallTraceDistance}, and the well known fact that the trace distance is preserved under unitary transform ($\TD(U\ket{\psi},U\ket{\phi}) = \TD(\ket{\psi},\ket{\phi})$):
      \begin{align}
        &\TD(\ket{\psi}_{\cT,\Bob,E}, (Z^{t m_{1-b}}_{\Bob,2}\ket{\psi}_{\cT,\Bob,E})) = \TD(\ket{t}\ket{\psi}_{\Bob,E}, \ket{t}Z^{t m_{1-b}}_{\Bob, 2}\ket{\psi}_{\Bob,E})\\
        \begin{split}
          &\leq \TD(\ket{\psi}_{\Bob,E}, \ket{x^*}\ket{\phi^*}) + \TD(\ket{x^*}\ket{\phi^*}, Z^{m_{1-b}}_2\ket{x^*}\ket{\phi^*}) \\
          &\quad+ \TD(Z^{m_{1-b}}_2 \ket{x^*}\ket{\phi^*}, Z^{m_{1-b}}_2\ket{\psi}_{\Bob,E})
        \end{split}\\
        &\leq \sqrt{\beta} + 0 + \TD(\ket{x^*}\ket{\phi^*}, \ket{\psi}_{\Bob,E})\\
        &\leq 2\sqrt{\beta} = 2t\sqrt{\beta}\label{eq:smaller2tsqbeta}
      \end{align}
    \end{subproof}
    We prove now that $\World_2 \approxRV_{\alpha} \World_3$, where $\alpha \eqdef \esp*[\ket{t}\ket{\psi}_{\Bob,E}\gets \xi_0(\sigma)]{2t\sqrt{\beta_\psi}}$.
    \begin{subproof}
      First, we remark that by stopping the worlds right before the $Z$ rotations, we can define two binary POVM\footnote{We slightly abuse notations, as technically a POVM is not a map but a set of projectors (one for each outcome), so we define $\pr{\xi_0(\rho) = 1} = \Tr(\xi_0\rho)$.} $\xi_0$ (taking as input (a purification of) $\sigma_\lambda$ and outputting the state $\ket{\psi}_{\cT,\Bob,E} = \ket{t}\ket{\psi_{\Bob,E}}$ defined above) and $\xi_1$ (performing the rest $Z^{m_b}$ rotation, the measurement in the $H$ basis, the adversary $\hat{\Alice}_2$ and $\Zenv$) such that $\World_3$ is the sequential composition of $\xi_0$ and $\xi_1$, and $\World_2$ is the sequential composition of $\xi_0$, $Z^{t m_{1-b}}_2$ and $\xi_1$. This way, we can write:
      \begin{align}
        &|\pr{\World_3 = 1} - \pr{\World_2 = 1}|\\
        \begin{split}
          &= \left|\pr*[\ket{t}\ket{\psi}_{\Bob,E} \gets \xi_0(\sigma)]{\xi_1(\ket{t}\ket{\psi}_{\Bob,E}) = 1} \right.\\
          &\quad\left.- \pr*[\ket{t}\ket{\psi}_{\Bob,E} \gets \xi_0(\sigma)]{\xi_1(\ket{t}(Z^{t m_{1-b}}_{\Bob,2}\ket{\psi}_{\Bob,E}) = 1}\right|\\
        \end{split}\\
        \begin{split}
          &= \left|\int_{\ket{t}\ket{\psi}_{\Bob,E}} \pr{\xi_0(\sigma) = \ket{t}\ket{\psi}_{\Bob,E}} \left(\pr*{\xi_1(\ket{t}\ket{\psi}_{\Bob,E}) = 1} \right.\right.\\
          &\qquad\left.\left.- \pr*{\xi_1(\ket{t}(Z^{t m_{1-b}}_{\Bob,2}\ket{\psi}_{\Bob,E}))}\right)\right|\\
        \end{split}\\
        \begin{split}
          &\leq \int_{\ket{t}\ket{\psi}_{\Bob,E}} \pr{\xi_0(\sigma) = \ket{t}\ket{\psi}_{\Bob,E}} \left|\pr*{\xi_1(\ket{t}\ket{\psi}_{\Bob,E}) = 1} \right.\\
          &\qquad\left.- \pr*{\xi_1\left(\ket{t}(Z^{t m_{1-b}}_{\Bob,2}\ket{\psi}_{\Bob,E})\right)=1}\right|\\
        \end{split}\\
        \begin{split}
          &=\underset{\subalign{\xi_0(\sigma) = \ket{t}\ket{\psi}_{\Bob,E}}}{\mathbb{E}}\left[\,\left|\pr*{\xi_1(\ket{t}\ket{\psi}_{\Bob,E}) = 1} \right.\right.\\
          &\qquad\left.\left.- \pr*{\xi_1(\ket{t}(Z^{t m_{1-b}}_{\Bob,2}\ket{\psi}_{\Bob,E}))=1}\right|\,\right]\label{eq:espWithPr}
        \end{split}
      \end{align}
      Moreover, it is well known that for any state $\rho$ and $\sigma$, $\TD(\rho,\sigma) = \max_P \Tr{P(\rho-\sigma)} = \max_P \pr{P(\rho) = 1}- \pr{P(\sigma)=1}$ (see e.g.~\cite[eq.~9.22]{NC10_QuantumComputationQuantum}), where the maximum is taken over any POVM. Since $\xi_1$ is a POVM, we have therefore
      \begin{align}
        &|\pr*{\xi_1(\ket{t}\ket{\psi}_{\Bob,E}) = 1} - \pr*{\xi_1(\ket{t}(Z^{t m_{1-b}}_{\Bob,2}\ket{\psi}_{\Bob,E}))=1}|\\
        &\leq \TD(\ket{\psi}_{\cT,\Bob,E},Z^{t m_{1-b}}_{\Bob,2}\ket{\psi}_{\cT,\Bob,E})\\
        &\leqref{eq:smaller2tsqbeta} 2t\sqrt{\beta_\psi}
      \end{align}
      (Note that $\beta$ is different for any value of $\ket{\psi}_{\cT,\Bob,E}$, hence the notation $\beta_\psi$)
      
      By injecting that into \cref{eq:espWithPr}, we get
      \begin{align}
        |\pr{\World_3 = 1} - \pr{\World_2 = 1}| \leq \esp*[\ket{t}\ket{\psi}_{\Bob,E}\gets \xi_0(\sigma)]{2t\sqrt{\beta_\psi}} = \alpha
      \end{align}
    \end{subproof}
    Finally, the last step is to prove that $\alpha$ is negligible, by reducing it to the probability of finding a collision.
    \begin{subproof}
      We already shown above that given $\ket{t}\ket{\psi}_{\Bob,E}$, there is a procedure $P_h$ to find a collision with probability greater than $t\beta_\psi$ (but the initial state might depend on $h$). From that, we can define the algorithm that first runs $\xi_0(\sigma_\lambda)$ ($\sigma_\lambda$ is now independent of $h$, and $h$ is sampled according to $\Gen(1^\lambda)$ in $\xi_0$ as expected), then $P_h$. The probability success of this procedure is therefore:
      \begin{align}
        \alpha' \eqdef \esp*[\ket{t}\ket{\psi}_{\Bob,E}\gets \xi_0(\sigma)]{t\beta_\psi}
      \end{align}
      Because $h$ is collision resistant (\cref{def:collisionResistance}), we have $\alpha' = \negl$. Moreover, by defining $f(x) = \sqrt{2x}$, $f$ is concave, and therefore using Jensen's inequality we get:      
      \begin{align}
        \alpha
        &= \esp*[\ket{t}\ket{\psi}_{\Bob,E}\gets \xi_0(\sigma)]{2t\sqrt{\beta_\psi}}\\
        &= \esp*[\ket{t}\ket{\psi}_{\Bob,E}\gets \xi_0(\sigma)]{2\sqrt{t\beta_\psi}}\\
        &\leq 2\sqrt{\esp*[\ket{t}\ket{\psi}_{\Bob,E}\gets \xi_0(\sigma)]{t\beta_\psi}}\\
        &=2\sqrt{\alpha}\\
        &=\negl
      \end{align}
      Therefore, since we know that $\World_2 \approxRV_{\alpha} \World_3$ and we just proved that $\alpha$ is negligible, we get that $\World_2 \approxRV \World_3$.
    \end{subproof}
  \end{subproof}
  
  Finally, it is easy to see that $\World_3 = \World_4$ since they are actually exactly the same quantum map, except that we attribute the operations to different parties. By defining $\Sim_{\hat{\Alice},\Zenv}$ as the block composed of all elements on the left of the ideal resource, we have $\World_5 = \Ideal^{\sigma,\Fot}_{\tilde{\Pi},\Sim_{\hat{\Alice}},\Zenv}$.

  By transitivity, we have $\World_0 \approxRV \World_4$, i.e.\ $\Real^{\sigma}_{\Pi,\hat{\Alice},\Zenv} \approx \Ideal^{\sigma,\Fot}_{\tilde{\Pi},\Sim_{\hat{\Alice}},\Zenv}$ which concludes the proof.
\end{proofE}

\textEnd{This small lemma is useful to prove \cref{thm:realizesOT}:}

\begin{lemmaE}[][all end]\label{lem:smallTraceDistance}
  Let $\ket{\phi}$ and $\ket{\phi'}$ be two normalized orthogonal pure states and $\beta \in [0,1]$. We consider the normalized state $\ket{\psi} = \sqrt{1-\beta}\ket{\phi} + \sqrt{\beta} \ket{\phi'}$. Then, $\TD(\ket{\phi}, \ket{\psi}) = \sqrt{\beta}$.
\end{lemmaE}
\begin{proofE} 
  This is a direct characterization of the fact that for any pure states $\ket{\phi}$ and $\ket{\psi}$, $\TD(\ket{\phi}, \ket{\phi}) = \sqrt{1-|\braket{\psi | \phi}|^2}$ (see e.g.~\cite[eq.~(9.173)]{Wil17_ClassicalQuantumShannon}):
  \begin{align}
    \TD(\ket{\phi}, \ket{\psi})
    &= \sqrt{1-|\braket{\phi | \psi}|^2}
    = \sqrt{1-|\bra{\phi}(\sqrt{1-\beta}\ket{\phi} + \sqrt{\beta} \ket{\phi'})|^2}\\
    &= \sqrt{1-|\sqrt{1-\beta}|^2} = \sqrt{\beta}
  \end{align}
\end{proofE}

\section{(NI)ZKoQS and $k$-out-of-$n$ string OT}
\pratendSetLocal{category=zkoqs} 

\subsection{ZKoQS}

The main contribution in our main protocol (\cref{protoc:2messOT}) is to provide a method to prove (potentially non-interactively) a statement on a received quantum state without revealing much information beside the fact that this statement is true: we call this property (Non-Interactive) Zero-Knowledge proofs on Quantum State ((NI)ZKoQS), by analogy with their classical analogue. While we have not yet introduced formally this definition in order to provide a self-contained OT protocol and proof, we will address this issue here.

NIZKoQS were introduced in~\cite{CGK21_NonDestructiveZeroKnowledgeProofs}, but the protocol we present here is using a very different approach. While \cite{CGK21_NonDestructiveZeroKnowledgeProofs} can be used to prove more advanced properties on the obtained quantum state, it also has multiple drawbacks that were left as open questions:
\begin{itemize}
\item First, while their protocol is purely classical, their approach is fundamentally \emph{incompatible with statistical security} (like other potential approaches based on quantum multi-party computing~\cite{DNS12_ActivelySecureTwoParty,DGJ+20_SecureMultipartyQuantum,KKL+23_AsymmetricQuantumSecure}, since these protocols build upon classical MPC, which are not only impossible to do with statistical security~\cite{Lo97_InsecurityQuantumSecure}, but they also require OT, which is one application of ZKoQS). A malicious unbounded verifier/receiver can always fully describe the received state. On the other hand, with our approach we can get statistical security for both parties (not as the same time).
\item Secondly, \cite{CGK21_NonDestructiveZeroKnowledgeProofs} relies on lattice based cryptography (\LWE{}), living in Cryptomania, and the protocol is really \emph{costly} to implement in practice as the parameters used in the \LWE{} instance lead to very large functions. On the other side, our approach only relies on hash functions, does not exploit any structure or trapdoors, and is therefore much more efficient.
\end{itemize}

Note that the definition of ZKoQS introduced in~\cite{CGK21_NonDestructiveZeroKnowledgeProofs} is slightly too restrictive for our setting as their notion of quantum language does not allow states to be $\eps$-close to the quantum language, the states cannot be entangled with an adversary, they omit the step where the description is given back to the sender (which is important when the protocol is used in other protocols), and their adversaries are \QPT{}. For this last reason, we introduce different notations inspired by classical ZK proofs: when the prover is unbounded (resp.\ bounded) we say that we have a ZK \emph{proof} (resp.\ \emph{argument}) on quantum states, denoted ZKPoQS (resp.\ ZKAoQS). When the verifier is unbounded, we say that we have a statistical ZKoQS (S-ZKoQS). Note than when the protocol in Non-Interactive (a single message from the prover to the verifier), we replace the ``ZK'' with ``NIZK'' in these acronyms. We formalize now these concepts.

\begin{inAppendixIfPublishedEnv}
  \begin{figure}
    \begin{subfigure}[t]{0.45\textwidth}\centering
      \begin{ZX}[mytext/.style={}]
        |[minimum height=1.5em]|(x,w)\dar[->] & [3mm] |[minimum height=1.5em]| x \dar[->] \\[2mm]
        |[draw,inner sep=2mm]| \P \rar[<->] & |[draw,,inner sep=2mm]| \V \dar[->]     \\[2mm]
        |[minimum height=1.5em]|              & |[minimum height=1.5em]| \top/\bot
      \end{ZX}
      \caption{Usual setting of classical ZK protocols: we typically have $x \in \lang_w$, with $\lang_w \eqdef \{x \mid x \cR w\}$.}
    \end{subfigure}\hfill%
    \begin{subfigure}[t]{0.45\textwidth}\centering
      \begin{ZX}
        |[minimum height=1.5em]|(x,w)\dar[->] & [3mm] |[minimum height=1.5em]|        \\[2mm]
        |[draw,inner sep=2mm]| \P \rar[<->] & |[draw,,inner sep=2mm]| \V \dar[->] \\[2mm]
        |[minimum height=1.5em]|              & |[minimum height=1.5em]| x/\bot
      \end{ZX}
      \caption{Alternative equivalent setting of classical ZK protocols.}
    \end{subfigure}\hfill\\%
    \begin{subfigure}[t]{0.45\textwidth}\centering
      \begin{ZX}
        |[minimum height=1.5em]| w\dar[->]             & [3mm] |[minimum height=1.5em]|        \\[2mm]
        |[draw,inner sep=2mm]| \P \rar[<->] \dar[->] & |[draw,,inner sep=2mm]| \V \dar[->] \\[2mm]
        |[minimum height=1.5em]| x                     & |[minimum height=1.5em]| x/\bot
      \end{ZX}
      \caption{Another alternative setting of classical ZK protocols (e.g.\ if $w$ contains $x$). We expect $x \in \lang_w$.}
    \end{subfigure}\hfill%
    \begin{subfigure}[t]{0.45\textwidth}\centering
      \begin{ZX}
        |[minimum height=1.5em]|\omega\dar[->]         & [3mm] |[minimum height=1.5em]|                                                                                        \\[2mm]
        |[draw,inner sep=2mm]| \P \rar[<->] \dar[->] & |[draw,inner sep=2mm]| \V \dar[->,decorate, decoration={snake,segment length=.7mm, amplitude=.3mm,post length=1mm}] \\[2mm]
        |[minimum height=1.5em]| \omega_s              & |[minimum height=1.5em]| \rho /\bot
      \end{ZX}
      \caption{Setting of ZK on Quantum States: $\rho$ is the (quantum) equivalent of $x$, and $(\omega,\omega_s)$ can be seen as a (partial, see (cf.\ \cref{rk:choiceOmega})) classical description of $\rho$: we expect $\rho \in \lang_{\omega,\omega_s}$.}
    \end{subfigure}%
    \caption{Parallel between classical ZK and ZK on Quantum States: while classically all the above definitions are mostly equivalent, quantumly we cannot send $\rho$ (the quantum equivalent of $x$) as an input since the laws of physics forbid us from extracting any information from $\rho$ without altering it. See also \cref{rk:classicalVsQuantum} and \cref{rk:choiceOmega} for the justification of the choice of $\omega$ and $\omega_s$.}
    \label{fig:classicalZKvsquantumZK}
  \end{figure}
\end{inAppendixIfPublishedEnv}

\paragraph{Quantum language.}

First, we define a quantum language (we draw a parallel with classical ZK in \cameraVsOther{the pictures in the full version}{\cref{fig:classicalZKvsquantumZK}}, and illustrate this with an example in \cref{example:semicol}), which is informally speaking a set $\lang_\cQ$ of bipartite quantum states on two registers $\V$ and $\P$ that characterizes all states that a malicious adversary might be able to obtain (the register $\V$ being controlled by the honest verifier, and $\P$ by the malicious prover and/or the environment\footnote{Sometimes, we will write $(\P,\Zenv)$ instead of $\P$ to denote a more precise cut between the two sub-registers owned by the prover and the environment.}). Moreover, we also provide additional information on the honest expected behavior, via sets of (bipartite\footnote{Contrary to $\lang_\cQ$ that must represent all states potentially obtainable by a malicious party (hence the need of a second register), here $\lang_\omega$ are only used to denote the states obtainable by honest parties, and can therefore often be seen as a set of states on a single register owned by the verifier. The reason we define it as a bipartite state here is that we might later be interested by the generation of truly bipartite states like graph states.}) quantum states $\lang_\omega \subseteq \lang_\cQ$: when the prover is given as input a \emph{class} $\omega$ (the quantum equivalent\footnote{Note that classically, we can see a witness in two different ways: it can be used to efficiently verify that $x \in \lang$, but more abstractly it can be seen as a way to partition $\lang$ into multiple $\lang_w$'s: in an honest setting, given $w$, we expect to have $x \in \lang_w$, where $\lang_w = \{x \mid x \cR w\}$. Quantumly, we will use this second point of view, as given $\omega$ (the quantum equivalent of $w$) we expect in an honest setting to have $\rho \in \lang_\omega$, even if $\omega$ cannot be used directly to verify that property once $\rho$ is generated because of the laws of physics.} of witnesses), we expect the final state to belong to $\lang_\omega$. Because there might be many states in $\lang_\omega$, the prover will also output a subclass $\omega_s$ to further describe the final state, interpreted as ``the verifier obtained a state belonging to $\lang_{\omega, \omega_s} \subseteq \lang_\omega \subseteq \lang_\cQ$''. 

\begin{example}\label{example:semicol} For instance, one might be interested in $\lang_\cQ$ defined as the set of states where the registers $\V$ contains exactly two qubits, where at least one of them is non-entangled with any other qubit and collapsed in the computational basis (think ``even if the prover is malicious, any state obtained by the verifier belongs to $\lang_\cQ$, i.e.\ contains at least one qubit collapsed in the computational basis). For the honest behavior, we can for instance define $\lang_{0,0} = \{\ket{+}\ket{0}, \ket{+}\ket{1}\}$, $\lang_{0,1} = \{\ket{-}\ket{0}, \ket{-}\ket{1}\}$, $\lang_{1,0} = \{\ket{0}\ket{+}, \ket{1}\ket{+}\}$, $\lang_{1,1} = \{\ket{0}\ket{-}, \ket{1}\ket{-}\}$, $\lang_0 = \lang_{0,0} \cup \lang_{0,1}$ and $\lang_1 = \lang_{1,0} \cup \lang_{1,1}$: this way, if the prover gets input $0$ and outputs $1$, the verifier is expected to output a state in $\lang_{0,1} = \{\ket{-}\ket{0}, \ket{-}\ket{1}\}$: the class $\omega$ represents the position of the state in the Hadamard basis, and the sub-class $\omega_s$ represents the value encoded in this state.
\end{example}

\begin{remark}[On the choice of definition of $\omega$ and $\omega_s$]\label{rk:choiceOmega}
  Note that $(\omega,\omega_s)$ only partially describes the state (in our example above, we remove the description of the state in the computational basis) as otherwise we are unable to prove the security of the scheme (but the lost information on $\rho$ is anyway of no interest since it is discarded in the OT protocol). One might also ask why $\omega_s$ is sent as an output and is not part of the input $\omega$: while in some cases it might be possible to move everything inside the input $\omega$ and remove $\omega_s$ (e.g. if we got a $\ket{+}$ instead of a $\ket{-}$ the prover could send another message ``apply an additional $Z$ gate'' to flip the encoded qubit), but this comes at the cost of an additional message. In most applications, the exact value of $\omega_s$ does not really matter as it is only a random key, while saving an additional round of communication is important.
\end{remark}

\begin{definition}[Quantum Language]\label{def:quantumLanguage}
  Let $E^{\V,\P} = \cup_{(n,m) \in \N^2} \linearOp_\circ(\cH_n\otimes \cH_m)$ be the set of finite dimensional quantum states on two registers. A \emph{quantum language} $(\lang_\cQ, \cC, \cC_s, \{\lang_{\omega, \omega_s}\}_{\omega \in \cC, \omega_s \in \cC_s})$ is characterized by a set $\lang_\cQ \subseteq E^{\V, \P}$ of bipartite quantum states\footnote{$\lang_\cQ$ represents informally the set of states that any malicious party can generate, where the first register is the output of the verifier and the second register corresponds to registers potentially controlled by an adversary. Since only $\lang_\cQ$ is needed to characterize the security of a protocol, it is sometimes called directly the quantum language.}, a set $\cC \subseteq \{0,1\}^*$ of \emph{classes} (or \emph{witnesses}) motivated above, a set $\cC_s \subseteq \{0,1\}^*$ of \emph{sub-classes}, and for any $\omega \in \cC$, $\omega_s \in \cC_s$, a set $\lang_{\omega, \omega_s}$ of bipartite quantum states called \emph{quantum sub-classes}. We also define for any $\omega$, $\lang_\omega = \cup_{\omega_s \in \cC_s} \lang_{\omega, \omega_s}$ (some of these sets might be empty in case $\omega$ is not a valid class), and require $\cup_\omega \lang_\omega \subseteq \lang_\cQ$. Moreover, for any set of quantum states $\lang$, we define $\rho \in_{\eps} \lang \Leftrightarrow \exists \sigma \in \lang, \TD(\rho, \sigma) \leq \eps$, and $\rho \notin_{\eps} \lang \Leftrightarrow \neg (\rho \in_{\eps}\lang)$.
\end{definition}

\paragraph{ZKoQS.}

We introduce now ZKoQS, that morally provides three guarantees, similar to classical ZK\cameraVsOther{}{ (cf.\ \cref{fig:classicalZKvsquantumZK})}:
\begin{itemize}
\item \textbf{Correctness}: if the parties are honest, the prover is given a class $\omega$ and ends up with the partial (cf.\ \cref{rk:choiceOmega}) description $(\omega,\omega_s)$ of the state $\rho$ obtained by the verifier, i.e.\ such that $\rho \in \lang_{\omega,\omega_s} \subseteq \lang_{\omega} \subseteq \lang_\cQ$.
\item \textbf{Soundness}: if the sender is malicious, the honest receiver still ends up with a state $\rho \in \lang_\cQ$.
\item \textbf{Zero-Knowledge}: if the verifier is malicious, they cannot learn the value of the class/witness $\omega$.
\end{itemize}

\begin{example}
  To continue our above \cref{example:semicol}, the correctness guarantees that given an input bit $\omega \in \{0,1\}$, the $\omega$-th qubit of $\rho$ is $H\ket{\omega_s}$ while the other qubit is in the computational basis (we lose the information of the encoded value). The soundness mostly guarantees that even if the sender is malicious, the received quantum state contains at least one qubit collapsed in the computational basis. The ZK property guarantees that a malicious verifier cannot learn $\omega$, the expected position of the qubit in the Hadamard basis.
\end{example}

Note that the formal definition is given with respect to a ``simulator'', simulating the whole protocol (and not anymore a single malicious party as usual), including in the soundness and correctness part (while usually simulators are only used in the ZK part). While we could define it without any simulator to get a more restricted definition (and during a first read, it might actually be easier to replace the simulator with the original process), simulators are helpful for multiple reasons to make the definition more useful:
\begin{itemize}
\item \textbf{In zero-knowledge}: the typical ZK definitions already use simulators to denote the fact the we can simulate the view of the malicious verifier without access to the witness… Therefore it should come at no surprise that we also use a simulator in the ZK property.
\item \textbf{In soundness}: In a real protocol, a malicious prover might be able to produce states negligibly close (in trace distance) to the quantum language $\lang_\cQ$, but not strictly speaking \emph{in} $\lang_\cQ$. One might be tempted to introduce an approximate notion $\rho \in_\eps \lang_\cQ$ taking into account trace distance to fix this issue, unfortunately it is not sufficient as this definition does not take into account states that are statistically speaking far from $\lang_\cQ$, but computationally speaking ``close'' to $\lang_\cQ$… Indeed, sometimes provers might actually be able to produce states far (in trace distance) from any state in $\lang_\cQ$, but because they are computationally bounded, they are unable to exploit that fact. This kind of false ``attack'' can actually be done against our protocol if the function $h$ is not injective (explaining why we require $h$ to be injective when considering an unbounded malicious receiver), by simply running the ZK protocol in superposition\footnote{Of course by still measuring the classical transcript to send to the verifier.}: in that case the output state might be relatively close to a $\ket{+}$ or $\ket{-}$ if $h$ is well balanced (while we expect the state to be close to $\ket{0}$ or $\ket{1}$), but a computationally bounded receiver cannot exploit this property as they need to compute all preimages of $h$ to know if we are close to $\ket{+}$ or $\ket{-}$. Simulator are therefore useful in the soundness definition to capture this ``computational distance'', and discard ineffective attacks.
\item \textbf{In correctness}: Perhaps surprisingly, we also use a simulator in the correctness definition. While this might not be useful when considering only a game-based security notion, we need simulator to prove for instance statements like ``If a protocol $\Pi$ realises a given functionality, then this protocol is a ZKoQS protocol'' (see e.g.\ \cref{thm:FpmImpliesZKoQS}). Without further details on $\Pi$, the correctness of $\Pi$ only tells us that $\Pi$ is indistinguishable from a functionality that produces states in $\lang_\cQ$, but it does not mean that $\Pi$ itself produces such states, hence the need of a simulator.
\end{itemize}
We formalize the notion of ZKoQS:

\begin{definition}[{Zero-Knowledge Proof on Quantum State (ZKoQS)}]\label{def:NIZKoQS}~\\
  Let $\lang \eqdef (\lang_\cQ, \cC, \cC_s, \{\lang_{\omega, \omega_s}\}_{\omega \in \cC, \omega_s \in \cC_s})$ be a quantum language (\cref{def:quantumLanguage}). We say that a protocol $\Pi = (\P, \V)$ is a ZKoQS protocol for $\lang$, where $\P$ takes as input a class $\omega \in \cC$ and outputs a sub-class $\omega_s \in \cC_s$ and\footnote{$\rho^\P$ will actually not be necessary in our main application, but we still include it in case it turns out to be useful in future applications.} a quantum state $\rho^\P$, and $\V$ takes no input and outputs a bit $a$, that is equal to $1$ if $\V$ does not abort, together with a quantum state $\rho^\V$ (potentially entangled with $\rho^\P$),  if the following properties are respected:
  \begin{itemize}
  \item \textbf{Correctness}: There exists a poly-time simulator $\Sim$ and a negligible function $\eps$ such that $(\P \interacts \V) \approxRVC \Sim$, and for any $\omega$ such that $\lang_\omega \neq \emptyset$:
    \begin{align}
      \pr[((\omega_s, \rho^\P),(a, \rho^{\V})) \gets \Sim(\omega)]{a = 1 \land \rho^{\V,\P} \in \lang_{\omega, \omega_s}} = 1\label{eq:correctnessZKoQS}
    \end{align}
  \item \textbf{Soundness}: For any malicious prover $\hat{\P} = \{\hat{\P}_\lambda\}_{\lambda \in \N}$, ($\QPT{}$ for ZKAoQS, unbounded for ZKPoQS) there exists a simulator $\Sim_{\hat{\P}} = \{\Sim_{\lambda, \hat{\P}}\}_{\lambda \in \N}$ (running in time polynomial in the runtime of $\hat{\P}$) such that $(\hat{\P} \interacts \V) \approxRVC \Sim_{\hat{\P}}$ ($\approxRVS$ for ZKPoQS), and such that there exists a negligible function $\eps$ such that for any sequence of bipartite state $\{\sigma^{\P, \Zenv}_\lambda\}_{\lambda \in \N}$ and $\lambda \in \N$:
    \begin{align}
      \pr[(\rho^\P,(a, \rho^{\V}),\rho^\Zenv) \gets (\Sim_{\lambda,\hat{\P}}^{\P} \otimes I^{\Zenv}) \otimes \sigma^{\P,\Zenv}_\lambda]{a = 1 \land \rho^{\V,(\P,\Zenv)} \notin \lang_\cQ} \leq \eps(\lambda)\label{eq:soundnessZKoQS}
    \end{align}
  \item \textbf{Quantum Zero-Knowledge}: For any malicious verifier $\hat{\V} = \{\hat{\V}_\lambda\}_{\lambda \in \N}$ ($\QPT{}$ for ZKoQS, unbounded for S-ZKoQS), there exists a simulator $\Sim_{\hat{\V}}(b, \cdot)$ (where $b \in \{0,1\}$ indicates if $\lang_\omega$ is non-empty, and $\cdot$ represents an additionally quantum input from the environment), and an efficiently computable map $\xi_{\cdot}(\cdot)$ (such that $\forall \omega, \xi_\omega$ takes one quantum register as input and outputs a classical message in $\cC_s$ and a quantum state $\rho^\P$), both running in polynomial time in the runtime of $\hat{\V}$, such that for any $\omega \in \cC$:
    \begin{align}
      (\P(\omega) \interacts \hat{\V}) \approxRVC (\xi_\omega \otimes I)(\Sim_{\hat{\V}}(\lang_\omega \neq \emptyset))\label{eq:quantumZK}
    \end{align}
    ($\approxRVS$ for ZKPoQS)
  \end{itemize}
  It can sometimes be handy to cut the protocol into two phases: the honest verifier will output the state $\rho^\V$ at the end of the first \textsf{send} phase, wile the output of the honest prover will be delivered in a second \textsf{describe} phase (allowing the prover to describe the state outputted earlier by the verifier). A ZKoQS protocol where each phase consists of a single message is said to be \emph{non-interactive} (denoted NIZKoQS, we can similarly add the ``NI'' prefix to the previously seen notions, to get NIZKPoQS, S-NIZKoQS…). Finally, for a set of parties $S$, we write $\text{ZKoQS}_S$ to denote the fact that the protocol is S-ZKoQS if $\V \in S$ and ZKPoQS if $\P \in S$.
\end{definition}

Note that in ZK protocols, there is a notion of extractability, where a simulator can extract the witness $w$ from a valid transcript (not all ZK protocols are extractable). We could define a similar notion here allowing the simulator to extract $\omega$, but since $\lang_Q$ might contain states not belonging to any $\lang_\omega$ (potentially producible by malicious provers), we need to slightly update the definition of quantum language by also introducing a special ``malicious'' subclass $\bot$, so that $\lang_\cQ = \cup_{\omega} (\lang_\omega \cup \lang_{\omega,\bot})$, and such that the simulator in the soundness property can extract the $\omega$ of the state produced by a malicious adversary:

\begin{definition}[Extractability]
  A ZKoQS protocol is said to be \emph{extractable} with respect to $(\lang_{\omega,\bot})_{\omega \in \cC}$ ($\bot$ being a special subclass not belonging to $\cC_s$) such that $\lang_\cQ = \cup_{\omega} (\lang_\omega \cup \lang_{\omega,\bot})$, and such that the soundness property is turned into:
  \begin{itemize}
  \item \textbf{Extractability}: For any malicious prover $\hat{\P} = \{\hat{\P}_\lambda\}_{\lambda \in \N}$, ($\QPT{}$ for ZKAoQS, unbounded for ZKPoQS) there exists a simulator $\Sim_{\hat{\P}} = \{\Sim_{\lambda, \hat{\P}}\}_{\lambda \in \N}$ (running in time polynomial in the runtime of $\hat{\P}$) such that $(\hat{\P} \interacts \V) \approxRVC \Sim_{\hat{\P}}$ ($\approxRVS$ for ZKPoQS), and such that there exists a negligible function $\eps$ such that for any sequence of bipartite state $\{\sigma^{\P, \Zenv}_\lambda\}_{\lambda \in \N}$ and $\lambda \in \N$:
    \begin{align}
      \pr[(\rho^\P,(a, \rho^{\V}),\rho^\Zenv, \omega) \gets (\Sim_{\lambda,\hat{\P}}(\sigma^\P_\lambda)) \otimes \sigma^{\Zenv}_\lambda]{a = 1 \land \rho^{\V,(\P,\Zenv)} \notin (\lang_\omega \cup \lang_{\omega,\bot})} \leq \eps(\lambda)\label{eq:soundnessZKoQSExtract}
    \end{align}
  \end{itemize}
\end{definition}

\arxivOnly{
  \subsection{Frequently Asked Questions}

  We answer here some natural questions regarding ZKoQS and quantum languages to complete the previous discussions.
  \begin{itemize}
  \item \textbf{Are quantum languages linear?} Or said differently, if $\ket{\psi} \in \lang_\cQ$ and $\ket{\phi} \in \lang_\cQ$, are linear combinations of $\ket{\psi}$ and $\ket{\phi}$ part of $\lang_\cQ$? Not always: while is it possible to define a language stable by linear combination, this property might be undesirable. For instance, the quantum language given in \cref{example:semicol} does not have this property (or our OT protocol would be insecure): indeed, both $\ket{00}$ and $\ket{11}$ belong to $\lang_\cQ$ since they are collapsed states, but $\frac{1}{\sqrt{2}} (\ket{00} + \ket{11}) $ does not belong to $\lang_\cQ$ since no qubit is collapsed. Actually, if the prover could send such a Bell pair to the receiver, then it would be possible to learn $m_0 \xor m_1$ by simply computing the XOR of the measurement outcomes, breaking the OT protocol.
  \item \textbf{Can the adversary generate states negligibly close to $\lang_\cQ$ but not strictly in $\lang_\cQ$?} Yes. And this is not surprising: if an adversary deviates in an undetectable way (e.g. by rotating the state with a negligible angle), then this deviation is simply not detectable and would have no consequences in term of security. Note that this does not contradict the definition of soundness that states that the probability of accepting and outputting $\rho \notin \lang_\cQ$ is negligible, since the $\rho$ is outputted by the \emph{simulator}, not the adversary directly. So the adversary might always send a state $\rho'$ $\eps$-close to $\lang_\cQ$ while the simulator will always generate a state inside $\lang_\cQ$ (for instance by doing an undetectable measurement on $\rho'$ to project it back into $\lang_\cQ$).
  \item \textbf{Why do we define the above notions with respect to a simulator?} The answer to the previous question gives a first element of answer: this way we do not need to define a notion of being $\eps$-close to a language. We also give other elements of answer in the paragraph before \cref{def:NIZKoQS}.
  \item \textbf{Can we obtain ZKoQS for any quantum language?} No. For instance, if we define $\lang_\cQ'$ like $\lang_\cQ$ in the above example, but such that $\omega$ also contains $\omega_s$ (the encoded value), then any correct protocol would not be ZK: it is indeed always possible to learn some information on the encoded value by simply measuring the state after rotating it with an angle $\frac{\pi}{4}$. Similarly, if $\omega_s$ contains the encoded value of all states (and not just those that are in the Hadamard basis), our proof method does not work since this additional information might help the distinguisher. While we do prove that the set of quantum languages verifiable in a ZK way is non-trivial, characterizing this set precisely is an open question.
  \item \textbf{Why isn't ZKoQS unidirectional, like classical ZK?} Classically, the prover has no output, while quantumly they output an additional description $\omega_s$ of the obtained state. This is actually a \emph{choice that we made for efficiency reasons} ($\omega_s$ could also be part of the input, just like $x$ classically). Indeed, due to the fundamental non-deterministic nature of quantum computing, the state obtained by the verifier will be different at each run (in our case the encoded value is random): so if the prover wants a fixed encoded value, an additional correction message must be sent to the verifier, creating additional rounds of communications. But it seems like for most of the applications, we do not really need to fix the value of $\omega_s$, we just need to know its value: gaining unidirectionality at the cost of round efficiency was not worth it as it would complicate the construction and add rounds of communication, but there is nothing fundamental here.
  \end{itemize}
}

\subsection{Proof of partial measurement: a generic framework to get ZKoQS}

While the notion of ZKoQS (\cref{def:NIZKoQS}) does not explicitly mention functionalities, it is often handy to model a ZKoQS protocol inside an ideal functionality as it is easier to interpret it and use it inside other protocols. While it is not clear how to translate the ZKoQS definition into a functionality, we provide below a few ideal functionalities that ``imply'' ZKoQS. We will first see what is a ZKoQS ideal functionality, then we will see a class of functionalities that are ZKoQS, and we will show that our protocol realizes a particular case of these functionalities.

\begin{definition}[{ZKoQS ideal functionality}]\label{def:NIZKoQSFunc}
  Let $(\lang_\cQ, \cC, \cC_s, \{\lang_{\omega, \omega_s}\}_{\omega \in \cC, \omega_s \in \cC_s})$ be a quantum language (\cref{def:quantumLanguage}). We say that an ideal functionality $\cF$ is a ZKoQS (resp.\ $\text{ZKoQS}_S$) ideal functionality for $\lang_\cQ$ iff for any protocol $\Pi = (\P, \V)$ that quantum standalone realizes $\cF$ (resp.\ $\CSQSA{S}$-realizes $\cF$), $\Pi$ is a ZKoQS protocol (resp.\ $ZKoQS_S$ protocol) for $\lang_\cQ$ (\cref{def:NIZKoQS}). 
\end{definition}

The most natural class of ideal functionalities leading to ZKoQS are the ones in which the functionality applies an operation (a partial measurement) on an arbitrary input to enforce some structures on the output state:

\begin{definition}[Partial measurement $\Fpm{M,f_0}$]\label{def:Fpm}
  Let $M \eqdef \{M_m\}_{m \in \cM}$ be a collection of measurement operators\footnote{They are the most generic way to represent a measurement.} (i.e.\ operators such that $\sum_m M_m^\dagger M_m = I$ \cite[Sec.~2.2.3]{NC10_QuantumComputationQuantum}), implementable in quantum polynomial time, and let $f_0 \colon \cM \rightarrow \cC_s$ be an efficiently computable function\footnote{Informally, $f_0$ is used to filter some information on the measurement outcome $m$ during an honest protocol.}. Then, we define the \emph{proof of partial measurement} functionality $\Fpm{M,f_0}$ as follows:
  \begin{itemize}
  \item $\Fpm{M,f_0}$ receives a state $\rho$ from the prover's interface, together with an abort bit $a$.
  \item If $a = \bot$, it sends $\bot$ to both parties and stops.
  \item Otherwise, $\Fpm{M,f_0}$ measures $\rho$ using $M$, obtaining an outcome $m \in \cM$ and a post-measured state
    \begin{align}
      \rho' \eqdef \xi_m(\rho) \eqdef \frac{M_m \rho M_m^\dagger}{\Tr(M_m^\dagger M_m\rho)}
    \end{align}
  \item It sends $\rho'$ to the verifier, and waits back for a message $f$, such that either $f = \bot$ (in which case the functionality sends $\bot$ to the prover to abort and stops), $f = \top$ (in which case the ideal functionality redefines $f \eqdef f_0$), or $f$ is an efficiently computable function $f \colon \cM \rightarrow \{0,1\}^*$.
  \item Finally, it sends $f(m)$ to the prover.
  \end{itemize}
\end{definition}

We would like to prove that this functionality is a ZKoQS functionality, but not all such functionalities are ZKoQS (in particular, if the post-measured state contains information on $\omega$, it has no chance of being ZK). For this reason, we expect our functionality to have an additional property, intuitively saying that we can postpone the actual measurement \emph{after} sending the quantum state. While this might seem counter intuitive, this can actually be realized exploiting entanglement, and similar techniques were used in previous works to prove security of protocols~\cite{DFPR14_ComposableSecurityDelegated}.

\begin{definition}[Postponable measurement operator]\label{def:postponableOperator}
  A measurement operator $M$ outputting a quantum state and a classical measurement outcome is said to be postponable with respect to a collection of sampling procedures $\{G_\omega\}_{\omega \in A}$ outputting a quantum state if there exist a bipartite state $\rho^{\V,F}$ and a quantum map $M'$ taking as input a bipartite system and outputting a measurement outcome $m'$ such that for all $\omega \in A$, $M G_\omega \approxRVS (I^\V \otimes M')(\rho^{\V,F} \otimes G_\omega)$:
  \tikzset{for fit/.style={minimum height=1.5em,inner xsep=3mm}}
  \begin{align}
    \begin{ZX}[mysnake/.style={
      }]
      |[draw,inner sep=2pt,minimum height=1.5em]| G_\omega \rar[mysnake] & [2mm] |[draw,inner sep=2pt,minimum height=1.5em]| M \rar[yshift=1mm] \rar[yshift=-1mm, double] & [2mm] \zxN{}
    \end{ZX} =
    \begin{ZX}[execute at end picture={
        \node[fit=(top)(bot),draw,thick,fill=white, inner sep=0pt, "center:$\rho$"]{};
        \node[fit=(Mtop)(Mbot),draw,thick,fill=white, inner sep=0pt, "center:$M'$"]{};
      }]
      |[for fit,a=top]| \ar[rr]                                   & [2mm]                           & [2mm] \zxN{} \\
      |[for fit,a=bot]| \ar[r]                                    & |[for fit,a=Mtop]| \rar[double] & \zxN{}       \\
      |[draw,inner sep=2pt,minimum height=1.5em]| G_\omega \ar[r] & |[for fit,a=Mbot]|
    \end{ZX}
  \end{align}
\end{definition}

We prove now that such a functionality is a ZKoQS functionality for a given quantum language and appropriately defined dummy ideal parties:
\begin{theoremE}[$\Fpm{}$ implies ZKoQS][end,text link=]\label{thm:FpmImpliesZKoQS}
  Let $E^{\V_0,\P} = \cup_{(n,m) \in \N^2} \linearOp_\circ(\cH_n\otimes \cH_m)$ be the set of finite dimensional quantum states on two registers $\V_0$ and $\P$. Let $\cC$ and $\cC_s$ be two sets, and for any $\omega \in \cC$, let $E_{\omega} \subseteq E^{\V_0,\P}$ be a set of bipartite quantum states. Let $M \eqdef \{M_m\}_{m \in \cM}$ be a collection of measurement operators (and $\xi_m$ as defined in \cref{def:Fpm}), and $f_0 \colon \cM \rightarrow \cC_s$ be a function. We define for any $\omega \in \cC$ and $\omega_s \in \cC_s$:
\begin{align}
  \lang_{\omega, \omega_s} &\eqdef \{ \rho^{\V,\P} \mid \exists \rho_0^{\V_0,\P} \in E_{\omega}, m \in \cM, \text{ s.t. } \omega_s = f_0(m), \rho^{\V,\P} = \xi_m(\rho_0^{\V_0,\P})\}\\
  \lang_\omega &\eqdef \cup_{\omega_s} \lang_{\omega,\omega_s}\\
  \begin{split}
  \lang_\cQ &\eqdef \{ (\xi_m \otimes \hat{\xi}_{f_0(m)})\rho^{\V,(\P,\Zenv)} \mid \rho \in E^{\V_0,(\P,\Zenv)}, m \in \cM, m \neq \bot, \\ &\qquad\hat{\xi}_{f_0(m)} \text{ being an arbitrary CPTP map depending on $f_0(m)$.} \}
  \end{split}
\end{align}
  Then, let $\tilde{\P}$ and $\tilde{\V}$ be any poly-time ideal parties, such that:
  \begin{itemize}
  \item If $E_\omega = \emptyset$, $\tilde{\P}(\omega)$ sends the abort bit $a=\bot$ to the functionality and outputs $\bot$. Otherwise, $\tilde{\P}(\omega)$ produces a state in $E_{\omega}$ according to an arbitrary sampling procedure $G$, sends the register $\V_0$ to the ideal functionality, and outputs the $\omega_s$ given back from the functionality together with the register $\P$.
  \item If $\tilde{\V}$ receives $\bot$ from the functionality, it outputs $a = \bot$ and stop. Otherwise, it outputs the state $\rho'$ given by the functionality together with a bit $a = \top$ and sends back to the functionality $f = \top$.
  \end{itemize}
  Then, if $M$ are postponable measurement operators with respect to $\{G_\omega\}_{\omega, \lang_\omega \neq \emptyset}$ (\cref{def:postponableOperator}), $\Fpm{M,f_0}$ is a ZKoQS protocol (actually $\text{ZKoQS}_S$ for any set $S$, see \cref{def:NIZKoQSFunc}) for the language $\lang_\cQ$ previously defined.
\end{theoremE}
\begin{proof}[Sketch of proof]
  The proof mostly derives from the definitions, and from the fact that having postponable operators allows us to push the part of the ideal functionality that depends on the secret after the interaction with the adversary, preserving the ZK property. We refer to the \cameraVsOther{full security proof in the full version~\cite{CMS23_ObliviousTransferZeroKnowledge}}{\hyperref[proof:prAtEnd\pratendcountercurrent]{full security proof} in \pratendSectionlikeCref{}} for more details.
\end{proof}
\begin{proofE}
  Let $S$ be any subset of parties, and let us assume the above assumptions. Let $\Pi = (\P, \V)$ be a protocol that $\CSQSA{S}$-realizes $\Fpm{M,f_0}$ with the above dummy ideal parties. We prove that $\Pi$ is a $\text{ZKoQS}_S$ protocol, by proving first the completeness.
  \begin{subproof}
    First, since $\Pi$ $\CSQSA{S}$-realizes $\Fpm{M,f_0}$, by defining $\Sim(\omega) \eqdef \tilde{\P}(\omega) \interactsF{\Fpm{M,f_0}} \tilde{\V}$, we have in particular $(\P \interacts \V) \approxRVC \Sim$ and $\Sim$ runs in poly-time since $M$ is efficiently implementable. Let $\omega$ such that $\lang_\omega \neq \emptyset$. Because $\lang_\omega \neq \emptyset$, $\tilde{\P}(\omega)$ produces, by definition, a state $\rho_0^{\V_0,\P}$ in $E_\omega$, and since $\tilde{\V}$ sends $f = \top$, $\tilde{\P}$ outputs $\omega_s = f_0(m)$, where $m$ is the measurement outcome of the ideal functionality. Also, by definition of $\Fpm{M,f_0}$, the post-measured state is $\rho^{\P,\V} = \xi_m(\rho_0^{\V_0,\P})$. Therefore, by definition of $\lang_{\omega, \omega_s}$, the output state $\rho^{\P,\V}$ belongs to $\lang_{\omega, \omega_s}$ with probability $1$, and $\tilde{\V}$ always outputs $a=1$. Therefore, \cref{eq:correctnessZKoQS} is true, completing the proof of completeness.
  \end{subproof}
  We proceed now with the soundness:
  \begin{subproof}
    Let $\hat{\P} = \{\hat{\P}_\lambda\}_{\lambda \in \N}$ be a malicious prover (unbounded if $\P \in S$, in which case all symbols $\approxRVC$ should be replaced with $\approxRVS$ in the rest of this proof). Because $\Pi$ $\CSQSA{S}$-realizes $\Fpm{M,f_0}$, there exists $\Sim'_{\hat{\P}}$ such that $(\hat{\P} \interacts \V) \approxRVC \Sim'_{\hat{\P}} \interactsF{\Fpm{M,f_0}} \tilde{\V}$. So let us define $\Sim_{\hat{\P}}(\sigma^\P_\lambda) = (\Sim'_{\hat{\P}}(\sigma^\P_\lambda) \interactsF{\Fpm{M,f_0}} \tilde{\V})$: we then have $(\hat{\P} \interacts \V) \approxRVC \Sim_{\hat{\P}}$. Moreover, since $\lang_\cQ$ corresponds to all the bipartite states that can be obtained when applying $\xi_m$ for some $m$ on the first register, and an arbitrary deviation depending on $f_0(m)$ on the other register (remember that $\Fpm{M,f_0}$ gives back $f_0(m)$ to the adversary, so $\hat{\xi}_{f_0(m)}$ would be defined as the second part of $\Sim{\hat{\P}}$, after waiting for the answer of the functionality), \cref{eq:soundnessZKoQS} is true (even with probability $1$), completing the proof.
  \end{subproof}
  We finish now with the quantum ZK property:
  \begin{subproof}
    Let $\hat{\V} = \{\hat{\V}_\lambda\}_{\lambda \in \N}$ be a malicious verifier (unbounded if $\V \in S$, in which case all symbols $\approxRVC$ should be replaced with $\approxRVS$ in the rest of this proof). Because $\Pi$ $\CSQSA{S}$-realizes $\Fpm{M,f_0}$, there exists $\Sim'_{\hat{\V}}$ such that $(\P \interacts \hat{\V}) \approxRVC (\tilde{\P} \interactsF{\Fpm{M,f_0}} \Sim'_{\hat{\V}})$. Then, we define $\Sim_{\hat{\V}}(b)$ as follows. First, if $b = \bot$, it forwards the input of the environment to $\Sim'_{\hat{\V}}$ together with $\bot$ (pretending to be the functionality $\Fpm{M,f_0}$). Since $\tilde{\P}(\omega)$ aborts iff $\lang_\omega = \emptyset$, we actually have for any $\omega$ such that $\lang_\omega = \emptyset$, $\tilde{\P}(\omega) \interactsF{\Fpm{M,f_0}} \Sim'_{\hat{\V}} = (\xi_\omega \otimes I)(\Sim_{\hat{\V}}(\lang_\omega \neq \emptyset)$ where $\xi_\omega$ is given $b$ as input and outputs $\bot$ if $b = \bot$: by transitivity, \cref{eq:quantumZK} is true in that case. Otherwise, if $b = \top$, since $M$ are postponable measurement operators with respect to $G$, there exists by definition $\rho^{\V,F}$ and a quantum map $M'$ such that $M G \approxRVS (I^\V \otimes M')(\rho^{\V,F} \otimes G)$. Therefore, $\tilde{\P} \interactsF{\Fpm{M,f_0}} \Sim'_{\hat{\V}}$ is equivalent to:
    \begin{enumerate}
    \item Generate $\rho^{\V,F}$ and send $\rho^\V$ to $\Sim'_{\hat{\V}}$ (pretending to be $\Fpm{M,f_0}$) to get a function $f$ and an output state $\rho'$.
    \item Run $M'(\rho^F \otimes G(\omega))$ to get outcome $m'$, send back $f_\omega(m')$ to $\tilde{\P}(\omega)$ and outputs the final state of $\tilde{\P}(\omega)$.
    \end{enumerate}
    However, if we define $\Sim_{\hat{\V}}$ to do the first step, and $\xi_\omega$ to be the second and third step (when $b = 1$), we have $(\tilde{\P} \interactsF{\Fpm{M,f_0}} \Sim'_{\hat{\V}}) = (\xi_\omega \otimes I)\Sim_{\hat{\V}}$, therefore \cref{eq:quantumZK} is respected, concluding the proof.
  \end{subproof}
\end{proofE}

While the above results show that we can obtain a ZKoQS protocol from any protocol realizing the functionality $\Fpm{M,f_0}$ (where $M$ must be postponable), we show in the next section how we can realize such a functionality to prove that a state was partially collapsed (measured in the computational basis) without revealing the position of the collapsed qubit. We will then see that, as a corollary, there exists a ZKoQS protocol for the quantum language of ``semi-collapsed'' states.

\subsection{Protocol to prove that a state has been semi-collapsed}

We prove now that we can realize the functionality below, that informally measures a set $T$ of qubits (the measured qubits, chosen by the prover, being constraint to respect $\Pred(T) = \top$, for an arbitrary predicate $\Pred$), randomly rotates the other one, and provides the resulting state to the verifier.

\begin{definition}[{Semi-collapsing functionality $\Fsemicol{\Pred}$}]\label{def:fsemicol}
  Let $n \in \N$, and $\Pred \colon \cP([n]) \rightarrow \{\top,\bot\}$ be an efficiently computable predicate on the subsets of $[n]$. We define the \emph{semi-collapsing functionality} $\Fsemicol{\Pred}$ as $\Fpm{M,f_0}$ (\cref{def:Fpm}), where:
  \begin{itemize}
  \item $M$ is the measurement operator that receives a quantum state on two registers, measures (destructively) the first register\footnote{Informally this register contains the subset of qubits in the second register to measure and a (typically random) sequence of $Z$ rotations to apply on the remaining qubits. Since the first operation of $M$ is to measure them, we can (and will) also consider them as classical inputs.} in the computational basis to get (an encoding of) $T \subseteq [n]$ and a sequence of bits $(r^{(i)})_{i \in [n] \setminus T}$, checks if $\Pred(T) = \top$: if not it outputs $m = \bot$ and a dummy quantum state $\ket{\bot}$. Otherwise, it measures (non-destructively) in the computational basis all qubits in the second register whose index belongs to the set of ``target'' qubits $T$, getting outcomes $\{m^{(j)}\}_{j \in T}$, and for any $i \in [n] \setminus T$, it applies $Z^{r^{(i)}}$ on the $i$-th qubit. Finally it outputs $m = (T, (m^{(j)})_{j \in T}, (r^{(i)})_{i \in [n] \setminus T})$ and the post-measured state.
  \item If $m = \bot$, $f_0(m) = \bot$, otherwise if $m = (T, (m^{(j)})_{j \in T}, (r^{(i)})_{i \in [n] \setminus T})$, $f_0(m) = (r^{(i)})_{i \in [n] \setminus T})$.
  \end{itemize}
  We also consider the following dummy ideal parties:
  \begin{itemize}
  \item $\tilde{P}(T, \rho)$ samples\footnote{Note that this sequence of rotations in only needed for correctness as in the real protocol the non-measured qubits will be arbitrarily rotated.} uniformly at random a sequence of bits $(r^{(i)})_{i \in [n] \setminus T}$, sends $a = \Pred(T)$ and $\ketbra{T,(r^{(i)})_{i \in [n] \setminus T}} \otimes \rho$ to the ideal functionality $\Fsemicol{\Pred}$, and forwards the received message from the functionality.
  \item $\tilde{V}$ checks if it received $a = \bot$ from the functionality, or if the received quantum state is $\ket{\bot}$. If so it sends back $f = \bot$ to the functionality and aborts, and otherwise it sets $f = \top$ for the functionality and outputs the quantum state to the environment.
  \end{itemize}
\end{definition}

We prove now that we can realize the functionality $\Fsemicol{\Pred}$:
\begin{theoremE}[{Realization of $\Fsemicol{\Pred}$}][text link=,end]\label{thm:fsemicol}
  Let $\{h_k\}_{k \in \cK}$ be a family of collision resistant functions sampled by $\Gen$, having the hardcore second-bit property (\cref{def:firstBitHardcore}). Let $\Pi_h = (\P_h, \V_h)$ be a protocol\footnote{As a reminder, this protocol is sampling and distributing a function $h$ according to $\Gen$, and can either be done without communication in the CRS model (or heuristically if we replace $h$ with a well known collision-resistant hash function), or with one message in the plain model.} $\CSQSA{S_h}$ realizing $\cFcrs{\Gen}$ and $\Pizk = (\Azk,\Bzk)$ be a protocol that \CSQSA{S} realizes the ZK functionality $\Fzk{\cR}$, where $(h^{(c)}_d)_{c \in [n], d \in \{0,1\}} \cR (T, (w^{(c)}_d)_{c \in [n], d \in \{0,1\}}) \Leftrightarrow \Pred(T) = \top \land \forall c,d, h(d \| w_d^{(c)}) = h_d^{(c)}$ and $\forall c \in T, \exists c$ such that $w_d^{(c)}[1] = 1$.
  
  Then, the protocol $\Pisemicol$ (\publishedVsArxiv{\cref{protoc:fsemicol}\cameraVsOther{}{, see also the equivalent picture in \cref{protoc:fsemicolPicture}}}{\cref{protoc:fsemicol}}) $\CSQSA{S'}$-realizes $\Fsemicol{\Pred}$ for any $S'$ such that:
  \begin{itemize}
  \item $S' \subseteq S \cap S_h$,
  \item $\{ \P \} \in S'$ only if $h$ has the statistical hardcore second-bit property,
  \item $\{ \V \} \in S'$ only if for any $k \in \cK$, $h_k$ is injective (i.e.\ statistically collision resistant).
  \end{itemize}
\end{theoremE}
\begin{proof}[Sketch of proof]
  Part of the proofs of this theorem are generalizations of \cref{thm:realizesOT}. Some care must be taken to show that the distributions in the honest case (ideal world versus real world) are really indistinguishable, we do so by computing the appropriate density matrices. There is also a slight difference as here we measure the state instead of applying a rotation, but it turns out that measuring is indistinguishable from rotating a state and discarding the rotation angle. We refer to the \cameraVsOther{full security proof in the full version~\cite{CMS23_ObliviousTransferZeroKnowledge}}{\hyperref[proof:prAtEnd\pratendcountercurrent]{full security proof} in \pratendSectionlikeCref{}} for more details.
\end{proof}
\begin{proofE}
  \textbf{Case 1: correctness (no corrupted party).}
  
  We first prove the correctness (when all parties are honest):
  
  \begin{subproof}
    First, if $\Pred(T) = \bot$, the parties abort in both the ideal and real worlds. Otherwise, like in \cref{thm:realizesOT}, the ZK proof in the real world succeeds by the completeness of the ZK protocol, and the first measurement of Bob will not disturb the state, so we can remove indistinguishably the ZK proof and the first measurement of Bob (we should technically define new worlds as we did before, but we omit them for conciseness as we already applied similar arguments earlier). Similarly, since for any $j \in T$, the first register of the state $\rho^{(j)}$ is already measured, the Hadamard basis measurement on its second register does not alter the state, so we can remove indistinguishably all operations involving $\rho^{(j)}$'s except for the measurements in the computational basis by $\P$. Therefore, we can concentrate now on $\rho^{(i)}$ for $i \in [n] \setminus T$, and show that the random rotation $Z^{r^{(i)}}$ followed by the map $x \rightarrow x,w_x^{(i)}$, the Hadamard measurement on the second register, and the update of $r^{(i)}$ is statistically indistinguishable from a single random $Z^{r^{(i)}}$.
    \begin{subproof}
      Let $i \in [n] \setminus T$. We can show this property in two ways: either by doing the computation directly on density matrices (which is a bit long, but we do it below for completeness), or remark that the map and Hadamard measurement commute with the $Z^{r^{(i)}}$, and that the map and Hadamard measurement performs a $Z^{\langle s^{(b)}, w_0^{(b)} \rangle}$ rotation: therefore we can sample $r^{(i)}$ after knowing the outcome of the measurement, and since the distribution $r^{(i)}$ is indistinguishable from the distribution $c \xor r^{(i)}$ for any constant $c$, we can take $c = \langle s^{(b)}, w_0^{(b)} \rangle$ to cancel the rotation applied earlier by the Hadamard, making it virtually equal to a single rotation $Z^{r^{(i)}}$.

      We provide now an \textbf{alternative proof}, more verbose but certainly more formal, by directly computing the appropriate density matrices. If we consider a purification\footnote{Technically $\rho^{(i)}$ could be entangled with other $\rho^{(i')}$, so this purification might contain elements in $\rho^{(i')}$. This is not an issue as soon as we apply this transformation sequentially, on a single $i$ at a time.} of $\rho^{(i)}$, there exists two vectors $\ket{\psi_0}$ and $\ket{\psi_1}$ such that $\rho^{(i)} = \alpha_0\ket{0}\ket{\psi_0}+\alpha_1 \ket{1}\ket{\psi_1}$. Moreover, if we consider the density operator of this whole process, we have after the sampling of $a$ (put on the first register) and $Z^a$ rotation:
      \begin{align}
        \begin{split}
          &\frac{1}{2} (\ket{0}(\alpha_0\ket{0}\ket{\psi_0}+\alpha_1 \ket{1}\ket{\psi_1}))((\alpha_0^*\bra{\psi_0}\bra{0}+\alpha_1^* \bra{\psi_1}\bra{1})\bra{0})\\
          + &\frac{1}{2} (\ket{1}(\alpha_0\ket{0}\ket{\psi_0}-\alpha_1 \ket{1}\ket{\psi_1}))((\alpha_0^*\bra{\psi_0}\bra{0}-\alpha_1^* \bra{\psi_1}\bra{1})\bra{1})
        \end{split}
      \end{align}
      Then, after the $x \mapsto x, w_x^{(c)}$ operation, and the Hadamard (omitting the normalisation factor), we get:
      \begin{align}
        \begin{split}
          &\sum_{s^{b}}\sum_{s^{\prime (i)}}(\frac{1}{2} (\ket{0}(\alpha_0(-1)^{\langle s^{(i)}, w_0^{(i)} \rangle}\ket{0}\ket{\psi_0}\ket{s^{(i)}}+\alpha_1 (-1)^{\langle s^{(i)}, w_1^{(i)} \rangle}\ket{1}\ket{\psi_1}\ket{s^{(i)}}))\\
          &((\alpha_0^*(-1)^{\langle s^{\prime (i)}, w_0^{(i)} \rangle}\bra{s^{\prime (i)}}\bra{\psi_0}\bra{0}+\alpha_1^* (-1)^{\langle s^{\prime (i)}, w_1^{(i)} \rangle}\bra{s^{\prime (i)}}\bra{\psi_1}\bra{1})\bra{0})\\
          &+ \frac{1}{2} (\ket{1}(\alpha_0(-1)^{\langle s^{(i)}, w_0^{(i)} \rangle}\ket{0}\ket{\psi_0}\ket{s^{(i)}}-\alpha_1(-1)^{\langle s^{(i)}, w_1^{(i)} \rangle} \ket{1}\ket{\psi_1}\ket{s^{(i)}}))\\
          &((\alpha_0^*(-1)^{\langle s^{\prime (i)}, w_0^{(i)} \rangle}\bra{s^{\prime (i)}}\bra{\psi_0}\bra{0}-\alpha_1^*(-1)^{\langle s^{\prime (i)}, w_1^{(i)} \rangle} \bra{s^{\prime (i)}}\bra{\psi_1}\bra{1})\bra{1}))
        \end{split}
      \end{align}
      However, since the output of $\P$ XOR to $b$ the value $\langle s^{(i)}, w_0^{(i)} \xor w_1^{(i)} \rangle$, the final density matrix representing this process is as follows (to obtain this, we factored out the $(-1)^{\langle s^{(i)}, w_0^{(i)} \rangle}$ that gets canceled as a global phase, we rename $\alpha_{s^{(i)}} \eqdef \langle s^{(i)}, w_0^{(i)} \xor w_1^{(i)} \rangle$, and we XOR the first register with $\alpha_{s^{(i)}}$):
      \begin{align}
        \begin{split}
          &\frac{1}{2} \sum_{s^{b}}\sum_{s^{\prime (i)}}((\ket{\alpha_{s^{(i)}}}(\alpha_0\ket{0}\ket{\psi_0}\ket{s^{(i)}}+\alpha_1 (-1)^{\alpha_{s^{(i)}}}\ket{1}\ket{\psi_1}\ket{s^{(i)}}))\\
          &((\alpha_0^*\bra{s^{\prime (i)}}\bra{\psi_0}\bra{0}+\alpha_1^* (-1)^{\alpha_{s^{\prime (i)}}}\bra{s^{\prime (i)}}\bra{\psi_1}\bra{1})\bra{\alpha_{s^{(i)}}})\\
          &+ (\ket{1 \xor \alpha_{s^{(i)}}}(\alpha_0\ket{0}\ket{\psi_0}\ket{s^{(i)}}-\alpha_1(-1)^{\alpha_{s^{(i)}}} \ket{1}\ket{\psi_1}\ket{s^{(i)}}))\\
          &((\alpha_0^*\bra{s^{\prime (i)}}\bra{\psi_0}\bra{0}-\alpha_1^*(-1)^{\alpha_{s^{\prime (i)}}} \bra{s^{\prime (i)}}\bra{\psi_1}\bra{1})\bra{1 \xor \alpha_{s^{(i)}}}))
        \end{split}
      \end{align}
      Moreover, the last register $s^{(i)}$ is discarded, so we can trace is out:
      \begin{align}
        \begin{split}
          &\frac{1}{2} \sum_{s^{b}}((\ket{\alpha_{s^{(i)}}}(\alpha_0\ket{0}\ket{\psi_0}+\alpha_1 (-1)^{\alpha_{s^{(i)}}}\ket{1}\ket{\psi_1}))\\
          &((\alpha_0^*\bra{\psi_0}\bra{0}+\alpha_1^* (-1)^{\alpha_{s^{(i)}}}\bra{\psi_1}\bra{1})\bra{\alpha_{s^{(i)}}})\\
          &+ (\ket{1 \xor \alpha_{s^{(i)}}}(\alpha_0\ket{0}\ket{\psi_0}-\alpha_1(-1)^{\alpha_{s^{(i)}}} \ket{1}\ket{\psi_1}))\\
          &((\alpha_0^*\bra{\psi_0}\bra{0}-\alpha_1^*(-1)^{\alpha_{s^{(i)}}} \bra{\psi_1}\bra{1})\bra{1 \xor \alpha_{s^{(i)}}}))
        \end{split}
      \end{align}
      We have now two cases: if $w_0^{(i)} \xor w_1^{(i)} = 0 \dots 0$, then $\alpha_{s^{(i)}} = 0$ and we actually see that the second register is not even entangled with the first one, so the Hadamard measurement does not change the first qubit, therefore we only apply a random $Z^{a}$ on it and output $a$, like in the ideal world. Now, if $w_0^{(i)} \xor w_1^{(i)} \neq 0$, then the the number of $s^{(i)}$ such that $\alpha_{s^{(i)}} = 0$ is exactly equal to the number of cases where $\alpha_{s^{(i)}} = 1$:
      \begin{subproof}
        To see that, since $w_0^{(i)} \xor w_1^{(i)} \neq 1$, there exists one position where they differ: then, just flipping the bit of $s^{(i)}$ at that position will also flip the value of $\alpha_{s^{(i)}}$, providing a simple way to partition $s^{(i)}$'s in two sets of equal size, each set having the same value of $\alpha_{s^{(i)}}$.
      \end{subproof}
      Therefore, we can sum over $\alpha_{s^{(i)}}$ instead of $s^{b}$ (this adds a fixed constant (thanks to the argument we just mentioned) factor that we ignore for simplicity):
      \begin{align}
        \begin{split}
          &\frac{1}{2} \sum_{\alpha_{s^{(i)}} \in \{0,1\}}((\ket{\alpha_{s^{(i)}}}(\alpha_0\ket{0}\ket{\psi_0}+\alpha_1 (-1)^{\alpha_{s^{(i)}}}\ket{1}\ket{\psi_1}))\\
          &\qquad((\alpha_0^*\bra{\psi_0}\bra{0}+\alpha_1^* (-1)^{\alpha_{s^{(i)}}}\bra{\psi_1}\bra{1})\bra{\alpha_{s^{(i)}}})\\
          &\qquad+ (\ket{1 \xor \alpha_{s^{(i)}}}(\alpha_0\ket{0}\ket{\psi_0}-\alpha_1(-1)^{\alpha_{s^{(i)}}} \ket{1}\ket{\psi_1}))\\
          &\qquad((\alpha_0^*\bra{\psi_0}\bra{0}-\alpha_1^*(-1)^{\alpha_{s^{(i)}}} \bra{\psi_1}\bra{1})\bra{1 \xor \alpha_{s^{(i)}}}))
        \end{split}
        \\
        \begin{split}
          &=
            \frac{1}{2} ((\ket{0}(\alpha_0\ket{0}\ket{\psi_0}+\alpha_1 \ket{1}\ket{\psi_1}))\\
          &\qquad((\alpha_0^*\bra{\psi_0}\bra{0}+\alpha_1^* \bra{\psi_1}\bra{1})\bra{0})\\
          &\qquad+ (\ket{1}(\alpha_0\ket{0}\ket{\psi_0}-\alpha_1 \ket{1}\ket{\psi_1}))\\
          &\qquad((\alpha_0^*\bra{\psi_0}\bra{0}-\alpha_1^* \bra{\psi_1}\bra{1})\bra{1})\\
          &\qquad+ (\ket{1}(\alpha_0\ket{0}\ket{\psi_0}-\alpha_1 \ket{1}\ket{\psi_1}))\\
          &\qquad((\alpha_0^*\bra{\psi_0}\bra{0}-\alpha_1^*\bra{\psi_1}\bra{1})\bra{1})\\
          &\qquad+ (\ket{0}(\alpha_0\ket{0}\ket{\psi_0}+\alpha_1 \ket{1}\ket{\psi_1}))\\
          &\qquad((\alpha_0^*\bra{\psi_0}\bra{0}+\alpha_1^* \bra{\psi_1}\bra{1})\bra{0})
            )
        \end{split}
        \\
        \begin{split}
          &=
            \frac{1}{2} ((\ket{0}(\alpha_0\ket{0}\ket{\psi_0}+\alpha_1 \ket{1}\ket{\psi_1}))\\
          &\qquad((\alpha_0^*\bra{\psi_0}\bra{0}+\alpha_1^* \bra{\psi_1}\bra{1})\bra{0})\\
          &\qquad+ (\ket{1}(\alpha_0\ket{0}\ket{\psi_0}-\alpha_1 \ket{1}\ket{\psi_1}))\\
          &\qquad((\alpha_0^*\bra{\psi_0}\bra{0}-\alpha_1^* \bra{\psi_1}\bra{1})\bra{1})\\
          &\qquad+ (\ket{1}(\alpha_0\ket{0}\ket{\psi_0}-\alpha_1 \ket{1}\ket{\psi_1}))\\
          &\qquad((\alpha_0^*\bra{\psi_0}\bra{0}-\alpha_1^*\bra{\psi_1}\bra{1})\bra{1})\\
          &\qquad+ (\ket{0}(\alpha_0\ket{0}\ket{\psi_0}+\alpha_1 \ket{1}\ket{\psi_1}))\\
          &\qquad((\alpha_0^*\bra{\psi_0}\bra{0}+\alpha_1^* \bra{\psi_1}\bra{1})\bra{0})
            )
        \end{split}
        \\
        \begin{split}
          &=
            (\ket{0}(\alpha_0\ket{0}\ket{\psi_0}+\alpha_1 \ket{1}\ket{\psi_1}))\\
          &\qquad((\alpha_0^*\bra{\psi_0}\bra{0}+\alpha_1^* \bra{\psi_1}\bra{1})\bra{0})\\
          &\qquad+ (\ket{1}(\alpha_0\ket{0}\ket{\psi_0}-\alpha_1 \ket{1}\ket{\psi_1}))\\
          &\qquad((\alpha_0^*\bra{\psi_0}\bra{0}-\alpha_1^* \bra{\psi_1}\bra{1})\bra{1})
        \end{split}
      \end{align}
      We see (once we renormalize this state) that this is exactly the density matrix of the ideal world that applies a random $Z^a$ operation on the qubit and outputs $a$ in the first register.   
    \end{subproof}
  \end{subproof}
  
  \textbf{Case 2: malicious verifier.}
  We prove now the equivalence of the ideal and real worlds if the adversary corrupts the verifier.
  \begin{subproof}
    We consider now the case where the adversary $\cA = \hat{\V}$ corrupts the verifier. The proof of this section is quite close to the second case of the proof of \cref{thm:realizesOT}, so we will be quicker here. First, as before we cut $\hat{\V}$ into multiple parts ($\hat{\V}_0$ running against $\A_h$ to generate $h$, $\hat{\V}_1$ will be the circuit run when receiving $\bot$ (since the interaction will stop there in that case, $\hat{\V}_1$ only outputs a state for the environment), otherwise $\hat{\V}_2$ will be the circuit playing the ZK proof, and $\hat{\V}_3$ receiving the quantum state and outputting a final state and the measurements back to $\A$). Then, similarly to what we did before we define the simulator $\Sim_{\hat{\V}}$ as follows: first, the simulator will simulate the protocol $\Pi_h$ by running $\Sim_{h,\hat{B}_0}$ to get $h$. Then, if the output of the functionality $\Fsemicol{\Pred}$ is $\bot$, it will run $\hat{\V}_1$ and forward the output of $\hat{\V}_1$ to the environment. Otherwise, it samples all $w_c^{(d)}$ starting with a $0$, computes the hashes $\{h^{(c)}_d\}_{c \in [n], d \in \{0,1\}}$, and runs the ZK simulator $\Sim_{zk,\hat{B}_2}((h^{(c)}_d)_{c,d})$. Then, for all $c \in [n]$, it will sample $r^{(c)} \sample \{0,1\}$ and send the states $\rho^{(c)} = \ket{0}\ket{w_0^{(c)}} + (-1)^{r^{(c)}} \ket{1}\ket{w_1^{(c)}}$. After receiving the $\{s^{(c)}\}_{c \in \{0,1\}}$, it defines $f(T,(m^{(j)})_{j \in T},(r^{(i)})_{i \in [n] \setminus T}) \eqdef (r^{(i)} \xor \bigoplus_k s^{(i)}[k] (w^{(i)}_0 \xor w^{(i)}_1)[k])_{i \in [n] \setminus T}$ and sends $f$ to the functionality.

    To prove that we have $\P \interacts \hat{\V} \approxRVC \tilde{\P} \interactsF{\Fsemicol{\Pred}} \Sim_{\hat{\V}}$ ($\approxRVC$ are replaced with $\approxRVS$ if $\{\V\} \in S$), we can first study the case where the classical message $T$ sent to the prover is such that $\Pred(T) = \bot$. In that case, by definition of $\P$, $\P$ will abort, and send $\bot$ to $\hat{\V}$, which is also exactly what $\Sim_{\hat{\V}}$ is doing once $\tilde{\P}$ sent the abort bit forwarded by the functionality to the simulator. Now, we focus on $T$ such that $\Pred(T) = \top$, and therefore $\hat{\V}_1$ is never called. We can design as before a series of games, where we also cut $\P$ is a part that runs $\Pi_h$, a part $\P_0$ that checks $\Pred$, a part $\P_1$ that measures the state and samples $w$'s, a part $\P_2$ that runs the ZK proof, and the last part $\P_3$ that runs the rest of the protocol (we refer to \cref{thm:realizesOT} for more details on the steps that are almost identical).
    \begin{itemize}
    \item We start from $\P \interacts \hat{\V}$. Then, we replace $\P_0$ interacting with $\hat{B}_0$ with $\Sim_{h,\hat{B}_0}$ to get $h$. Both worlds are indistinguishable since $\Pi_h$ $\CSQSA{S_h}$ realizes $\cFcrs{\Gen}$.
    \item Then, since the statement proven is true, we can replace $\P_2$ interacting with $\V_2$ with $\Sim_{zk,\V_2}$ that only takes as input the hashes. This is indistinguishable since $\Pizk$ $\CSQSA{S}$ realizes $\Fzk{}$.
    \item Then, we can new sample all $w$'s such that they start with a $0$, which is indistinguishable thanks to the hardcore second-bit property (\cref{def:firstBitHardcore}) of $h$.
    \item Then, we sample $r^{(i)}$ for all $i \in [n]$, and instead of measuring the qubits for $i \in T$, we rotate them according to $Z^{r^{(i)}}$. This is statistically indistinguishable since neither $l$ nor $r^{(i)}$ are reused anywhere else, and are therefore discarded. But one can easily see that measuring and discarding the outcome is statistically equivalent to applying $Z^{r^{(i)}}$ and discarding $r^{(i)}$. This can easily be seen diagrammatically (using the doubling formalism), or via simple computations:
      \begin{subproof}
        If we purify a state as $\alpha_0 \ket{0}\ket{\psi_0} + \alpha_1 \ket{1}\ket{\psi_1}$, then applying a random $Z$ phase on the first qubit gives the density matrix:
        \begin{align}
          \begin{split}
            &\frac{1}{2} ((\alpha_0 \ket{0}\ket{\psi_0} + \alpha_1 \ket{1}\ket{\psi_1})(\alpha_0^* \bra{\psi_0}\bra{0} + \alpha_1^* \bra{\psi_1}\bra{1})\\
            &+ (\alpha_0 \ket{0}\ket{\psi_0} - \alpha_1 \ket{1}\ket{\psi_1})(\alpha_0^* \bra{\psi_0}\bra{0} - \alpha_1^* \bra{\psi_1}\bra{1}))\\
          \end{split}
          \\
            &= |\alpha_0|^2 \ket{0}\ket{\psi_0}\bra{\psi_0}\bra{0} + |\alpha_1|^2 \ket{1}\ket{\psi_1}\bra{\psi_1}\bra{1}\label{eg:measuringIsRotating}
        \end{align}
        and this second line corresponds to the density matrix of a (non-destructive) measurement in the computational basis.
      \end{subproof}
    \item Finally, by reorganizing the elements and using the functionality with the appropriately chosen function $f$ defined above to pick the appropriate $r^{(i)} \xor \langle s^{(i)}, w_0^{(i)} \xor w_1^{(i)} \rangle$ only for $i \in [n] \setminus T$, we obtain the ideal world concluding the proof.
    \end{itemize}
  \end{subproof}

  \textbf{Case 3: malicious prover.}
  We prove now the equivalence of the ideal and real worlds if the adversary corrupts the prover.
  \begin{subproof}
    We consider now the case where the adversary $\cA = \hat{\P}$ corrupts the prover. The proof of this section is quite close to the last case of the proof of \cref{thm:realizesOT}, so we will be quicker here. First, we can divide $\hat{\P}$ and $\V$ into multiple parts: $\hat{\P}_0$ will interact with $\V_0$ that will play the protocol $\V_h$ to obtain $h$, then $\hat{\P}_1$ will interact with $\V_1$ to run the ZK protocol, and finally $\hat{\P}_2$ will interact with $\V_2$ for the remaining part of the protocol.

    Now, we can, as before, define a series of indistinguishable worlds.
    \begin{itemize}
    \item First, we start from the ideal world, and we replace $\hat{\P}_0$ and $\V_0$ with the simulator $\Sim_{h,\hat{\P}_0}$ interacting with $\Fh{\Gen}$. This is indistinguishable since $\Pi_h$ realises $\Fh{\Gen}$.
    \item Similarly, we replace $\hat{\P}_1$ and $\V_1$ with the simulator $\Sim_{zk,\hat{\P}_1}$ interacting 
    \item Then, since the simulator has now access to the set $T$, it can measure all the states in $T$ once the quantum test passes. To show that it is indistinguishable, we use the exact same argument as the one made in case 3 of \cref{thm:realizesOT} to show that $\World_2 \approxRV \World_3$. The only differences is that now we measure more states (but we can apply sequentially the same argument for one state at a time, and since the number of states is polynomial the distinguishing probability is still negligible: note that in the current proof the ZK contains more statements, but in particular the statements needed in \cref{thm:realizesOT} are fulfilled). The second difference is that we show that the trace distance between $\rho$ and $Z_1\rho Z_1^\dagger$ is smaller than $2t\sqrt{\beta}$ (\cref{eq:smaller2tsqbeta}), while here we want to show that the trace distance between $\rho$ and the measured $\rho$ (non-destructively and in the computational basis) denoted $\rho'$ is negligible. However, we showed in \cref{eg:measuringIsRotating} that rotating by a random $Z^a$ (and discarding the $a$) is strictly equivalent to measuring (non-destructively) and discarding the outcome. Therefore:
      \begin{align}
        \TD(\rho, M\rho)
        &= \TD(\rho, \frac{1}{2} \rho + \frac{1}{2} Z_1\rho Z_1^\dagger)\\
        &\leq \frac{1}{2}(\TD(\rho, \rho) + \TD( \rho, Z_1\rho Z_1^\dagger)\\
        &\leq \frac{1}{2} 2t\sqrt{\beta} \leq 2t\sqrt{\beta}
      \end{align}
      which allows us to conclude.
    \item Then, because the state is collapsed before applying the Hadamard basis measurement on its second register, it is still collapsed after applying this measurement: so we can indistinguishably apply a second measurement in the computational basis after.
    \item Then, we can just attribute the operations to the appropriate parties and $\Fsemicol{\Pred}$ to obtain the ideal world: the simulator will run all the tasks, except that it sends $\bot$ to $\Fsemicol{\Pred}$ if the ZK protocol aborted, and otherwise forwards $T$ and sets $\forall i \in [n] \setminus T, r_i = 0$, while $\Fsemicol{\Pred}$ performs the second measurement in the computational basis on qubits in $T$ (since $r_i$’s are all equal to $0$, $Z^{r_i}$ is the identity), as expected by definition. Since all these attribution does not change the global map performed by all parties, this is statistically indistinguishable, concluding the proof of security.
    \end{itemize}   
  \end{subproof}
\end{proofE}

\publishedVsArxiv{
  \begin{protocol}[!htbp]
    \caption{ZKoQS protocol to realize $\Fsemicol{\Pred}$}\label{protoc:fsemicol}
    \textbf{Inputs}: The prover $\P$ gets $T \subseteq [n]$, a subset of qubits to measure, and the quantum state $\rho^{(1),\dots,(n)}$ to partially measure, the verifier $\V$ gets no input.\\
    \textbf{Assumption}: $\Pred$ is an efficiently computable predicate on subsets of $[n]$, $(\P_{zk},\V_{zk})$ is a $n$-message ZK protocol (\cref{def:Fzk}), $h$ is a collision-resistant (\cref{def:collisionResistance}) and second-bit hardcore (\cref{def:firstBitHardcore}) function distributed using $\Fh{}$ (\cref{def:Fh}), either non-interactively via a CRS, heuristically using a fixed hash function, or sent by the verifier, adding an additional message (\cref{lem:cFcrPlainModel}).\\
    \textbf{Protocol}:
    \begin{enumerate}
    \item \textbf{The prover} checks if $\Pred(T) = \top$, and abort and send $\bot$ to $\V$ otherwise. The, she samples $\forall d \in \{0,1\}, i \in [n] \setminus T, w^{(i)}_d \sample \{0\} \times \{0,1\}^n$ and for each $j \in T$, she measures (non destructively) $\rho^{(j)}$ to get outcome $l$, and samples $w^{(j)}_l \sample \{0\} \times \{0,1\}^n$ $l \in \{0,1\}$ and $w^{(j)}_{1-l} \sample \{1\} \times \{0,1\}^n$. Then, for each $(c,d) \in T \times \{0,1\}$ she defines $h_d^{(c)} \assign h(d \| w_d^{(c)})$. Then, she sends $(h^{(c)}_d)_{c \in [n], d \in \{0,1\}}$ to the verifier (if the ZK protocol is non-interactive she can send it later in a single message with the NIZK proof and the quantum states) and runs the ZK protocol $\P_{zk}$ with the verifier (running $\P_{zk}$) to prove that:
      \begin{align}
        &\exists T \subseteq [n], (w_d^{(c)})_{c \in T, d \in \{0,1\}}, \forall c, d, h_d^{(c)} = h(d\| w_d^{(c)}) \\
        \text{and }& \forall j \in T, \exists d \text{ s.t. } w_d^{(j)}[1] = 1, \text{ and } \Pred(T) = \top
      \end{align}
      Then, she samples for each $i \in [n] \setminus T$, $r^{(i)} \sample \{0,1\}$, and applies $Z^{r^{(i)}}\rho^{(i)}$. Finally, for each $c \in [n]$, she applies on $\rho^{(c)}$ the unitary mapping $\ket{x} \mapsto \ket{x}\ket{w_x^{(c)}}$ (we call $\rho_1^{(1),\dots,(n)}$ the resulting state) and she sends $\rho_1^{(1),\dots,(n)}$ to the verifier.
    \item \textbf{The verifier} aborts if the prover aborted or if it received a wrong ZK proof. Then, it applies on each qubit $c$ the unitary $\ket{x}\ket{w} \mapsto \ket{x}\ket{w}\ket{w[1] \neq 1 \land \exists d, h(x\|w) = h_d^{(c)}}$, and measures the last auxiliary register, checking if they are all equal to $1$. If not, he aborts (and sends an abort message to the prover), otherwise he measures for each $c \in [n]$ the second registers of $\rho^{(c)}$ (getting outcomes $s^{(c)}$) in the Hadamard basis. Finally, it outputs the remaining (first) qubit of each $\rho_1^{(c)}$, and sends $(s^{(c)})_{c \in [n]}$ to the prover.
    \item \textbf{The prover} computes $\omega_s \assign (r^{(i)} \xor \bigoplus_k s^{(i)}[k] (w^{(i)}_0 \xor w^{(i)}_1)[k])_{i \in [n] \setminus T}$ and outputs $\omega_s$.
    \end{enumerate}
  \end{protocol}
}{}

\inAppendixIfPublished{
  \begin{protocol}[!htbp]
    \caption{ZKoQS protocol to realize $\Fsemicol{\Pred}$}\publishedVsArxiv{\label{protoc:fsemicolPicture}}{\label{protoc:fsemicol}}
    \begin{autoFit}
      \pseudocodeblock{
        \mathbf{\P(T \subseteq [n], \rho^{(1),\dots,(n)})} \< \< \mathbf{\V} \\[][\hline]
        \\[-.5\baselineskip]
        \pclinecomment{$T$ is a subset of qubits to measure.}\\
        \text{Run $\P_h$ to obtain $h$.} \< \sendmessage{<->}{length=1.6cm} \< \text{Run $\V_h$ to obtain $h$.}\begin{tikzpicture}[overlay,remember picture]
          \node[name=start,anchor=north,align=left,text width=8cm,blue,rounded corners,draw,yshift=-5mm,xshift=-7mm]{$h$ is a collision-resistant (\cref{def:collisionResistance}) and second-bit hardcore (\cref{def:firstBitHardcore}) function distributed using $\Fh{}$ (\cref{def:Fh}), either non-interactively via a CRS, heuristically using a fixed hash function, or sent by the verifier, adding an additional message (\cref{lem:cFcrPlainModel})};
          \draw[blue,->] (start) -- (-2mm,-1mm);
        \end{tikzpicture}\\
        \text{If $\Pred(T) = \bot$ abort and send $\bot$ to $\V$.}\\
        \forall d \in \{0,1\}, i \in [n] \setminus T, w^{(i)}_d \sample \{0\} \times \{0,1\}^n\\
        \forall j \in T, l \sample \text{Measure non destructively $\rho^{(j)}$}\\
        \t w^{(j)}_l \sample \{0\} \times \{0,1\}^n\\
        \t w^{(j)}_{1-l} \sample \{1\} \times \{0,1\}^n\\
        \pclinecomment{Compute the characterization}\\
        \pclinecomment{of the languages:}\\
        \forall (c,d) \in T \times \{0,1\}, h_d^{(c)} \assign h(d \| w_d^{(c)})\\
        \pi \assign \text{(NI)ZK proof that:}\begin{tikzpicture}[overlay,remember picture]
          \node[name=startPi,align=left,text width=5cm,anchor=west,yshift=1mm,xshift=5cm,blue,rounded corners,draw]{If the ZK proof is interactive, then we actually run the ZK protocol (before sending the quantum state) instead of sending the proof (of course this adds additional rounds of communication).};
          \draw[blue,->](startPi) -- (2mm,1mm);
        \end{tikzpicture}\\
        \t \exists T \subseteq [n], (w_d^{(c)})_{c \in T, d \in \{0,1\}},\\
        \t \t \forall c, d, h_d^{(c)} = h(d\| w_d^{(c)}) \\
        \t \t \text{and } \forall j \in T, \exists d \text{ s.t. } w_d^{(j)}[1] = 1,\\
        \t \t \text{and } \Pred(T) = \top.\\
        \forall i \in [n] \setminus T, r^{(i)} \sample \{0,1\}\\
        \text{$\forall i \in [n] \setminus T$, applies $Z^{r^{(i)}}\rho^{(i)}$}\\
        \forall c \in [n], \text{Apply on $\rho^{(c)}$: } \ket{x} \mapsto \ket{x}\ket{w_x^{(c)}}\\
        \text{(call $\rho_1^{(1),\dots,(n)}$ the resulting state)}\\
        \< \sendmessage{->}{style={transform canvas={yshift=-3mm}},length=1.6cm,topstyle={overlay},top={$\forall (c \in [n],d \in \{0,1\}): h^{(c)}_d, \pi, \rho_1^{(1),\dots,(n)}$}}\\
        \<\< \text{If Alice sent an abort message $\bot$, } \pcreturn \bot\\
        \<\< \text{Check (or run if interactive proof) $\pi$.}\\
        \<\<\text{$\forall c$, apply on $\rho_1^{(c)} \otimes \ketbra{0}{0}$ the unitary}\\
        \<\<\t \text{implementing:}\\
        \<\<\t x,w \mapsto w[1] \neq 1 \land \exists d, h(x\|w) = h_d^{(c)},\\
        \<\<\t \text{measure the last (output) register}\\
        \<\<\t \text{and check that the outcome is $1$.}\\
        \<\<\forall c, \text{measure the second register of $\rho_1^{(c)}$}\\
        \<\<\t \text{in the Hadamard basis (outcome $s^{(c)}$).}\pclb
        \pcintertext[dotted]{End of the send procedure}
        \<\< \pcreturn \text{Remaining (first) qubit of each $\rho_1^{(i)}$.}\\
        \<\sendmessage{<-}{length=1.6cm,topstyle={overlay},top={$\forall c, s^{(c)}$}}\\
        \text{Compute:} \\
        \omega_s \assign (r^{(i)} \xor \langle s^{(i)}, w^{(i)}_0 \xor w^{(i)}_1\rangle)_{i \in [n] \setminus T}\\
        \pcreturn \omega_s
      }
    \end{autoFit}
  \end{protocol}
}

We will see that the $\Fsemicol{\Pred}$ functionality can be used to trivially get more advanced OT protocols, notably string OT and $k$-out-of-$n$ OT for any $k$ and $n$. But first, we prove that it is a ZKoQS functionality for the quantum language of ``semi-collapsed'' states with respect to a predicate $\Pred$. Informally, we define the quantum language of \emph{semi-collapsed} states as the set of states such that there exists a subset $T$ of qubits such that $\Pred(T) = \top $, and such that all qubits in $T$ are collapsed, i.e.\ measured in the computational basis and equal to $\ket{0}$ or $\ket{1}$ (therefore not entangled with any other system). Moreover, the identity of the set $T$ of collapsed qubits stays hidden to a malicious verifier, and in an honest protocol the non-collapsed qubits are either a $\ket{+}$ or a $\ket{-}$, this description being known to the prover.

\arxivOnly{\begin{remark}
  Note that the predicate $\Pred$ might (implicitly\footnote{In which case, the witness must be used to generate the ZK proof.}) depend on an additional secret classical witness (like a password, a signature provided by some trusted parties, or any NP statement) only known by the prover. This can allow the prover to prove even more advanced statements, like ``Either all states are collapsed, or I am the owner of this bitcoin wallet and a single state is collapsed'', which can for instance be useful to obtain ``anonymous authorized OT'' (i.e.\ an OT protocol where only parties knowing the witness can participate, while the sender never knows if the receiver knows the witness or not).
\end{remark}}

\begin{definition}[Semi-collapsed states $\langSemCol{\Pred}$]\label{def:langSemCol}
  The quantum language $\langSemCol{\Pred}$ of \emph{semi-collapsed states} relative to a predicate $\Pred \colon \cP([n]) \rightarrow \{\top,\bot\}$ on the subsets of qubits is composed of the classes $\cC = \cP([n])$ (denoting the set of collapsed qubits), the sub-classes $\cC_s = \{ s \in \{0,1\}^* \mid |s| \leq n\}$ (denoting the description of the non-collapsed qubits), and the quantum (sub-)classes defined as follows, for any $T \in \cC$ and $\omega_s \in \cC_s$:
  \begin{itemize}
  \item $\lang_{T,\omega_s}$ is the empty set if $\Pred(T) = \false$ or if $|\omega_s| \neq |T|$, and otherwise is the set of all $n$-qubits states where qubits in $T$ are either $\ket{0}$ or $\ket{1}$, and other qubits $i$ ($i \in \{1,\dots,|T|\}$ is the index of the qubits in $[n] \setminus T$) are equal to $\ket{+}$ if $\omega_s[i] = 0$ and $\ket{-}$ otherwise.
  \item $\langSemCol{\Pred}$ is the set of bipartite states on registers $\P$ and $\V$ such that $\V$ contains $n$ qubits, and such that there exists $T \subseteq [n]$ such that $\Pred(T) = \bot$ and for any $i \in T$, $i$-th qubit of register $\V$ is not entangled with any other qubit and either $\ket{0}$ or $\ket{1}$.
  \end{itemize}
\end{definition}

\begin{corollaryE}[ZKoQS for semi-collapsed states][end]\label{cor:ZKoQSForSemiCol}
  Let $G'(T)$ be the procedure that samples $(r^{(i)})_{i \in [n]} \sample \{0,1\}$ and outputs the quantum state $\bigotimes_i H^{\delta_{i \notin T}}\ket{a^{(i)}}$ (i.e.\ all qubits in $T$ are in the computational basis, others are in the Hadamard basis).
  
  The functionality $\Fsemicol{\Pred}$ (where the ideal party $\tilde{\P}$ is slightly updated\footnote{$\Fsemicol{\Pred}$ can be used for any input quantum state, but for the ZKoQS we need to consider a particular case where the initial state is picked by the party instead of by the environment. The reason is that in ZKoQS protocols, an honest prover is only given as input a class.}: instead of receiving $T$ and $\rho$, it receives $T \subseteq [n]$, and samples $\rho \gets G'(T)$, before continuing as usual) is a ZKoQS ideal functionality (\cref{def:NIZKoQSFunc}) for the quantum language $\langSemCol{\Pred}$ (\cref{def:langSemCol}).
   
  In particular, if we consider the protocol where the honest prover gets as input $T$, picks $\rho \gets G'(T)$, and runs \cref{protoc:fsemicol}, this protocol is a ZKoQS protocol for the quantum language $\langSemCol{\Pred}$.
\end{corollaryE}
This is mostly a corollary of \cref{thm:FpmImpliesZKoQS}. The only non-trivial statement is to prove that the measurement is postponable: this can be done by teleporting the state without applying any correction.%
\begin{proofE}
  The first statement is a quite direct application of \cref{thm:FpmImpliesZKoQS}, where we define:
  \begin{itemize}
  \item $M$ like in \cref{def:fsemicol},
  \item for any $T \in [n] = \cC$, if $\Pred(T) = \bot$ then $E_T = \emptyset$, otherwise:
    \begin{align}
      E_{T} \eqdef \{\ket{T} \otimes \ket{r} \otimes (\ket{l}^{(T)} \ket{+}^{[n] \setminus T})\}_{l \in \{0,1\}^{|T|}, r \in \{0,1\}^{|[n]\setminus T|}}
    \end{align}
  \item and for any $T \subseteq [n]$, $G_T$ is defined as the procedure that runs $\rho \gets G'(T)$, samples $(r^{i})_{i \in [n]\setminus T} \sample \{0,1\}^{n-|T|}$, and outputs $\ket{T}\ket{(r^{i})_{i \in [n]\setminus T}}\otimes \rho$ (note that it is basically the operation performed by the ideal dummy party).
  \end{itemize}
  The only non-trivial check is to show that $M$ are postponable measurement operators (\cref{def:postponableOperator}) with respect to $\{G_T\}_{T, \lang_T \neq \emptyset} = \{G_T\}_{T, \Pred(T) = \top}$. The idea is to do teleportation without corrections, exploiting the fact that the uncorrected corrections just flip the encoded bit without changing its basis. (Note that to avoid dirty matrix computations, we give a more intuitive proof in the ZX calculus (but it is of course possible to derive the same proof with the usual matrix algebra). We refer curious readers unfamiliar with the ZX calculus to \cite{van20_ZXcalculusWorkingQuantum}.) More precisely, we define $\rho^{\V,F}$ as the system containing $n$ Bell pairs $\ket{00}+\ket{11}$ shared between registers $\V$ and $F$ (the Bell pairs being between the $\V^{(i)}$ and $F^{(i)}$ for all $i$). Then, we define $M'$ as follows:
  \begin{itemize}
  \item $M'$ takes as input the register $F$ containing Bell pairs, and a state $\rho_0$ (sampled by $G_T$).
  \item Then, it runs $M\rho_0$, to get a post-measured state $\rho_1$ and a measurement outcome
    \begin{align}
      (T, (m^{(j)})_{j \in T}, (r^{(i)})_{i \in [n] \setminus T})
    \end{align}
  \item Then, for all $i\in [n]$, it performs a Bell measurement (i.e.\ it does a projection on $\{\ket{0x}+(-1)^{z}\ket{1\bar{x}}\}_{x \in \{0,1\}, z \in \{0,1\}}$) between the $i$-th qubit of $F$ and the $i$-th qubit of $\V$ to get outcomes $(x^{(i)},z^{(i)})$.
  \item Finally, it outputs the outcome measurement $(T, (x^{(j)} \xor m^{(j)})_{j \in T}, (z^{(i)} \xor r^{(i)})_{i \in [n] \setminus T})$
  \end{itemize}
  We prove now that $M G_T \approxRVS (I^\V \otimes M')(\rho^{\V,F} \otimes G_T)$. Note that both $M$ and $M'$ work separately on each system composed of the $i$-th qubit of each register, we can therefore just consider each system $i \in [n]$ separately. Let $T \in \cP([n])$ such that $\Pred(T) = \top$, and $i \in [n]$. Then $G_T$ outputs on qubit $i$ the state $H^{\delta_{i \notin T}}\ket{0}$ with probability $1/2$ and $H^{\delta_{i \notin T}}\ket{1}$ otherwise. We do two cases: if $i \notin T$, then $G_T$ generated a state $H\ket{a}$ (we omit the index, $H\ket{a}$ being represented as $\zx{\zxZ{a\pi} \zxNR{}}$ in the ZX calculus), and $M$ (and therefore $M'$) will perform a random $Z^{r}$ flip on it (represented as $\zx{\zxNL \zxZ{r\pi} \zxNR{}}$), we can easily see (manually or with the ZX calculus) that the final state at the end of the procedure $(I^\V \otimes M')(\rho^{\V,F} \otimes G_T)$ is $Z^{r \xor z}H\ket{a}$:
  \begin{align}
    \begin{ZX}[
      execute at end picture={
        \zxNamedBox[fit margins={right=2.2mm}]{(measX)(measZ)}{below:\footnotesize Bell measurement}
        \zxNamedBox[fit margins={bottom=2pt,top=2pt,left=1.5mm}][green!80!black]{(bellA)(bellB)}{left:\footnotesize $\rho^{\V,F} =$ Bell pair: }
      }
      ]
      \zxN[a=bellA]{} \dar[C] \ar[rrrr] & & & & \zxN{} \\[\zxwCol]
      \zxN[a=bellB]{} \ar[rrr] &&&\zxX[a=measX]{x\pi} \dar[C-]&\\
      & \zxZ{a\pi} \rar& \zxZ{r\pi} \rar & \zxZ[a=measZ]{z\pi}
    \end{ZX}
    \eqAboveWidth{(S)}
    \begin{ZX}
      \zxN[a=bellA]{} \dar[C] \ar[rr] & & \zxN{} \\[\zxwCol]
      \zxN[a=bellB]{} \ar[r] &\zxX[a=measX]{x\pi} \dar[C-]&\\
      & \zxZ{(a\xor r \xor z)\pi} 
    \end{ZX}  
    \eqAboveWidth{(C)}
    \begin{ZX}
      \zxZ{(a\xor r \xor z)\pi} \zxNR{}
    \end{ZX}  
    \eqAboveWidth{(S)}
    \begin{ZX}
      \zxZ{a\pi} \rar & \zxZ{(r \xor z)\pi} \zxNR{}
    \end{ZX}  
  \end{align}
  (the first equality comes from the spider fusion rule that allows to merge spiders of the same color, adding the angles modulo $2\pi$, the second rule being a particular case of the copy rule and the ``only topology matters'' principle that states that one can deform a graph without changing its interpretation). Note that since $r$ is sampled uniformly at random, and since the outcome of the measurement $z$ is independent of the value of $r\pi$ and uniformly distributed (this is easy to see as the norm of this state (computable by computing the trace using the discarding ground operation that trivially absorbs\footnote{Note that here we removed all scalars and global phases as they are just constant re-normalisation factors.} all terms: $\zx{\zxZ{(a\xor r \xor z)\pi} \rar & \zxGround{}} = \zx{\zxEmptyDiagram} $) is independent of the value of all the variables). Therefore, by defining $r' \eqdef r \xor z$, $r'$ is sampled uniformly at random, and the final state corresponds to a random rotation $Z^{r'}$ of the original qubit, exactly like in the original $M$ operation, concluding the proof\footnote{If wants to be even more formal, we can actually compute the exact density matrix of the whole process, including the outcome of the measurement, using either standard linear algebra, or diagrammatically using the discard construction~\cite{CJPV21_CompletenessGraphicalLanguages} or using the doubling formalism~\cite{CK17_PicturingQuantumProcesses} to represent density matrices, but this lead to the same result.}.

  Now, if $i \in T$, then $M'$ will measure the state $\ket{a}$ provided by $G_T$, but since it is anyway already in the computational basis this measurement has no effect (except revealing the value of $a$). Then, the Bell measurement similarly gives:
  \begin{align}
    \begin{ZX}
      \zxN[a=bellA]{} \dar[C] \ar[rrr] & & & \zxN{} \\[\zxwCol]
      \zxN[a=bellB]{} \ar[rr] &&\zxX[a=measX]{x\pi} \dar[C-]&\\
      & \zxX{a\pi} \rar & \zxZ[a=measZ]{z\pi}
    \end{ZX}
    \eqAboveWidth{(C)}
    \begin{ZX}
      \zxN[a=bellA]{} \dar[C] \ar[rrr] & & & \zxN{} \\[\zxwCol]
      \zxN[a=bellB]{} \ar[rr] &&\zxX[a=measX]{x\pi} \dar[C-]&\\
      & \zxX{a\pi} \rar & \zxN{}
    \end{ZX}
    \eqAboveWidth{(S)}
    \begin{ZX}
      \zxX{(a \xor x)\pi} \zxNR{}
    \end{ZX}
  \end{align}
  where $\zx{\zxX{(a \xor x)\pi} \zxNR{}} = \ket{a \xor x}$. We can similarly see that the outcome $x$ is independent of $a$. Therefore, since $a$ and $x$ are uniformly at random, the final state is a state sampled uniformly at random in the computational basis, exactly like in $M G_T$, concluding this part of the proof.

  Finally, the second statement is trivial to check: since we just shown that the functionality $\Fsemicol{\Pred}$ is a ZKoQS, all protocols realizing it are ZKoQS protocols. However, we already know from \cref{thm:fsemicol} that \cref{protoc:fsemicol} realizes $\Fsemicol{\Pred}$, with a slightly different dummy ideal party, say $\tilde{\P}_0$. However, since we obtain the new ideal dummy party and new protocol by doing exactly the same pre-processing ($\tilde{\P} = \tilde{\P}_0(G_T)$ and  $\P = \P_0(G_T)$, where $\P_0$ is the original party in \cref{protoc:fsemicol}), the distinguishing probability between the new real world and ideal world is lower, otherwise a distinguisher could apply $G_T$ to attack the original protocol known to be secure.
\end{proofE}

\subsection{\texorpdfstring{\ZKstatesQIP{S}{k}}{ZKstatesQIP} and \texorpdfstring{\ZKstatesQMA{S}}{ZKstatesQMA}: ZKoQS from a complexity theory point of view}\label{sec:aComplexityView}

While we defined ZKoQS using a ``cryptographic'' definition, we can also consider them from the point of view of complexity theory. While classically, complexity classes involve a verifier taking an input $x$ potentially belonging to a given classical language $\lang$, and outputting a single accept bit (this is not an issue as the input $x$ can anyway be copied by the verifier if it needs to be used later), for quantum languages this definition turns out to be hard (or even impossible) to use as the verification procedure will alter the input state. (\cite{KA04_ComplexityQuantumLanguages} does something along that line, but needs to send many copies of the input state, which is of little interest in cryptography as it leads to polynomial security.) To overcome this issue, it is therefore natural to say that the quantum state belonging to the quantum language must be an \emph{output} of the verifier. This is the successful point of view that we took above, and a similar approach has also been used before in~\cite{RY22_InteractiveProofsSynthesizing} to quantify the complexity to produce a given state by defining a complexity class \stateQIP{}. However, the class \stateQIP{} only captures how hard it is to generate a given state, but it does not capture any notion of privacy against a malicious verifier. The following definition addresses this issue:

\begin{definition}[\ZKstatesQIP*{S}{k} and \ZKstatesQMA*{S}]\label{def:ZKstatesQIP}
  Let $\lang$ be a quantum language (\cref{def:quantumLanguage}), $k \in \N$ be a number of exchanged messages, $S \in \{\emptyset, \P, \V\}$ be a subset of parties allowed to be unbounded, and $\setup$ be a given setup assumption (e.g. CRS, Random Oracle, or plain-model). We say that $\lang$ belongs to the complexity class $\ZKstatesQIP*{S}[\setup]{k}$ if there exists a $\text{ZKoQS}_S$ protocol for $\lang$, secure assuming the setup assumption $\setup$, whose \textsf{send} phase consists of $k$ exchanged messages (note that we might omit $S$, $\setup$, or $k$ if we do not want to constraint this parameter).
  
  Similarly, we define $\ZKstatesQMA*{S}[\setup] = \ZKstatesQIP*{S}[\setup]{1}$ to capture non-interactive protocols.
\end{definition}

\publishedVsArxiv{For comparison with other works that introduced \textsf{stateQIP}, see \cameraVsOther{the full version~\cite{CMS23_ObliviousTransferZeroKnowledge}}{\cref{rk:linkWithQMA}}.}{}
\inAppendixIfPublished{
  \begin{remark}\label{rk:linkWithQMA}
    Note that \cite{RY22_InteractiveProofsSynthesizing} defines multiple complexity classes like \textsf{stateQIP} (update: similarly, \cite{DGLM23_QuantumMerlinArthurProof}, that was uploaded online a day after our own work, defines \textsf{stateQMA}): it is therefore natural to want to compare \textsf{stateQIP} and $\ZKstatesQIP{}{}$, similarly to the result $\textsf{stateQIP} = \textsf{statePSPACE}$ presented in \cite{MY23_StateQIPStatePSPACE}. However, note that since \cite{RY22_InteractiveProofsSynthesizing} and \cite{DGLM23_QuantumMerlinArthurProof} mostly care about the complexity required to create quantum states, there is no notion of hiding or witnesses\footnote{Actually, in \cite{DGLM23_QuantumMerlinArthurProof}, they do define witnesses but in a different way, as they send the witness directly to the verifier: thus, their notion of witness corresponds rather to the transcript of the proof in our case, and should not be understood as an information that must be hidden to the verifier like our own notion of witness (a.k.a.\ class).}: expressed with our terminology, their quantum languages have a single element (for a fixed, public, $n$) $\ket{\psi_n}$. Said differently, the verifier knows in advance the state $\ket{\psi_n}$ that will be generated with the help of the prover. On the other side, we have no reasons to introduce in $\ZKstatesQIP{}{}$ an asymptotic parameter $n$ denoting the size of the quantum state obtained by the verifier, as ZKoQS already makes sense for a fixed $n = 2$ (but of course, nothing prevents $\lang_{\cQ}$ from containing states of various sizes). However, it might be possible to generalize the definition of \cite{RY22_InteractiveProofsSynthesizing} by replacing $(\ket{\psi_n})_{n \in \N}$ with $\lang_{\cQ}$, where the parameter $n$ could represent the size of the state obtained by the verifier, or to consider a sequence of quantum languages $(\lang_{\cQ,n})_{n \in \N}$, in order to define $\mathsf{state\underline{\mathsf{s}}QIP}$ (the additional $\mathsf{s}$ being used to denote the fact that the verifier might produce a state among multiple, valid, candidates). However, properly generalizing \cite{RY22_InteractiveProofsSynthesizing} and defining $\mathsf{state\mathbf{s}QIP}$/$\mathsf{statesQMA}$ is out of the scope of this paper.
  \end{remark}
}

We prove now that $\langSemCol{\Pred}$ belongs to these classes:

\begin{corollaryE}[{$\langSemCol{\Pred}$ is in \mbox{$\ZKstatesQMA{}[\mathsf{RO}]$}}][end]\label{cor:existingZKstatesQMA}
  For any predicate $\Pred$, the quantum language $\langSemCol{\Pred}$ belong to $\ZKstatesQMA{}[\mathsf{RO}]$ (where $\mathsf{RO}$ stands for Random Oracle model). Moreover, assuming the hardness of \LWE{} (see~\cite{HSS11_ClassicalCryptographicProtocols} for the exact assumptions), $\langSemCol{\Pred}$ belongs to $\ZKstatesQIP{}[\mathsf{pm}]{}$ (where $\mathsf{pm}$ stands for plain-model).

  More generally, assuming the existence of a $k$-message ZK protocol $\CSQSA{S}$ realizing $\Fzk{}$ for any \NP{} statement assuming a setup $\setup$, $\langSemCol{\Pred}$ belong to $\ZKstatesQIP{S}[\setup]{k}$.
\end{corollaryE}
These statements can be proven using \cref{cor:ZKoQSForSemiCol}, together with the constructions of \cite{Unr15_NonInteractiveZeroKnowledgeProofs} and \cite{HSS11_ClassicalCryptographicProtocols}.
\begin{proofE}
  The last statement is a direct consequence of \cref{cor:ZKoQSForSemiCol} and of the definition of $\ZKstatesQMA{S}[\setup]{k}$. The first statement is obtained by instantiating the ZK protocol using the NIZK construction of~\cite{Unr15_NonInteractiveZeroKnowledgeProofs} secure in the Random Oracle model (we prove in \cref{sec:unr15_is_composable} that their definition can be translated in a the quantum standalone framework). The second statement is obtained by instantiating the ZK protocol using the construction of~\cite{HSS11_ClassicalCryptographicProtocols}, proven secure in the plain-model assuming the hardness of \LWE{} (this construction is already proven secure on the quantum standalone model, so no additional work is required).
\end{proofE}

\subsection{Applications to build string and $k$-out-of-$n$ OT protocols}

We prove in this section that the above functionality $\Fsemicol{\Pred}$ actually allows us to have string OT or $k$-out-of-$n$ OT. But first, we show that we can realize this functionality:

\begin{theoremE}[][end]\label{thm:semicolImpliesOtPred}
  Let $\Pred$ be a predicate on subsets of $[n]$. Assuming the existence of a protocol $\Pisemicol* = (\Asemicol, \Bsemicol)$ that $\CSQSA{S}$-realises $\Fsemicol{\Pred}$, there \cref{protoc:langSemColToOT} $\CSQSA{S}$-realises $\FotPred$.
\end{theoremE}
This is a generalisation of the last part of the proof of \cref{thm:realizesOT}.
\begin{proofE}
  This is a straightforward generalisation of the last part of the proof of \cref{thm:realizesOT}. We start with correctness.

  \paragraph{Case 1: correctness (no corrupted party).}
  Since $\Pisemicol*$ $\CSQSA{S}$-realises $\Fsemicol{\Pred}$, we can indistinguishably replace $\Asemicol*$ and $\Bsemicol*$ with dummy ideal adversaries interacting with $\Fsemicol{\Pred}$. Then, we show that this world is indistinguishable from the real world: if the input $B$ is such that $\Pred(T) = \bot$, all parties would abort like in the real world. Otherwise, if $\Pred(T) = \top$, then the functionality measures states not in $B$… but since these states are already measured, it left them unchanged. It also rotates the qubits in $B$ (in the Hadamard basis) them by $Z^{z^{(i)}}$. Therefore, for all $i \in B$, the $i$-th qubit becomes $\rho^{(i)} = Z^{r^{(i)} \xor s^{(i)}}\ket{+}$. After the rotation performed by $\Bob$, we get $Z^{r^{(i)} \xor s^{(i)} \xor m_i}\ket{+}$, and therefore the measurement in the Hadamard basis gives $z^{(i)} = r^{(i)} \xor s^{(i)} \xor m_i$. Since $\Alice$ outputs for each $i \in B$, $r^{(i)} \xor s^{(i)} \xor z^{(i)} = m_i$, the output of $\Alice$ is exactly the same as in the ideal world, making both worlds indistinguishable, concluding the correctness proof.

  \paragraph{Case 2: malicious receiver Alice.}
  If $\hat{\Alice}$ is malicious, we can cut $\hat{\Alice}$ in two parts, $\hat{\Alice}_0$ interacting with $\Bsemicol*$ and $\hat{\Alice}_1$ doing the rest of the computation. Since $\Pisemicol*$ $\CSQSA*{S}$-realises $\Fsemicol{\Pred}$, there exists a simulator $\Sim_{\hat{A}_0}$ such that $\hat{\Alice}_0 \interacts \Bsemicol* \approxRVC \Sim_{\hat{A}_0} \interactsF{\Fsemicol{\Pred}} \tilde{\Bsemicol*}$ (or $\approxRVS$ if $\P \in S$). We can therefore indistinguishably replace the first one with the last one. Since $\tilde{\Bsemicol*}$ measures qubits not in $B$, the rotation on these qubits performed by Bob has no effect: as a result, we can remove them indistinguishably. Now, neither $\Alice$ nor $\Bob$ depends on $m_i$ for $i \notin B$, we can therefore move them into the final simulator, using the $(m_i)_{i \in B}$ provided by $\FotPred$ otherwise (note that if $\Pred(B) = \bot$, $\FotPred$ would not provide these values and abort, but this is not an issue since anyway $\Fsemicol{\Pred}$ also aborts in that case). This new world is therefore equal to the ideal world, and indistinguishable from the previous world, concluding the proof of security.

  \paragraph{Case 3: malicious sender Bob.}
  If $\hat{\Bob}$ is malicious, we can cut $\hat{\Bob}$ in two parts, $\hat{\Bob}_0$ interacting with $\Asemicol*$ and $\hat{\Bob}_1$ doing the rest of the protocol. Since $\Pisemicol*$ $\CSQSA*{S}$-realises $\Fsemicol{\Pred}$, there exists a simulator $\Sim_{\hat{B}_0}$ such that $\Asemicol* \interacts \hat{\Bob}_0 \approxRVC \tilde{\Bsemicol*} \interactsF{\Fsemicol{\Pred}} \Sim_{\hat{B}_0}$ (or $\approxRVS$ if $\P \in S$). We can therefore indistinguishably replace the first one with the last one. Then, since the qubits in $B$ are already in the computational basis, the functionality can skip the measurements of these qubits without being detected. Similarly, as we already saw it earlier, since sampling a qubit in $\ket{0}$ or $\ket{1}$ is indistinguishable from sampling a qubit in $\ket{+}$ or $\ket{-}$ (the encoded value being discarded), we can indistinguishably apply an $H$ gate on these qubits to turn them back into qubits in the Hadamard basis, and also compute $r^{(i)} \xor s^{(i)} \xor z^{(i)}$ for these qubits (it will be discarded anyway). Then, except for the return procedure of $\P$ that discards some terms, all the steps are independent of $m_i$'s: we can therefore move everything into our final simulator, let it send the $r^{(i)} \xor s^{(i)} \xor z^{(i)}$ to the ideal functionality $\FotPred$ that will be in charge of discarded elements not it $T$. This last step is indistinguishable as we only moved some operations, ending this proof of security.  
\end{proofE}

\begin{protocol}[!htbp]
  \caption{Protocol to compile a ZKoQS protocol $(\Asemicol*,\Bsemicol*)$ for the quantum language $\langSemCol{\Pred}$ into a predicate OT protocol.}\label{protoc:langSemColToOT}\centering
  \begin{autoFit}
    \publishedVsArxiv{\def\myarrowlength{.5cm}}{\def\myarrowlength{3.3cm}}%
    \pseudocodeblock{
      \textbf{Alice($B \subseteq \{0,1\}^n$)} \< \< \textbf{Bob($(m_1,\dots,m_n) \in \{0,1\}^n$)} \\[][\hline]
      \\[-.5\baselineskip]
      \text{If $\Pred(B) = \bot$, abort.}\\
      \forall i \in [n], r^{(i)} \gets \{0,1\}\\
      \rho \eqdef \otimes_{i \in [n]} H^{\delta_{i \in B}}\ket{r^{(i)}}\\
      (s^{(i)})_{i \in B} \gets \Asemicol*(B,\rho) \< \sendmessage{<->}{length=\myarrowlength,topstyle={overlay},top={}} \< \rho \gets \Bsemicol* \\
      \text{Abort if the previous step aborted.} \< \< \text{If the previous step aborted, abort.}\\
      \<\< \forall c, \text{apply $Z^{m_c}$ on $\rho^{(c)}$ and measure it}\\
      \<\<\t \text{in the Hadamard basis (outcome $z^{(c)}$).}\\
      \pcreturn (r^{(i)} \xor s^{(i)} \xor z^{(i)})_{i \in B} \<\sendmessage{<-}{topstyle={overlay},length=\myarrowlength,top={$\forall c, z^{(c)}$}}\\
    }
  \end{autoFit}
\end{protocol}

\begin{corollaryE}[][normal]\label{cor:stringOT}
  By choosing appropriate values for $\Pred$ like in \cref{def:Fotpred}, the protocol \cref{protoc:langSemColToOT} realizes the string OT functionality $\FotStr$ and the $k$-out-of-$n$ OT functionality $\FotKoutN$.
\end{corollaryE}
\begin{proofE}
  This is a direct consequence of \cref{thm:semicolImpliesOtPred} and of the definition of $\FotKoutN$ and $\FotStr$. 
\end{proofE}

\section{Composability of \cite{Unr15_NonInteractiveZeroKnowledgeProofs}}\label{sec:unr15_is_composable}
\pratendSetLocal{category=unruhComposable}

We show now that the online extractable NIZK protocol from~\cite{Unr15_NonInteractiveZeroKnowledgeProofs} quantum stand-alone realizes the $\Fzk*{\cR}$ functionality in~\cref{def:Fzk}, when the RO assumption is made. This is needed to instantiate \cref{cor:stringOT} with a concrete ZK protocol.

\cameraVsOther{}{
  The polynomial-time QIM prover $\P{}$ and verifier $\V{}$ from \cite[Fig.~1]{Unr15_NonInteractiveZeroKnowledgeProofs} have access to two random oracles, $G$ and $H$, which can be queried in superposition by both parties (for simplicitly we will just refer to a single oracle $H$). We will denote the polynomial-time two-party protocol by $\Pizk^{H}=(\P{},\V{})$ to stress the interaction between two machines and the trusted random oracles $\P{}\interactsF{H}\V{}$. Note that a single message is sent from the prover $\P$ to the verifier $\V$, leading to a so-called non-interactive protocol.

The protocol $\Pizk^{H}$ is proven to be complete, zero-knowledge and (even simulation-sound) online-extractable. We recall the definitions in \cref{def:unruhZK} for clarity (note that we assume they also hold against non-uniform adversaries).

\textEnd{
  \begin{definition}\label{def:unruhZK}
    \begin{enumerate}
    \item \textbf{Completeness} (\cite[Def.~1]{Unr15_NonInteractiveZeroKnowledgeProofs}): $\Pizk^H$ is \emph{complete} iff for any quantum-polynomial-time oracle algorithm $A$ and advice $(\sigma_\lambda)_{\lambda \in \N}$,
      \begin{align}
      	\begin{multlined}[t]
        \Pr\Bigl[(x,w)\in\cR\land y= 0 \Big| H\gets\mathsf{ROdist}, (x,w)\gets A^H(\sigma_\lambda),\\
        y\gets\OUT_\V\langle\P(x,w)\interactsF{H}\V\rangle\Bigr] \leq \negl.\label{eq:un_complete}
        \end{multlined}
      \end{align}  
    \item \textbf{Zero-knowledge} (\cite[Def.~2]{Unr15_NonInteractiveZeroKnowledgeProofs}): $\Pizk^H$ is \emph{zero-knowledge} iff there exists a polynomial-time simulator $\Sim=(\Simin,\Sim_P)$ such that for every quantum-polynomial-time oracle algorithm $A$ and advice $(\sigma_\lambda)_{\lambda \in \N}$,
      \begin{align}
      	\begin{multlined}[t]
        \left|\pr[H\gets \mathsf{ROdist},z\gets A^{H,\P}(\sigma_\lambda)]{z=1}\right.\\
        -\left.\pr[H\gets \Simin,z\gets A^{H,\Sim_P}(\sigma_\lambda)]{z=1}\right|\leq\negl.
        \end{multlined}
      \end{align}
      Since in the quantum setting we cannot allow the simulator to learn the input for each query, because this can be done in superposition, here the simulator $S_{\text{init}}$ outputs a circuit describing a classical function representing the initial random oracle instead. We assume that both $\Simin$ and $\Sim_P$ have access to the polynomial upper bound on the runtime of $A$.
    \item \textbf{Online-extractability} (\cite[Def.~3]{Unr15_NonInteractiveZeroKnowledgeProofs}): $\Pizk^H$ is \emph{online extractable} with respect to $\Simin$ iff there exists a polynomial-time extractor $E$ such that for any quantum-polynomial-time oracle algorithm $A$ and advice $(\sigma_\lambda)_{\lambda \in \N}$,
      \begin{align}
      	\begin{multlined}[t]
        \Pr\Bigl[y=1\land (x,w)\not\in\cR \Big| H\gets \Simin(), (x,\pi)\gets A^H(\sigma_\lambda), y\gets \V^H(x,\pi),\\
        w\gets E(H,x,\pi)\Bigr]\leq\negl.
        \label{eq:online_extrac}
        \end{multlined}
      \end{align}
      We assume that both $\Simin$ and $E$ have access to the polynomial upper bound on the runtime of $A$.
    \end{enumerate}
  \end{definition}
Note that Unruh's original definition he considers only uniform adversaries, here we assume that the above conditions hold for the protocol when the adversary receives advice.
}

In the following we will prove that the protocol $\Pizk^{H}$ quantum stand-alone realizes the functionality $\Fzk*{\cR}$.
}

\begin{theoremE}[][end]\label{thm:Unr15Composable}
  Let $H$ be a random oracle. The non-interactive protocol $\Pizk^{H}=(\P{},\V{})$ from~\cite{Unr15_NonInteractiveZeroKnowledgeProofs} quantum stand-alone realizes the classical zero-knowledge functionality $\Fzk{\cR}$, were $x\in\cL\Leftrightarrow\exists w,\: x\cR w$.
\end{theoremE}
\begin{proofE}
  We define trivially the dummy parties $\tPizk = (\tP,\tV)$ that forward the inputs/outputs to/from $\Fzk*{\cR}$. We want to show that for any \mbox{(poly-)time} adversary $\cA$ there exists a \mbox{(poly-)time} simulator $\Sim$ such that, for any poly-time distinguisher $\Zenv$ and input state $\sigma_\lambda$, we have $\Real^{\sigma}_{\Pizk,\cA,\Zenv}\approxRV \Ideal^{\sigma,\Fzk{\cR}}_{\tPizk,\Sim_\cA,\Zenv}$. We will split the proof depending on the parties that the static adversary $\cA$ corrupts (nobody, the prover $\P$ or the verifier $\V$).
  
  Note that in the non-interactive protocol $\Pizk^{H}$ the prover rejects if it receives a non-valid witness $(x,w)\not\in\cR$, which is essential.
  
  \paragraph{Case 1: correctness (no corrupted party).}
  
  For any bipartite input state $\sigma_\lambda^{\P,\Zenv}\in\text{D}(\S_\lambda\otimes\R_\lambda)$, with $\R_\lambda$ an arbitrary reference system, we want to show that for any environment $\Zenv$ the probability of distinguishing $\Pizk^{H}$ from the ideal functionality $\tPizk$ is negligible 
  \begin{align}
  	\begin{split}
    &\left|\pr[ H\gets\ROd, y\gets\OUT_\V\langle\P(\sigma^{\P}_\lambda)\interactsF{H}\V\rangle]{\Zenv(y,\sigma^{\Zenv}_\lambda)=1}\right.\\
    &\left.-\pr[H\gets\ROd, y\gets\OUT_\V\langle\tP(\sigma^\P_\lambda)\interactsF{\cF}\tV\rangle]{\Zenv(y,\sigma^\Zenv_\lambda)=1}\right|\leq\negl.
    \end{split}\label{eq:correct}
  \end{align}

  In order to prove the above inequality we are interested in rewriting the probability of the distinguisher outputting $1$ in terms of the input/output of the interaction. Since the prover expects a classical message, we can model $\sigma_\lambda^{\P,\Zenv}$ as a quantum instrument:
  \begin{align}
  	\sigma_\lambda^{\P,\Zenv} = \sum_{x,w} p_{x,w}\ketbra{x,w}{x,w} \otimes \sigma_{\lambda,x,w}.
  \end{align}
  This allows to write the LHS of~\cref{eq:correct} as
  \begin{align}
  	\MoveEqLeft[3] \sum_{x,w}p_{x,w}\pr[H\gets\ROd, y\gets\OUT_\V\langle\P(x,w)\interactsF{H}\V\rangle]{\Zenv(y,\sigma_{\lambda,x,w})=1}\\
  	={}&\begin{multlined}[t]
  		\sum_{b\in\{0,1\}}\sum_{x,w}\Bigl(p_{x,w}\pr{\Zenv(b,\sigma_{\lambda,x,w})=1}\\
  		\cdot\left.\pr[H\gets\ROd, y\gets\OUT_\V\langle\P(x,w)\interactsF{H}\V\rangle]{y=b}\right)
  	    \end{multlined}\label{eq:developed}
  \end{align}

  For the dummy protocol $\tPizk$ the above equation has a very simple form since if $(x,w)\in\cR$ (resp. $(x,w)\not\in\cR$), then $\tPizk$ will output $1$ (resp. $0$) with probability $1$. Therefore, for any input state $\sigma_\lambda^{\P,\Zenv}$ we can simplify the RHS of~\cref{eq:correct} to
  \begin{align}
  	\sum_{(x,w)\in\cR}p_{x,w}\pr{\Zenv(1,\sigma_{\lambda,x,w})=1}+\sum_{(x,w)\not\in\cR}p_{x,w}\pr{\Zenv(0,\sigma_{\lambda,x,w})=1}
  \end{align}
  
  For the honest protocol $\Pizk^{H}$, note that for invalid witnesses $(x,w)\not\in\cR$ the honest prover $\P^{H}$ will also always reject, therefore for $(x,w)\not\in\cR$:
  \begin{align}
  	\pr[H\gets\ROd, y\gets\OUT_\V\langle\P(x,w)\interactsF{H}\V\rangle]{y=1}=0,
  \end{align}  
  thus for an arbitrary mixture of invalid witnesses $\sum_{(x,w)\not\in\cR}q_{x,w}\ketbra{x,w}{x,w}$ we will have that
  \begin{align}
  	\sum_{(x,w)\not\in\cR}q _{x,w}\pr[H\gets\ROd, y\gets\OUT_\V\langle\P(x,w)\interactsF{H}\V\rangle]{y=1}=0,
  	\label{eq:mix_nor_0}
  \end{align}
  and consequently
  \begin{align}
  	\sum_{(x,w)\not\in\cR}q _{x,w}\pr[H\gets\ROd, y\gets\OUT_\V\langle\P(x,w)\interactsF{H}\V\rangle]{y=0}=1.
  	\label{eq:mix_nor_1}
  \end{align} 
  
  The case of valid witnesses is also easy as we know from completeness~\cref{eq:un_complete} that given any input state $(x,w)\in\cR$, if we pick the constant algorithm $A:\sigma_\lambda^\P\mapsto(x,w)$, then
  \begin{align}
  	\pr[H\gets\ROd,y\gets\OUT_\V\langle\P(x,w)\interactsF{H}\V\rangle]{y=0}\leq\negl,
  \end{align}
  thus for an arbitrary mixture of valid witnesses $\sum_{(x,w)\in\cR}q_{x,w}\ketbra{x,w}{x,w}$ we will have that 
  \begin{align}
  	\sum_{(x,w)\in\cR}q _{x,w}\pr[H\gets\ROd, y\gets\OUT_\V\langle\P(x,w)\interactsF{H}\V\rangle]{y=0}\leq\negl,
  	\label{eq:mix_r_0}
  \end{align}
  and consequently
  \begin{align}
  	\sum_{(x,w)\in\cR}q _{x,w}\pr[H\gets\ROd, y\gets\OUT_\V\langle\P(x,w)\interactsF{H}\V\rangle]{y=1}\geq1-\negl.
  	\label{eq:mix_r_1}
  \end{align}

  We can combine the above~\cref{eq:mix_r_0,eq:mix_r_1,eq:mix_nor_0,eq:mix_nor_1} to obtain the desired inequality~\cref{eq:correct} for any input $\sigma_\lambda^{\P,\Zenv}$ and any distinguisher $\Zenv$ by noting that for any received input $b\in\{0,1\}$ and advice $\sigma_{\lambda,x,w}$, the probability of the distinguisher outputing $1$ is bounded
  \begin{align}
  	\pr{\Zenv(b,\sigma_{\lambda,x,w})=1}\leq 1,
  \end{align}
  and therefore by developing~\cref{eq:correct} in terms of the advice as in~\cref{eq:developed} we can bound the difference by
  \begin{align}
  	\begin{split}
  	\leq&\left|\sum_{(x,w)\in\cR}p_{x,w}\pr[H\gets\ROd, y\gets\OUT_\V\langle\P(x,w)\interactsF{H}\V\rangle]{y=0}\right|\\
  	&+\left|\sum_{(x,w)\in\cR}p_{x,w}\left(\pr[H\gets\ROd, y\gets\OUT_\V\langle\P(x,w)\interactsF{H}\V\rangle]{y=1}-1\right)\right|\\
  	&+\left|\sum_{(x,w)\not\in\cR}p_{x,w}\left(\pr[H\gets\ROd, y\gets\OUT_\V\langle\P(x,w)\interactsF{H}\V\rangle]{y=0}-1\right)\right|\\
  	&+\left|\sum_{(x,w)\not\in\cR}p_{x,w}\pr[H\gets\ROd, y\gets\OUT_\V\langle\P(x,w)\interactsF{H}\V\rangle]{y=1}\right|\\
  	\leq& \negl,
  	\end{split}
  \end{align}  

  \paragraph{Case 2: malicious Alice.}
  
  If the adversary corrupts the prover $\cA=\hat{\P}$, we will use online-extractability to construct the desired simulator.
  \begin{framed}
    \noindent \textbf{Simulator} $\Sim_\mathcal{\cA}\eqdef(S,E,\Simin)$: Prover is corrupted.
    \begin{enumerate}
    \item $S$ initializes $\hat{\P}$ with whatever input state it receives.
    \item $S$ obtains $(x,\pi)$ from $\hat{\P}$.
    \item $S$ initializes $E$ with $(x,\pi)$ and the description of the oracle $H$ given by the simulator $\Simin$.
    \item $E(H,x,\pi)$ extracts a witness $w$ or an abort message $\bot$ and sends it to $S$. \label{item:witness_extraction}
    \item $S$ sends $(x,w)$ or $\bot$ to $\Fzk*{\cR}$.
    \end{enumerate}
  \end{framed}
  We will now prove that no distinguisher can differentiate between the real protocol with corrupted Alice and the ideal functionality with the above simulator. This proof relies on the closeness of the following hybrid worlds (see\cameraVsOther{the full version~\cite{CMS23_ObliviousTransferZeroKnowledge}}{~\cref{fig:worlds0_2,fig:worlds2_4}} for a graphical depiction):
  \begin{itemize}
  \item $\World_0 \eqdef \Ideal^{\sigma,\Fzk{\cR}}_{\tPizk,\Sim_\cA,\Zenv}$ is the ideal world. Consists of one output by the functionality (forwarded by the dummy verifier) which is accepting if the witness obtained by the extractor in~\cref{item:witness_extraction} is valid, i.e., $(x,w)\in\cR$.
  \item $\World_1$ is like $\World_0$ except that we substitute the dummy verifier by a merge of Unruh's verifier $\V$ and the simulator. In particular, we replace $\tV$ by a verifier $\V_1$ that forwards the proof from the simulator to the extractor $E$. If the extractor provides a witness $w$, then $\V_1$ accepts $y=x$ and else aborts $y=\bot$. 
  \item $\World_2$ is like $\World_1$ except that we drop the extractor (as it is only being used to check the proof) and the dummy verifier $V_1$ which is only forwarding information, and we use Unruh's verifier $\V$ to perform the check of the proof received by the simulator $S$ instead.
  \item $\World_3$ is like $\World_2$ except that we drop the simulator $S$ as it is only forwarding the information to the verifier.
  \item $\World_4$ differs from $\World_3$ in that we replace the simulator $\Simin$ by the oracle $H$ that is simulating. Note that now $\World_4 \eqdef \Real^{\sigma}_{\Pizk,\cA,\Zenv}$. 
  \end{itemize}

  \begin{figure}
  	\centering
  	\begin{minipage}[b]{0.45\textwidth}
  		\includegraphics[scale=0.75]{figures/world_0.tex}
  		\caption{$\World_0$}
  	\end{minipage}
  	\hfill
  	\begin{minipage}[b]{0.45\textwidth}
  		\includegraphics[scale=0.75]{figures/world_1.tex}
  		\caption{$\World_1$}
  	\end{minipage}
  \label{fig:worlds0_2}
  \end{figure}

  The similarity of the first two worlds, $\World_0\approxRV\World_1$, is a consequence of online-extractability. More precisely, if we could distinguish these two worlds, there would exit an input $\sigma_\lambda^\P$ such that
  \begin{align}
  	\begin{split}
  	&\left|\pr[H\gets\Simin(),  y\gets\OUT_{\tV}\langle\Sim_\cA(\sigma_\lambda^\P)\interactsF{\cF} \tV\rangle]{\Zenv(y,\sigma^{\Zenv}_\lambda)=1}\right.\\
  	&\left.- \pr[H\gets\Simin(), y\gets\OUT_{V_1}\langle\Sim_\cA(\sigma_\lambda^\P)\interacts V_1\rangle]{\Zenv(y,\sigma^\Zenv_\lambda)=1}\right|>\negl.
  	\end{split}
  \end{align}
  In order to work with them jointly we expand them in terms of the verifier accepting/rejecting $y\in\{x,\bot\}$:
  \begin{align}
  	&\begin{multlined}[t]
  	\left|\pr[H\gets\Simin(),  y\gets\OUT_{\tV}\langle\Sim_\cA(\sigma_\lambda^\P)\interactsF{\cF} \tV\rangle]{\Zenv(y,\sigma^{\Zenv}_\lambda)=1}\right.\\
  	\left.- \pr[H\gets\Simin(), y\gets\OUT_{V_1}\langle\Sim_\cA(\sigma_\lambda^\P)\interacts V_1\rangle]{\Zenv(y,\sigma^\Zenv_\lambda)=1}\right|
  	\end{multlined}\\
    \begin{split}
  	={}&\left|\sum_{b\in\{x,\bot\}}\pr{\Zenv(b,\sigma^{\Zenv}_\lambda)=1}\pr[H\gets\Simin(),  y\gets\OUT_{\tV}\langle\Sim_\cA(\sigma_\lambda^\P)\interactsF{\cF} \tV\rangle]{y=b}\right.\\
  	&-\left.\sum_{b\in\{x,\bot\}}\pr{\Zenv(b,\sigma^\Zenv_\lambda)=1}\pr[H\gets\Simin(), y\gets\OUT_{V_1}\langle\Sim_\cA(\sigma_\lambda^\P)\interacts V_1\rangle]{y=b}\right|
  	\end{split}\\
    ={}&\begin{multlined}[t]
  	    \Big|\pr{\Zenv(x,\sigma^{\Zenv}_\lambda)=1}\nonumber\\
  	    \cdot\sum_{M\in\{\tV,V_1\}}(-1)^{\delta_V}\pr[H\gets\Simin(),  y\gets\OUT_{M}\langle\Sim_\cA(\sigma_\lambda^\P)\interactsF{\cF} M\rangle]{y=x}
  	    \end{multlined}\nonumber\\
  	&\begin{multlined}[t] +\pr{\Zenv(\bot,\sigma^\Zenv_\lambda)=1}\\
  	 \cdot\sum_{M\in\{\tV,V_1\}}(-1)^{\delta_V}\pr[H\gets\Simin(), y\gets\OUT_{M}\langle\Sim_\cA(\sigma_\lambda^\P)\interacts M\rangle]{y=\bot}\Big|
  	 \end{multlined}\\
    \begin{split}
  	\leq&\left|\sum_{M\in\{\tV,V_1\}}(-1)^{\delta(V)}\pr[H\gets\Simin(),  y\gets\OUT_{M}\langle\Sim_\cA(\sigma_\lambda^\P)\interactsF{\cF} M\rangle]{y=x}\right|\\
  	&+\left|\sum_{M\in\{\tV,V_1\}}(-1)^{\delta(V)}\pr[H\gets\Simin(),  y\gets\OUT_{M}\langle\Sim_\cA(\sigma_\lambda^\P)\interactsF{\cF} M\rangle]{y=\bot}\right|.
  	\end{split}
  \end{align}
  We write down the probabilities of each protocol accepting, $y=x$, to better visualize:
  \begin{align}
  	\MoveEqLeft[3] \pr[H\gets\Simin(),  y\gets\OUT_{\tV}\langle\Sim_\cA(\sigma_\lambda^\P)\interactsF{\cF} \tV\rangle]{y=x}\\
  	={}&\begin{multlined}[t]\Pr\Bigl[y=x\Big| H\gets\Simin(),  (x,\pi)\gets\hat{\P}^{H}(\sigma_\lambda^{\P}),w'\gets E(x,\pi,H),\\
  		y\gets\Fzk{}(x,w')\Bigr]
  		\end{multlined}\\
  	={}&\begin{multlined}[t] \Pr\Bigl[ w'\not=\bot\land (x,w')\in\cR \Big| H\gets\Simin(),  (x,\pi)\gets\hat{\P}^{H}(\sigma_\lambda^{\P}),\\
  		w'\gets E(x,\pi,H)\Bigr] \label{eq:world_1_exp1},
  		\end{multlined}\\
  	\MoveEqLeft[3] \pr[H\gets\Simin(), y\gets\OUT_{V_1}\langle\Sim_\cA(\sigma_\lambda^\P)\interacts V_1\rangle]{y=x}\\
    ={}&\begin{multlined}[t]
    	\Pr\Bigl[y=x\land w'\not=\bot\Big|H\gets\Simin(), (x,\pi)\gets\hat{\P}^{H}(\sigma_\lambda^{\P}),w'\gets E(x,\pi,H),\\
    	y\gets V(x,\pi)\Bigr]
        \end{multlined}\\
  	={}&\begin{multlined}[t] \Pr\Bigl[y=x\land w'\not=\bot\land (x,w')\in\cR\Big|H\gets\Simin(),  (x,\pi)\gets\hat{\P}^{H}(\sigma_\lambda^{\P}),\\
  	    w'\gets E(x,\pi,H),y\gets V(x,\pi)\Bigr]\label{eq:world_1_exp2}
  	    \end{multlined}\\
    &\begin{multlined}[t] +\Pr\Bigl[y=x\land w'\not=\bot\land (x,w')\not\in\cR\Big| H\gets\Simin(),  (x,\pi)\gets\hat{\P}^{H}(\sigma_\lambda^{\P}),\\
  	 w'\gets E(x,\pi,H),y\gets V(x,\pi)\Bigr], \label{eq:world_1_exp3}
  	 \end{multlined}
  \end{align}
  where in the last equality we just used the marginal probability expansion. Moreover, by online-extractability~\cref{eq:online_extrac}, we know that for all $\sigma_\lambda^\P$:
  \begin{align}
  	\Pr\Bigl[y=x \land w'\not=\bot\land (x,w')\not\in\cR \Big| H\gets\Simin(), (x,\pi)\gets\hat{\P}^{H}(\sigma_\lambda^\P),&\nonumber\\
  	y\gets\V(x,\pi), w'\gets E(x,\pi,H)\Bigr]& <\negl, \label{eq:online_ext_2}\\
    \Pr\Bigl[y=x \land w'\not=\bot\land (x,w')\in\cR\Big| H\gets\Simin(), (x,\pi)\gets\hat{\P}^{H}(\sigma_\lambda^\P), &\nonumber\\
    y\gets\V(x,\pi), w'\gets E(x,\pi,H)\Bigr]& >1-\negl,
  \end{align}
  thus bounding~\cref{eq:world_1_exp3}. Note that we can bound~\cref{eq:world_1_exp2} by~\cref{eq:world_1_exp1}, as the former adds one more restriction to the output. Therefore, the probabilities of each protocol accepting is nearly the same as
  \begin{align}
  	\MoveEqLeft[3] \left|\pr[H\gets\Simin(),  y\gets\OUT_{\tV}\langle\Sim_\cA(\sigma_\lambda^\P)\interactsF{\cF} \tV\rangle]{y=x}\right.\nonumber\\
  	&\quad -\left.\pr[H\gets\Simin(), y\gets\OUT_{V_1}\langle\Sim_\cA(\sigma_\lambda^\P)\interacts V_1\rangle]{y=x}\right|\\
  	\leq &\begin{multlined}[t] \left|\Pr\Bigl[ w'\not=\bot\land (x,w')\in\cR\Big|  H\gets\Simin(),  (x,\pi)\gets\hat{\P}^{H}(\sigma_\lambda^{\P}),\right.\\
  		w'\gets E(x,\pi,H)\Bigr]
  		\end{multlined}\\
  	&\begin{multlined}[t] -\Pr\Bigl[ y=x\land w'\not=\bot\land (x,w')\in\cR \Big| H\gets\Simin(),  (x,\pi)\gets\hat{\P}^{H}(\sigma_\lambda^{\P}),\\
  		\left.w'\gets E(x,\pi,H),y\gets V(x,\pi)\Bigr]\right| +\negl
     \end{multlined}\\
  	\leq&\negl.
  \end{align}
  This bound in enough to show the similarity of the worlds as
  \begin{align}
  	\begin{multlined}[t] \pr[H\gets\Simin(),  y\gets\OUT_{M}\langle\Sim_\cA(\sigma_\lambda^\P)\interactsF{\cF} M\rangle]{y=\bot}\\
  	=1-\pr[H\gets\Simin(),  y\gets\OUT_{M}\langle\Sim_\cA(\sigma_\lambda^\P)\interactsF{\cF} M\rangle]{y=x},
  	\end{multlined}
  \end{align}
  for both $M=\tV$ and $M=V_1$.
  
  \begin{figure}
  	\centering
  	\begin{minipage}[b]{0.4\textwidth}
  	    \includegraphics[scale=0.75]{figures/world_2.tex}
  		\caption{$\World_2$}
  	\end{minipage}
  	\hfill
  	\begin{minipage}[b]{0.25\textwidth}
  		\includegraphics[scale=0.75]{figures/world_3.tex}
  		\caption{$\World_3$}
  	\end{minipage}
  	\hfill
  	\begin{minipage}[b]{0.25\textwidth}
  		\includegraphics[scale=0.75]{figures/world_4.tex}
  		\caption{$\World_4$}
  	\end{minipage}
  \label{fig:worlds2_4}
  \end{figure}

  We can also prove $\World_1\approxRV\World_2$ using online extractability, as the verifier $V_1$ is only using the extractor to see if it does not abort. More precisely, following the same argument as before, and expanding the probability of accepting for the protocol from $\World_2$:
  \begin{align}
  	\MoveEqLeft[3] \pr[H\gets\Simin(), y\gets\OUT_{\V}\langle\Sim_\cA(\sigma_\lambda^\P)\interacts \V\rangle]{y=x}\\
  	={}&\pr[H\gets\Simin(), (x,\pi)\gets\hat{\P}^{H}(\sigma_\lambda^{\P}),y\gets\V(x,\pi)]{y=x}\\
  	={}&\pr[H\gets\Simin(), (x,\pi)\gets\hat{\P}^{H}(\sigma_\lambda^{\P}),y\gets\V(x,\pi)]{y=x\land (x,w')\in}\nonumber\\
  	&+\pr[H\gets\Simin(), (x,\pi)\gets\hat{\P}^{H}(\sigma_\lambda^{\P}),y\gets\V(x,\pi)]{y=x}.
  \end{align}
  In order to proof the equivalence it is enough to note that
  \begin{align}
  	\MoveEqLeft[3] \left|\pr[H\gets\Simin(), y\gets\OUT_{V_1}\langle\Sim_\cA(\sigma_\lambda^\P)\interacts V_1\rangle]{y=x}\right.\nonumber\\
  	&-\left.\pr[H\gets\Simin(), y\gets\OUT_{\V}\langle\Sim_\cA(\sigma_\lambda^\P)\interacts \V\rangle]{y=x}\right|\\
  	={}&\begin{multlined}[t] \left|\Pr\Bigl[ y=x\land w'\not=\bot \Big| H\gets\Simin(),  (x,\pi)\gets\hat{\P}^{H}(\sigma_\lambda^{\P}),w'\gets E(x,\pi,H),\right.\\
  		y\gets V(x,\pi)\Bigr]
  		\end{multlined}\nonumber\\
  	&-\left.\pr[H\gets\Simin(),  (x,\pi)\gets\hat{\P}^{H}(\sigma_\lambda^{\P}),y\gets V(x,\pi)]{y=x}\right|.\\
  	={}&\begin{multlined}[t]\Pr\Bigl[ y=x\land w'=\bot \Big| H\gets\Simin(),  (x,\pi)\gets\hat{\P}^{H}(\sigma_\lambda^{\P}),w'\gets E(x,\pi,H),\\
  		y\gets V(x,\pi)\Big]
  		\end{multlined}\\
  	={}&\begin{multlined}[t] \Pr\Bigl[ y=x\land w'=\bot\land (x,w')\not\in\cR \Big| H\gets\Simin(),  (x,\pi)\gets\hat{\P}^{H}(\sigma_\lambda^{\P}),\\
  		w'\gets E(x,\pi,H),y\gets V(x,\pi)]\label{eq:world_2_bound}
  		\end{multlined}\\
  	<&\negl,
  \end{align}
  where in~\cref{eq:world_2_bound} we used that $\{w'=\bot\}\subseteq\{(x,w')\not\in\cR\}$.  
  
  The rest of the relations are obvious since $\World_3$ is just $\World_2$ with a different routing of the messages -- the simulator $S$ is only redirecting the information. $\World_3\approxRV\World_4$ is changing the $\Simin$ by the oracle that it is simulating $H$.
  
  \paragraph{Case 3: malicious Bob.} If the adversary corrupts the verifier $\cA=\V{}$, the simulator from \cite{Unr15_NonInteractiveZeroKnowledgeProofs} from the zero-knowledge property will be enough. 
  
  Recall that in his description of the adversary it also encompasses the distinguisher, but by allowing adversaries that receive advice, it is equivalent to assuming an adversary that outputs a proof for the verifier. This is, Unruh's simulator $\Sim=(\Simin,\Sim_P')$ fulfills
  \begin{align}
  	\begin{multlined}[t] \left|\pr[H\gets\ROd,(x,\pi)\gets\P^H(\sigma_\lambda^\P),y\gets\hat{\V}(x,\pi,\sigma_\lambda^\V)]{\Zenv(y,\sigma_\lambda^\Zenv)=1}\right.\\
  		\left.-\pr[H\gets\Simin(),(x,\pi)\gets\Sim_P'(\sigma_\lambda^\P),y\gets\hat{\V}(x,\pi,\sigma_\lambda^\V)]{\Zenv(y,\sigma_\lambda^\Zenv)=1}\right|\\
  		<\negl.
  	\end{multlined}
  \end{align}
  Note that the simulator $\Sim_P'$ is replacing both the ideal functionality $\Fzk*{\cR}$ and the dummy verifier $\tP$ in the ideal world, i.e.\ aborts whenever $(x,w)\not\in\cR$ and runs $\Sim_P(x,\sigma_\lambda^\V)$ otherwise. However, we can easily modify this simulator to obtain the desired one in terms of the subsimulator $\Sim_P$.
  \begin{framed}
  	\noindent \textbf{Simulator} $\Sim_\mathcal{\cA}\eqdef(S,\Simin)$: Verifier is corrupted.
  	\begin{enumerate}
  		\item If $S$ does not receive an abort $\bot$ message from the ideal functionality $\Fzk*{\cR}$, redirects the input $x$ to the simulator $\Sim_P$. Else, it aborts.
  		\item $S$ receives $(x,\pi)$ from the simulator $\Sim_P(x)$.
  		\item $S$ sends $(x,\pi,\sigma_\lambda)$ to the verifier $\V$.
  		\item $S$ redirects the output of the verifier $\V$ to the distinguisher $\Zenv$.
  	\end{enumerate}
  \end{framed}
  It is clear that $\Zenv$ cannot distinguish between the proofs provided by the verifier and the simulator, as our simulator is just a rewiering of Unruh's simulator, see~\cref{fig:comparison_simulators}.
  \begin{figure}
	\centering
	 \begin{minipage}[b]{0.45\textwidth}
	 	\includegraphics[scale=0.7]{figures/unruh_bob.tex}
	 	\caption{Protocol for $\Sim_P'$.}
	 \end{minipage}
	 \hfill
	 \begin{minipage}[b]{0.45\textwidth}
	 	\includegraphics[scale=0.7]{figures/ideal_bob.tex}
	 	\caption{Ideal functionality with $\Sim_\mathcal{\cA}$.}
	 \end{minipage}
	 \caption{Construction of the simulator for adversarial verifier.}\label{fig:comparison_simulators}
  \end{figure}
\end{proofE}

\begin{corollaryE}
  In the random oracle model, assuming the existence of a collision-resistant and second-bit hardcore hash function (which holds if $h$ is modeled as a random oracle model, see discussion in \cref{thm:informalOT}), there exists a protocol realizing the string OT functionality $\FotStr$ and the $k$-out-of-$n$ OT functionality $\FotKoutN$.
\end{corollaryE}
\begin{proof}
  This is a direct consequence of \cref{cor:stringOT} and \cref{thm:Unr15Composable}, where \cite{Unr15_NonInteractiveZeroKnowledgeProofs} is used to instantiate the ZK protocol.
\end{proof}

\section{Acknowledgment}

The authors deeply thank Christian Schaffner for many insightful exchanges, together with Stacey Jeffery, Alex Grilo, Geoffroy Couteau and James Bartusek for precious discussions, and anonymous reviewers for many helpful comments and for pointing a mistake (now corrected) in a proof that generalizes our first result. This work is co-funded by the European Union (ERC, ASC-Q, 101040624) and supported by the Dutch National Growth Fund (NGF), as part of the Quantum Delta NL programme. \arxivOnly{Views and opinions expressed are however those of the author(s) only and do not necessarily reflect those of the European Union or the European Research Council. Neither the European Union nor the granting authority can be held responsible for them.}

\FloatBarrier

\newpage

\publishedVsArxiv{\printbibliography[segment=0,heading=bibintoc]}{\printbibliography[heading=bibintoc]}

\cameraVsOther{}{
  \newpage
  \appendix
  \ifdefined\separateAppendixReferences\newrefsegment\fi 
  \begin{center}
    \Huge{\textsc{Supplementary Material}}
  \end{center}
  \ifdefined\separateAppendixReferences\publishedOnly{\printbibliography[title=Appendix References,filter=appendixOnlyFilter]}\fi

@inproceedings{HSS11_ClassicalCryptographicProtocols,
  title = {Classical {{Cryptographic Protocols}} in a {{Quantum World}}},
  booktitle = {Advances in {{Cryptology}} – {{CRYPTO}} 2011},
  author = {Hallgren, Sean and Smith, Adam and Song, Fang},
  editor = {Rogaway, Phillip},
  date = {2011},
  series = {Lecture {{Notes}} in {{Computer Science}}},
  pages = {411--428},
  publisher = {{Springer}},
  location = {{Berlin, Heidelberg}},
  doi = {10.1007/978-3-642-22792-9_23},
  isbn = {978-3-642-22792-9},
  langid = {english}
}

@inproceedings{BS20_PostquantumZeroKnowledge,
  title = {Post-Quantum Zero Knowledge in Constant Rounds},
  booktitle = {Proceedings of the 52nd {{Annual ACM SIGACT Symposium}} on {{Theory}} of {{Computing}}},
  author = {Bitansky, Nir and Shmueli, Omri},
  date = {2020-06-22},
  series = {{{STOC}} 2020},
  pages = {269--279},
  publisher = {{Association for Computing Machinery}},
  location = {{New York, NY, USA}},
  doi = {10.1145/3357713.3384324},
  url = {https://doi.org/10.1145/3357713.3384324},
  urldate = {2021-12-08},
  isbn = {978-1-4503-6979-4}
}

@inproceedings{Unr12_QuantumProofsKnowledge,
  title = {Quantum {{Proofs}} of {{Knowledge}}},
  booktitle = {Advances in {{Cryptology}} – {{EUROCRYPT}} 2012},
  author = {Unruh, Dominique},
  editor = {Pointcheval, David and Johansson, Thomas},
  date = {2012},
  series = {Lecture {{Notes}} in {{Computer Science}}},
  pages = {135--152},
  publisher = {{Springer}},
  location = {{Berlin, Heidelberg}},
  doi = {10.1007/978-3-642-29011-4_10},
  isbn = {978-3-642-29011-4},
  langid = {english}
}

@article{Wat09_ZeroKnowledgeQuantumAttacks,
  title = {Zero-{{Knowledge}} against {{Quantum Attacks}}},
  author = {Watrous, John},
  date = {2009-01-01},
  journaltitle = {SIAM Journal on Computing},
  shortjournal = {SIAM J. Comput.},
  volume = {39},
  number = {1},
  pages = {25--58},
  publisher = {{Society for Industrial and Applied Mathematics}},
  issn = {0097-5397},
  doi = {10.1137/060670997},
  url = {https://epubs.siam.org/doi/abs/10.1137/060670997},
  urldate = {2021-12-08}
}

@inproceedings{Unr15_NonInteractiveZeroKnowledgeProofs,
  title = {Non-{{Interactive Zero-Knowledge Proofs}} in the {{Quantum Random Oracle Model}}},
  booktitle = {Advances in {{Cryptology}} - {{EUROCRYPT}} 2015},
  author = {Unruh, Dominique},
  editor = {Oswald, Elisabeth and Fischlin, Marc},
  date = {2015},
  series = {Lecture {{Notes}} in {{Computer Science}}},
  pages = {755--784},
  publisher = {{Springer}},
  location = {{Berlin, Heidelberg}},
  doi = {10.1007/978-3-662-46803-6_25},
  isbn = {978-3-662-46803-6},
  langid = {english}
}

@inproceedings{GL89_HardcorePredicateAll,
  title = {A Hard-Core Predicate for All One-Way Functions},
  booktitle = {Proceedings of the Twenty-First Annual {{ACM}} Symposium on {{Theory}} of Computing},
  author = {Goldreich, O. and Levin, L. A.},
  date = {1989-02-01},
  series = {{{STOC}} '89},
  pages = {25--32},
  publisher = {{Association for Computing Machinery}},
  location = {{New York, NY, USA}},
  doi = {10.1145/73007.73010},
  url = {https://doi.org/10.1145/73007.73010},
  urldate = {2022-12-07},
  isbn = {978-0-89791-307-2}
}

@misc{NC10_QuantumComputationQuantum,
  title = {Quantum {{Computation}} and {{Quantum Information}}: 10th {{Anniversary Edition}}},
  shorttitle = {Quantum {{Computation}} and {{Quantum Information}}},
  author = {Nielsen, Michael A. and Chuang, Isaac L.},
  year = {2010},
  month = dec,
  journal = {Higher Education from Cambridge University Press},
  publisher = {{Cambridge University Press}},
  doi = {10.1017/CBO9780511976667},
  url = {https://www.cambridge.org/highereducation/books/quantum-computation-and-quantum-information/01E10196D0A682A6AEFFEA52D53BE9AE},
  urldate = {2021-09-08},
  isbn = {9780511976667},
  langid = {english}
}

@unpublished{Wil17_ClassicalQuantumShannon,
  title = {From {{Classical}} to {{Quantum Shannon Theory}}},
  author = {Wilde, Mark M.},
  date = {2017},
  eprint = {1106.1445},
  eprinttype = {arxiv},
  primaryclass = {quant-ph},
  doi = {10.1017/9781316809976.001},
  url = {http://arxiv.org/abs/1106.1445},
  urldate = {2022-01-03},
  archiveprefix = {arXiv}
}

@unpublished{CGK21_NonDestructiveZeroKnowledgeProofs,
  title = {Non-{{Destructive Zero-Knowledge Proofs}} on {{Quantum States}}, and {{Multi-Party Generation}} of {{Authorized Hidden GHZ States}}},
  author = {Colisson, Léo and Grosshans, Frédéric and Kashefi, Elham},
  date = {2021-04-10},
  eprint = {2104.04742},
  eprinttype = {arxiv},
  primaryclass = {quant-ph},
  url = {http://arxiv.org/abs/2104.04742},
  urldate = {2021-04-13},
  archiveprefix = {arXiv}
}

@inproceedings{DFPR14_ComposableSecurityDelegated,
  title = {Composable {{Security}} of {{Delegated Quantum Computation}}},
  booktitle = {Advances in {{Cryptology}} – {{ASIACRYPT}} 2014},
  author = {Dunjko, Vedran and Fitzsimons, Joseph F. and Portmann, Christopher and Renner, Renato},
  editor = {Sarkar, Palash and Iwata, Tetsu},
  date = {2014},
  series = {Lecture {{Notes}} in {{Computer Science}}},
  pages = {406--425},
  publisher = {{Springer}},
  location = {{Berlin, Heidelberg}},
  doi = {10.1007/978-3-662-45608-8_22},
  isbn = {978-3-662-45608-8},
  langid = {english}
}

@unpublished{van20_ZXcalculusWorkingQuantum,
  title = {{{ZX-calculus}} for the Working Quantum Computer Scientist},
  author = {van de Wetering, John},
  options = {useprefix=true},
  date = {2020-12-27},
  eprint = {2012.13966},
  eprinttype = {arxiv},
  primaryclass = {quant-ph},
  url = {http://arxiv.org/abs/2012.13966},
  urldate = {2021-05-29},
  archiveprefix = {arXiv}
}

@article{CJPV21_CompletenessGraphicalLanguages,
  title = {Completeness of {{Graphical Languages}} for {{Mixed}}\&\#xa0;{{State Quantum Mechanics}}},
  author = {Carette, Titouan and Jeandel, Emmanuel and Perdrix, Simon and Vilmart, Renaud},
  date = {2021-12-21},
  journaltitle = {ACM Transactions on Quantum Computing},
  shortjournal = {ACM Transactions on Quantum Computing},
  volume = {2},
  number = {4},
  pages = {17:1--17:28},
  issn = {2643-6809},
  doi = {10.1145/3464693},
  url = {https://doi.org/10.1145/3464693},
  urldate = {2023-02-07}
}

@inproceedings{GMW87_HowPlayANY,
  title = {How to Play {{ANY}} Mental Game},
  booktitle = {Proceedings of the Nineteenth Annual {{ACM}} Symposium on {{Theory}} of Computing},
  author = {Goldreich, O. and Micali, S. and Wigderson, A.},
  date = {1987-01-01},
  series = {{{STOC}} '87},
  pages = {218--229},
  publisher = {{Association for Computing Machinery}},
  location = {{New York, NY, USA}},
  doi = {10.1145/28395.28420},
  url = {https://doi.org/10.1145/28395.28420},
  urldate = {2021-04-09},
  isbn = {978-0-89791-221-1}
}

@inproceedings{Yao82_ProtocolsSecureComputations,
  title = {Protocols for Secure Computations},
  booktitle = {23rd {{Annual Symposium}} on {{Foundations}} of {{Computer Science}} (Sfcs 1982)},
  author = {Yao, Andrew C.},
  date = {1982-11},
  pages = {160--164},
  issn = {0272-5428},
  doi = {10.1109/SFCS.1982.38},
  eventtitle = {23rd {{Annual Symposium}} on {{Foundations}} of {{Computer Science}} (Sfcs 1982)}
}

@inproceedings{Kil88_FoundingCrytpographyOblivious,
  title = {Founding Crytpography on Oblivious Transfer},
  booktitle = {Proceedings of the Twentieth Annual {{ACM}} Symposium on {{Theory}} of Computing},
  author = {Kilian, Joe},
  date = {1988-01-01},
  series = {{{STOC}} '88},
  pages = {20--31},
  publisher = {{Association for Computing Machinery}},
  location = {{New York, NY, USA}},
  doi = {10.1145/62212.62215},
  url = {https://doi.org/10.1145/62212.62215},
  urldate = {2023-02-09},
  isbn = {978-0-89791-264-8}
}

@inproceedings{PVW08_FrameworkEfficientComposable,
  title = {A {{Framework}} for {{Efficient}} and {{Composable Oblivious Transfer}}},
  booktitle = {Advances in {{Cryptology}} – {{CRYPTO}} 2008},
  author = {Peikert, Chris and Vaikuntanathan, Vinod and Waters, Brent},
  editor = {Wagner, David},
  date = {2008},
  series = {Lecture {{Notes}} in {{Computer Science}}},
  pages = {554--571},
  publisher = {{Springer}},
  location = {{Berlin, Heidelberg}},
  doi = {10.1007/978-3-540-85174-5_31},
  isbn = {978-3-540-85174-5},
  langid = {english}
}

@misc{Rab05_HowExchangeSecrets,
  title = {How {{To Exchange Secrets}} with {{Oblivious Transfer}}},
  author = {Rabin, Michael O.},
  date = {2005},
  number = {187},
  url = {https://eprint.iacr.org/2005/187},
  urldate = {2023-02-09}
}

@article{EGL85_RandomizedProtocolSigning,
  title = {A Randomized Protocol for Signing Contracts},
  author = {Even, Shimon and Goldreich, Oded and Lempel, Abraham},
  date = {1985-06-01},
  journaltitle = {Communications of the ACM},
  shortjournal = {Commun. ACM},
  volume = {28},
  number = {6},
  pages = {637--647},
  issn = {0001-0782},
  doi = {10.1145/3812.3818},
  url = {https://doi.org/10.1145/3812.3818},
  urldate = {2022-07-08}
}

@article{Wie83_ConjugateCoding,
  title = {Conjugate Coding},
  author = {Wiesner, Stephen},
  date = {1983-01-01},
  journaltitle = {ACM SIGACT News},
  shortjournal = {SIGACT News},
  volume = {15},
  number = {1},
  pages = {78--88},
  issn = {0163-5700},
  doi = {10.1145/1008908.1008920},
  url = {https://doi.org/10.1145/1008908.1008920},
  urldate = {2022-01-02}
}

@inproceedings{CGS02_SecureMultipartyQuantum,
  title = {Secure Multi-Party Quantum Computation},
  booktitle = {Proceedings of the Thiry-Fourth Annual {{ACM}} Symposium on {{Theory}} of Computing},
  author = {Crépeau, Claude and Gottesman, Daniel and Smith, Adam},
  date = {2002-05-19},
  series = {{{STOC}} '02},
  pages = {643--652},
  publisher = {{Association for Computing Machinery}},
  location = {{New York, NY, USA}},
  doi = {10.1145/509907.510000},
  url = {https://doi.org/10.1145/509907.510000},
  urldate = {2021-10-11},
  isbn = {978-1-58113-495-7}
}

@inproceedings{DGJ+20_SecureMultipartyQuantum,
  title = {Secure {{Multi-party Quantum Computation}} with a {{Dishonest Majority}}},
  booktitle = {Advances in {{Cryptology}} – {{EUROCRYPT}} 2020},
  author = {Dulek, Yfke and Grilo, Alex B. and Jeffery, Stacey and Majenz, Christian and Schaffner, Christian},
  editor = {Canteaut, Anne and Ishai, Yuval},
  date = {2020},
  series = {Lecture {{Notes}} in {{Computer Science}}},
  pages = {729--758},
  publisher = {{Springer International Publishing}},
  location = {{Cham}},
  doi = {10.1007/978-3-030-45727-3_25},
  isbn = {978-3-030-45727-3},
  langid = {english}
}

@article{KP17_MultipartyDelegatedQuantum,
  title = {Multiparty {{Delegated Quantum Computing}}},
  author = {Kashefi, Elham and Pappa, Anna},
  date = {2017-09},
  journaltitle = {Cryptography},
  volume = {1},
  number = {2},
  pages = {12},
  publisher = {{Multidisciplinary Digital Publishing Institute}},
  doi = {10.3390/cryptography1020012},
  url = {https://www.mdpi.com/2410-387X/1/2/12},
  urldate = {2021-12-14},
  issue = {2},
  langid = {english}
}

@inproceedings{LT22_ReviewStateArt,
  title = {Review of the State of the Art in Secure Multiparty Computation},
  author = {Laud, Peeter and Talviste, Riivo},
  date = {2022},
  eventtitle = {Cybernetica As},
  langid = {english}
}

@article{YAVV22_SurveyObliviousTransfer,
  title = {A {{Survey}} of {{Oblivious Transfer Protocol}}},
  author = {Yadav, Vijay Kumar and Andola, Nitish and Verma, Shekhar and Venkatesan, S.},
  date = {2022-09-13},
  journaltitle = {ACM Computing Surveys},
  shortjournal = {ACM Comput. Surv.},
  volume = {54},
  pages = {211:1--211:37},
  issn = {0360-0300},
  doi = {10.1145/3503045},
  url = {https://doi.org/10.1145/3503045},
  urldate = {2023-01-27},
  issue = {10s}
}

@inproceedings{Imp95_PersonalViewAveragecase,
  title = {A Personal View of Average-Case Complexity},
  booktitle = {Proceedings of {{Structure}} in {{Complexity Theory}}. {{Tenth Annual IEEE Conference}}},
  author = {Impagliazzo, R.},
  date = {1995-06},
  pages = {134--147},
  issn = {1063-6870},
  doi = {10.1109/SCT.1995.514853},
  eventtitle = {Proceedings of {{Structure}} in {{Complexity Theory}}. {{Tenth Annual IEEE Conference}}}
}

@incollection{BD18_TwoMessageStatisticallySenderPrivate,
  title = {Two-{{Message Statistically Sender-Private OT}} from {{LWE}}},
  booktitle = {Theory of {{Cryptography}}},
  author = {Brakerski, Zvika and Döttling, Nico},
  editor = {Beimel, Amos and Dziembowski, Stefan},
  date = {2018},
  series = {Lecture {{Notes}} in {{Computer Science}}},
  volume = {11240},
  pages = {370--390},
  publisher = {{Springer International Publishing}},
  location = {{Cham}},
  doi = {10.1007/978-3-030-03810-6_14},
  url = {https://link.springer.com/10.1007/978-3-030-03810-6_14},
  urldate = {2023-02-09},
  isbn = {978-3-030-03809-0 978-3-030-03810-6},
  langid = {english}
}

@inproceedings{Qua20_UCSecureOTLWE,
  title = {{{UC-Secure OT}} from {{LWE}}, {{Revisited}}},
  booktitle = {Security and {{Cryptography}} for {{Networks}}},
  author = {Quach, Willy},
  editor = {Galdi, Clemente and Kolesnikov, Vladimir},
  date = {2020},
  series = {Lecture {{Notes}} in {{Computer Science}}},
  pages = {192--211},
  publisher = {{Springer International Publishing}},
  location = {{Cham}},
  doi = {10.1007/978-3-030-57990-6_10},
  isbn = {978-3-030-57990-6},
  langid = {english}
}

@inproceedings{GLSV21_ObliviousTransferMiniQCrypt,
  title = {Oblivious {{Transfer Is}} in {{MiniQCrypt}}},
  booktitle = {Advances in {{Cryptology}} – {{EUROCRYPT}} 2021},
  author = {Grilo, Alex B. and Lin, Huijia and Song, Fang and Vaikuntanathan, Vinod},
  editor = {Canteaut, Anne and Standaert, François-Xavier},
  date = {2021},
  series = {Lecture {{Notes}} in {{Computer Science}}},
  pages = {531--561},
  publisher = {{Springer International Publishing}},
  location = {{Cham}},
  doi = {10.1007/978-3-030-77886-6_18},
  isbn = {978-3-030-77886-6},
  langid = {english}
}

@inproceedings{BCKM21_OneWayFunctionsImply,
  title = {One-{{Way Functions Imply Secure Computation}} in a {{Quantum World}}},
  booktitle = {Advances in {{Cryptology}} – {{CRYPTO}} 2021},
  author = {Bartusek, James and Coladangelo, Andrea and Khurana, Dakshita and Ma, Fermi},
  editor = {Malkin, Tal and Peikert, Chris},
  date = {2021},
  series = {Lecture {{Notes}} in {{Computer Science}}},
  pages = {467--496},
  publisher = {{Springer International Publishing}},
  location = {{Cham}},
  doi = {10.1007/978-3-030-84242-0_17},
  isbn = {978-3-030-84242-0},
  langid = {english}
}

@inproceedings{JLS18_PseudorandomQuantumStates,
  title = {Pseudorandom {{Quantum States}}},
  booktitle = {Advances in {{Cryptology}} – {{CRYPTO}} 2018},
  author = {Ji, Zhengfeng and Liu, Yi-Kai and Song, Fang},
  editor = {Shacham, Hovav and Boldyreva, Alexandra},
  date = {2018},
  series = {Lecture {{Notes}} in {{Computer Science}}},
  pages = {126--152},
  publisher = {{Springer International Publishing}},
  location = {{Cham}},
  doi = {10.1007/978-3-319-96878-0_5},
  isbn = {978-3-319-96878-0},
  langid = {english}
}

@inproceedings{CK88_AchievingObliviousTransfer,
  title = {Achieving Oblivious Transfer Using Weakened Security Assumptions},
  booktitle = {[{{Proceedings}} 1988] 29th {{Annual Symposium}} on {{Foundations}} of {{Computer Science}}},
  author = {Crepeau, C. and Kilian, J.},
  date = {1988-10},
  pages = {42--52},
  doi = {10.1109/SFCS.1988.21920},
  eventtitle = {[{{Proceedings}} 1988] 29th {{Annual Symposium}} on {{Foundations}} of {{Computer Science}}}
}

@misc{ABKK22_NewFrameworkQuantum,
  title = {A {{New Framework}} for {{Quantum Oblivious Transfer}}},
  author = {Agarwal, Amit and Bartusek, James and Khurana, Dakshita and Kumar, Nishant},
  date = {2022},
  number = {1191},
  url = {https://eprint.iacr.org/2022/1191},
  urldate = {2023-02-10}
}

@online{ELE_ZcashPrivacyprotectingDigital,
  title = {Zcash: {{Privacy-protecting}} Digital Currency},
  author = {{ELECTRIC COIN COMPANY}},
  url = {https://z.cash/},
  urldate = {2023-02-10},
  langid = {american},
  organization = {{Zcash}}
}

@misc{LMS21_PostQuantumZeroKnowledge,
  title = {Post-{{Quantum Zero Knowledge}}, {{Revisited}} (or: {{How}} to {{Do Quantum Rewinding Undetectably}})},
  shorttitle = {Post-{{Quantum Zero Knowledge}}, {{Revisited}} (Or},
  author = {Lombardi, Alex and Ma, Fermi and Spooner, Nicholas},
  date = {2021-11-23},
  number = {arXiv:2111.12257},
  eprint = {2111.12257},
  eprinttype = {arxiv},
  primaryclass = {quant-ph},
  publisher = {{arXiv}},
  doi = {10.48550/arXiv.2111.12257},
  url = {http://arxiv.org/abs/2111.12257},
  urldate = {2022-07-05},
  archiveprefix = {arXiv}
}

@inproceedings{AL20_SecureQuantumExtraction,
  title = {Secure {{Quantum Extraction Protocols}}},
  booktitle = {Theory of {{Cryptography}}},
  author = {Ananth, Prabhanjan and La Placa, Rolando L.},
  editor = {Pass, Rafael and Pietrzak, Krzysztof},
  date = {2020},
  series = {Lecture {{Notes}} in {{Computer Science}}},
  pages = {123--152},
  publisher = {{Springer International Publishing}},
  location = {{Cham}},
  doi = {10.1007/978-3-030-64381-2_5},
  isbn = {978-3-030-64381-2},
  langid = {english}
}

@inproceedings{GMR85_KnowledgeComplexityInteractive,
  title = {The Knowledge Complexity of Interactive Proof-Systems},
  booktitle = {Proceedings of the Seventeenth Annual {{ACM}} Symposium on {{Theory}} of Computing},
  author = {Goldwasser, S and Micali, S and Rackoff, C},
  date = {1985-12-01},
  series = {{{STOC}} '85},
  pages = {291--304},
  publisher = {{Association for Computing Machinery}},
  location = {{New York, NY, USA}},
  doi = {10.1145/22145.22178},
  url = {https://doi.org/10.1145/22145.22178},
  urldate = {2021-11-11},
  isbn = {978-0-89791-151-1}
}

@article{Lin13_NoteConstantRoundZeroKnowledge,
  title = {A {{Note}} on {{Constant-Round Zero-Knowledge Proofs}} of {{Knowledge}}},
  author = {Lindell, Yehuda},
  date = {2013-10-01},
  journaltitle = {Journal of Cryptology},
  shortjournal = {J Cryptol},
  volume = {26},
  number = {4},
  pages = {638--654},
  issn = {1432-1378},
  doi = {10.1007/s00145-012-9132-7},
  url = {https://doi.org/10.1007/s00145-012-9132-7},
  urldate = {2023-02-10},
  langid = {english}
}

@inproceedings{RY22_InteractiveProofsSynthesizing,
  title = {Interactive Proofs for Synthesizing Quantum States and Unitaries},
  booktitle = {13th Innovations in Theoretical Computer Science Conference, {{ITCS}} 2022, January 31 - February 3, 2022, Berkeley, {{CA}}, {{USA}}},
  author = {Rosenthal, Gregory and Yuen, Henry},
  editor = {Braverman, Mark},
  date = {2022},
  series = {{{LIPIcs}}},
  volume = {215},
  pages = {112:1--112:4},
  publisher = {{Schloss Dagstuhl - Leibniz-Zentrum für Informatik}},
  doi = {10.4230/LIPIcs.ITCS.2022.112},
  url = {https://doi.org/10.4230/LIPIcs.ITCS.2022.112},
  bibsource = {dblp computer science bibliography, https://dblp.org}
}

@inproceedings{KZ09_ZeroKnowledgeProofsWitness,
  title = {Zero-{{Knowledge Proofs}} with {{Witness Elimination}}},
  booktitle = {Public {{Key Cryptography}} – {{PKC}} 2009},
  author = {Kiayias, Aggelos and Zhou, Hong-Sheng},
  editor = {Jarecki, Stanisław and Tsudik, Gene},
  date = {2009},
  series = {Lecture {{Notes}} in {{Computer Science}}},
  pages = {124--138},
  publisher = {{Springer}},
  location = {{Berlin, Heidelberg}},
  doi = {10.1007/978-3-642-00468-1_8},
  isbn = {978-3-642-00468-1},
  langid = {english}
}

@inproceedings{DNS12_ActivelySecureTwoParty,
  title = {Actively {{Secure Two-Party Evaluation}} of {{Any Quantum Operation}}},
  booktitle = {Advances in {{Cryptology}} – {{CRYPTO}} 2012},
  author = {Dupuis, Frédéric and Nielsen, Jesper Buus and Salvail, Louis},
  editor = {Safavi-Naini, Reihaneh and Canetti, Ran},
  date = {2012},
  series = {Lecture {{Notes}} in {{Computer Science}}},
  pages = {794--811},
  publisher = {{Springer}},
  location = {{Berlin, Heidelberg}},
  doi = {10.1007/978-3-642-32009-5_46},
  isbn = {978-3-642-32009-5},
  langid = {english}
}

@misc{KKL+23_AsymmetricQuantumSecure,
  title = {Asymmetric {{Quantum Secure Multi-Party Computation With Weak Clients Against Dishonest Majority}}},
  author = {Kapourniotis, Theodoros and Kashefi, Elham and Leichtle, Dominik and Music, Luka and Ollivier, Harold},
  date = {2023-03-15},
  number = {arXiv:2303.08865},
  eprint = {2303.08865},
  eprinttype = {arxiv},
  primaryclass = {quant-ph},
  publisher = {{arXiv}},
  doi = {10.48550/arXiv.2303.08865},
  url = {http://arxiv.org/abs/2303.08865},
  urldate = {2023-09-14},
  archiveprefix = {arXiv}
}

@misc{BKS23_SecureComputationShared,
  title = {Secure {{Computation}} with {{Shared EPR Pairs}} ({{Or}}: {{How}} to {{Teleport}} in {{Zero-Knowledge}})},
  shorttitle = {Secure {{Computation}} with {{Shared EPR Pairs}} ({{Or}}},
  author = {Bartusek, James and Khurana, Dakshita and Srinivasan, Akshayaram},
  date = {2023},
  number = {2023/564},
  url = {https://eprint.iacr.org/2023/564},
  urldate = {2023-09-14}
}

@misc{KA04_ComplexityQuantumLanguages,
  title = {On the {{Complexity}} of {{Quantum Languages}}},
  author = {Kashefi, Elham and Alves, Carolina Moura},
  date = {2004-04-12},
  number = {arXiv:quant-ph/0404062},
  eprint = {quant-ph/0404062},
  eprinttype = {arxiv},
  publisher = {{arXiv}},
  url = {http://arxiv.org/abs/quant-ph/0404062},
  urldate = {2022-12-15},
  archiveprefix = {arXiv},
  langid = {english}
}

@book{CK17_PicturingQuantumProcesses,
  title = {Picturing {{Quantum Processes}}: {{A First Course}} in {{Quantum Theory}} and {{Diagrammatic Reasoning}}},
  shorttitle = {Picturing {{Quantum Processes}}},
  author = {Coecke, Bob and Kissinger, Aleks},
  date = {2017},
  publisher = {{Cambridge University Press}},
  location = {{Cambridge}},
  doi = {10.1017/9781316219317},
  url = {https://www.cambridge.org/core/books/picturing-quantum-processes/1119568B3101F3A685BE832FEEC53E52},
  urldate = {2021-11-17},
  isbn = {978-1-107-10422-8}
}

@inproceedings{BBCS92_PracticalQuantumOblivious,
  title = {Practical {{Quantum Oblivious Transfer}}},
  booktitle = {Advances in {{Cryptology}} — {{CRYPTO}} ’91},
  author = {Bennett, Charles H. and Brassard, Gilles and Crépeau, Claude and Skubiszewska, Marie-Hélène},
  editor = {Feigenbaum, Joan},
  date = {1992},
  series = {Lecture {{Notes}} in {{Computer Science}}},
  pages = {351--366},
  publisher = {{Springer}},
  location = {{Berlin, Heidelberg}},
  doi = {10.1007/3-540-46766-1_29},
  isbn = {978-3-540-46766-3},
  langid = {english}
}

@inproceedings{DFL+09_ImprovingSecurityQuantum,
  title = {Improving the {{Security}} of {{Quantum Protocols}} via {{Commit-and-Open}}},
  booktitle = {Advances in {{Cryptology}} - {{CRYPTO}} 2009},
  author = {Damgård, Ivan and Fehr, Serge and Lunemann, Carolin and Salvail, Louis and Schaffner, Christian},
  editor = {Halevi, Shai},
  date = {2009},
  series = {Lecture {{Notes}} in {{Computer Science}}},
  pages = {408--427},
  publisher = {{Springer}},
  location = {{Berlin, Heidelberg}},
  doi = {10.1007/978-3-642-03356-8_24},
  isbn = {978-3-642-03356-8},
  langid = {english}
}

@inproceedings{Unr10_UniversallyComposableQuantum,
  title = {Universally {{Composable Quantum Multi-party Computation}}},
  booktitle = {Advances in {{Cryptology}} – {{EUROCRYPT}} 2010},
  author = {Unruh, Dominique},
  editor = {Gilbert, Henri},
  date = {2010},
  series = {Lecture {{Notes}} in {{Computer Science}}},
  pages = {486--505},
  publisher = {{Springer}},
  location = {{Berlin, Heidelberg}},
  doi = {10.1007/978-3-642-13190-5_25},
  isbn = {978-3-642-13190-5},
  langid = {english}
}

@article{SMP22_QuantumObliviousTransfer,
  title = {Quantum Oblivious Transfer: A Short Review},
  shorttitle = {Quantum Oblivious Transfer},
  author = {Santos, Manuel B. and Mateus, Paulo and Pinto, Armando N.},
  date = {2022-07-07},
  journaltitle = {Entropy},
  shortjournal = {Entropy},
  volume = {24},
  number = {7},
  eprint = {2206.03313},
  eprinttype = {arxiv},
  primaryclass = {quant-ph},
  pages = {945},
  issn = {1099-4300},
  doi = {10.3390/e24070945},
  url = {http://arxiv.org/abs/2206.03313},
  urldate = {2022-12-28},
  archiveprefix = {arXiv},
  langid = {english}
}

@inproceedings{PS19_NoninteractiveZeroKnowledge,
  title = {Noninteractive {{Zero Knowledge}} for {{NP}} from ({{Plain}}) {{Learning}} with {{Errors}}},
  booktitle = {Advances in {{Cryptology}} – {{CRYPTO}} 2019},
  author = {Peikert, Chris and Shiehian, Sina},
  editor = {Boldyreva, Alexandra and Micciancio, Daniele},
  date = {2019},
  series = {Lecture {{Notes}} in {{Computer Science}}},
  pages = {89--114},
  publisher = {{Springer International Publishing}},
  location = {{Cham}},
  doi = {10.1007/978-3-030-26948-7_4},
  isbn = {978-3-030-26948-7},
  langid = {english}
}

@inproceedings{MS94_QuantumObliviousTransfer,
  title = {Quantum Oblivious Transfer Is Secure against All Individual Measurements},
  booktitle = {Proceedings {{Workshop}} on {{Physics}} and {{Computation}}. {{PhysComp}} '94},
  author = {Mayers, D. and Salvail, L.},
  date = {1994-11},
  pages = {69--77},
  doi = {10.1109/PHYCMP.1994.363696},
  eventtitle = {Proceedings {{Workshop}} on {{Physics}} and {{Computation}}. {{PhysComp}} '94}
}

@inproceedings{Yao95_SecurityQuantumProtocols,
  title = {Security of Quantum Protocols against Coherent Measurements},
  booktitle = {Proceedings of the Twenty-Seventh Annual {{ACM}} Symposium on {{Theory}} of Computing},
  author = {Yao, Andrew Chi-Chih},
  date = {1995-05-29},
  series = {{{STOC}} '95},
  pages = {67--75},
  publisher = {{Association for Computing Machinery}},
  location = {{New York, NY, USA}},
  doi = {10.1145/225058.225085},
  url = {https://doi.org/10.1145/225058.225085},
  urldate = {2023-02-16},
  isbn = {978-0-89791-718-6}
}

@inproceedings{BF10_SamplingQuantumPopulation,
  title = {Sampling in a {{Quantum Population}}, and {{Applications}}},
  booktitle = {Advances in {{Cryptology}} – {{CRYPTO}} 2010},
  author = {Bouman, Niek J. and Fehr, Serge},
  editor = {Rabin, Tal},
  date = {2010},
  series = {Lecture {{Notes}} in {{Computer Science}}},
  pages = {724--741},
  publisher = {{Springer}},
  location = {{Berlin, Heidelberg}},
  doi = {10.1007/978-3-642-14623-7_39},
  isbn = {978-3-642-14623-7},
  langid = {english}
}

@inproceedings{WW06_ObliviousTransferSymmetric,
  title = {Oblivious {{Transfer Is Symmetric}}},
  booktitle = {Advances in {{Cryptology}} - {{EUROCRYPT}} 2006},
  author = {Wolf, Stefan and Wullschleger, Jürg},
  editor = {Vaudenay, Serge},
  date = {2006},
  series = {Lecture {{Notes}} in {{Computer Science}}},
  pages = {222--232},
  publisher = {{Springer}},
  location = {{Berlin, Heidelberg}},
  doi = {10.1007/11761679_14},
  isbn = {978-3-540-34547-3},
  langid = {english}
}

@article{Lo97_InsecurityQuantumSecure,
  title = {Insecurity of Quantum Secure Computations},
  author = {Lo, Hoi-Kwong},
  date = {1997-08-01},
  journaltitle = {Physical Review A},
  shortjournal = {Phys. Rev. A},
  volume = {56},
  number = {2},
  pages = {1154--1162},
  publisher = {{American Physical Society}},
  doi = {10.1103/PhysRevA.56.1154},
  url = {https://link.aps.org/doi/10.1103/PhysRevA.56.1154},
  urldate = {2023-03-02}
}

@misc{DGLM23_QuantumMerlinArthurProof,
  title = {Quantum {{Merlin-Arthur}} Proof Systems for Synthesizing Quantum States},
  author = {Delavenne, Hugo and Gall, François Le and Liu, Yupan and Miyamoto, Masayuki},
  date = {2023-03-03},
  number = {arXiv:2303.01877},
  eprint = {2303.01877},
  eprinttype = {arxiv},
  primaryclass = {quant-ph},
  publisher = {{arXiv}},
  url = {http://arxiv.org/abs/2303.01877},
  urldate = {2023-03-06},
  archiveprefix = {arXiv},
  langid = {english}
}

@misc{MY23_StateQIPStatePSPACE,
  title = {{{stateQIP}} = {{statePSPACE}}},
  author = {Metger, Tony and Yuen, Henry},
  date = {2023-01-18},
  number = {arXiv:2301.07730},
  eprint = {2301.07730},
  eprinttype = {arxiv},
  primaryclass = {quant-ph},
  publisher = {{arXiv}},
  doi = {10.48550/arXiv.2301.07730},
  url = {http://arxiv.org/abs/2301.07730},
  urldate = {2023-03-06},
  archiveprefix = {arXiv}
}

@misc{CMS23_ObliviousTransferZeroKnowledge,
  title = {Oblivious {{Transfer}} from {{Zero-Knowledge Proofs}}, or {{How}} to {{Achieve Round-Optimal Quantum Oblivious Transfer}} and {{Zero-Knowledge Proofs}} on {{Quantum States}}},
  author = {Colisson, Léo and Muguruza, Garazi and Speelman, Florian},
  date = {2023-03-02},
  number = {arXiv:2303.01476},
  eprint = {2303.01476},
  eprinttype = {arxiv},
  primaryclass = {quant-ph},
  publisher = {{arXiv}},
  url = {http://arxiv.org/abs/2303.01476},
  urldate = {2023-03-03},
  archiveprefix = {arXiv},
  note = {To appear in ASIACRYPT 2023}
}

@inproceedings{Unr16_ComputationallyBindingQuantum,
  title = {Computationally {{Binding Quantum Commitments}}},
  booktitle = {Advances in {{Cryptology}} – {{EUROCRYPT}} 2016},
  author = {Unruh, Dominique},
  date = {2016},
  pages = {497--527},
  publisher = {{Springer, Berlin, Heidelberg}},
  doi = {10.1007/978-3-662-49896-5_18},
  url = {https://link.springer.com/chapter/10.1007/978-3-662-49896-5_18},
  urldate = {2022-07-18},
  eventtitle = {Annual {{International Conference}} on the {{Theory}} and {{Applications}} of {{Cryptographic Techniques}}},
  langid = {english}
}

@inproceedings{AQY22_CryptographyPseudorandomQuantum,
  title = {Cryptography from~{{Pseudorandom Quantum States}}},
  booktitle = {Advances in {{Cryptology}} – {{CRYPTO}} 2022},
  author = {Ananth, Prabhanjan and Qian, Luowen and Yuen, Henry},
  editor = {Dodis, Yevgeniy and Shrimpton, Thomas},
  date = {2022},
  series = {Lecture {{Notes}} in {{Computer Science}}},
  pages = {208--236},
  publisher = {{Springer Nature Switzerland}},
  location = {{Cham}},
  doi = {10.1007/978-3-031-15802-5_8},
  isbn = {978-3-031-15802-5},
  langid = {english}
}

  \section{Proofs of statements in preliminaries}
  \textEnd[category=preliminaries]{}
  \printProofs[preliminaries]

  \section{Proofs of security of the bit OT protocol}
  \textEnd[category=bitOT]{}
  \printProofs[bitOT]

  \section{Proof of the ZKoQS and $k$-out-of-$n$ string OT protocols}
  \textEnd[category=zkoqs]{}
  \printProofs[zkoqs]

  \section{Proof of the composability of \cite{Unr15_NonInteractiveZeroKnowledgeProofs}}
  \textEnd[category=unruhComposable]{}
  \printProofs[unruhComposable]
}
  
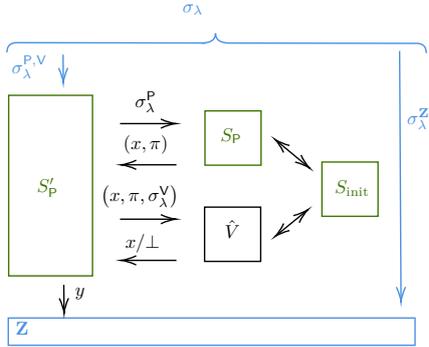
\end{document}